\documentclass[amsmath,amssymb,11pt]{article}
\usepackage{jheppub2}
\pdfoutput=1
\usepackage{graphicx,epsfig}
\usepackage{float}
\usepackage{subcaption}
\usepackage{subfloat}
\usepackage{graphicx}
\usepackage[utf8]{inputenc}
\usepackage{placeins}
\usepackage{hyperref}
\usepackage{caption}
\usepackage{color}
\usepackage{bm}
\usepackage{cleveref}
\usepackage{booktabs}
\usepackage{tabularx}
\usepackage{array}

\newcommand{\dn}{\mathrm{d}}

\newcommand{\im}{\mathbb{I}\mathrm{m}}

\newcommand{\figref}[1]{\hyperref[#1]{Fig.~\ref*{#1}}}

\newcommand*{\affmark}[1][*]{\textsuperscript{#1}}

\definecolor{rRGB}{RGB}{169.2, 155.0, 0.5}

\newcommand{\beq}{\begin{equation}}

\newcommand{\eeq}{\end{equation}}

 \newcommand{\be}{\begin{equation}}
 \newcommand{\ee}{\end{equation}}
 \newcommand{\bea}{\begin{eqnarray}}
 \newcommand{\eea}{\end{eqnarray}}

\usepackage{comment}
\usepackage{tikz}
\usetikzlibrary{positioning}
\usetikzlibrary{intersections}
\usetikzlibrary{fadings} 
\usetikzlibrary{arrows.meta} 
\usetikzlibrary{arrows}

\tikzfading[name=fade out,
inner color=transparent!0,
outer color=transparent!100]

\definecolor{cherryblossompink}{rgb}{1.0, 0.72, 0.77}
\definecolor{lightblue}{rgb}{0.68, 0.85, 0.9}

\usetikzlibrary{decorations.pathmorphing}
\usetikzlibrary{decorations.pathreplacing,decorations.markings}

\usetikzlibrary{backgrounds,automata}

\usepackage{comment}
\usepackage{tikz}
\usetikzlibrary{positioning}
\usetikzlibrary{intersections}
\usetikzlibrary{fadings} 
\usetikzlibrary{arrows.meta} 
\usetikzlibrary{arrows}

\tikzfading[name=fade out,
inner color=transparent!0,
outer color=transparent!100]

\definecolor{cherryblossompink}{rgb}{1.0, 0.72, 0.77}
\definecolor{lightblue}{rgb}{0.68, 0.85, 0.9}

\usetikzlibrary{decorations.pathmorphing}
\usetikzlibrary{decorations.pathreplacing,decorations.markings}

\usetikzlibrary{backgrounds,automata}

\definecolor{rRGB}{RGB}{169.2, 155.0, 0.5}

\title{Quasinormal modes and quantum black hole interiors}

\author{Chiara Coviello,\affmark[1]}
\author{Ansh Gupta,\affmark[1]}
\author{Robie A. Hennigar,\affmark[2]}
\author{Kai Shi,\affmark[1]}
\author{and Andrew Svesko\affmark[3,4]}
 \affiliation{\affmark[1]Department of Physics, King’s College London,
 Strand, London, WC2R 2LS, UK\\
\affmark[2]Centre for Particle Theory, Department of Mathematical Sciences, Durham University, Durham
DH1 3LE, UK\\
 \affmark[3]Department of Mathematics, King’s College London,
 Strand, London, WC2R 2LS, UK\\
 \affmark[4]The Center of Gravity, Niels Bohr Institute, University of Copenhagen\\
 Blegdamsvej 17, DK-2100 Copenhagen Ø, Denmark}

\emailAdd{chiara.coviello@kcl.ac.uk}
\emailAdd{ansh.gupta@kcl.ac.uk}
\emailAdd{robie.a.hennigar@durham.ac.uk}
\emailAdd{kai.shi@kcl.ac.uk}
\emailAdd{Andrew.Svesko@nbi.ku.dk}

\abstract{Quasinormal modes (QNMs) of black hole perturbations in the large overtone limit, namely, asymptotic QNMs, are highly sensitive to a black hole's interior geometry. We probe the singularity structure of a class of exact (2+1)-dimensional quantum black hole solutions to semi-classical gravity by computing their asymptotic QNM spectra. We do this using methods of complex analysis in which the radial coordinate is analytically continued to the complex plane. The leading order QNM frequencies all fit the generic form $\omega=(\text{offset})+n(\text{gap})$, for overtone number $n$, which we also confirm numerically.
 Under a general assumption about the global Stokes topology of the complexified radial coordinate, we show how to reconstruct the scaling exponent of more general (spacelike or timelike) black hole singularities from the offset. We then compute subleading corrections to the asymptotic QNMs, from which we provide a robust method for extracting the scaling of the metric function near the singularity. Our findings exemplify the transition between ``Kasner eons'', successive regimes encountered on approach to a spacelike singularity, as non-perturbative quantum effects become dominant. 
 For charged quantum black holes, the asymptotic QNMs exhibit a crossover between two regimes in which the neutral and charge contributions to the metric function dominate, respectively. Within the crossover region, the QNMs develop an oscillatory behavior that may be a signature of the inner horizon.

}

\begin{document}

\maketitle

\section{Introduction}

Black holes remain important laboratories to explore quantum aspects of gravity. Indeed, quantum black holes have distinct signatures relative to classical black holes. Black hole interiors, in particular, are highly sensitive to 
% become significantly altered due to
quantum effects. The main purpose of this article is to probe the interiors of a class of quantum black holes via their highly damped, i.e., asymptotic, quasinormal mode (QNM) spectra. 

\vspace{2mm}

\noindent \textbf{Probing the interior of a black hole.} 
QNMs characterize the way a black hole vibrates when perturbed. Quasinormal mode frequencies carry unique and intrinsic information about the parameters of a black hole which, during the ringdown phase of a black hole collision, can be directly detected via gravitational wave spectroscopy. This makes QNMs valuable probes to test theories beyond classical general relativity, cf. \cite{Blazquez-Salcedo:2016enn,Cano:2020cao,Pierini:2021jxd,Pierini:2022eim,Moura:2021eln,Berti:2025hly}. Further, within the QNM spectrum of a black hole lies information about its structure inside its horizon. In this way, QNMs also provide, at least indirectly, a means to probe the interior geometry of black holes. This follows from the fact that highly damped modes, also known as asymptotic QNMs, are sensitive to the small scale structure of the (analytically continued) spacetime. 

To wit, the propagation of QNMs of a spherically symmetric black hole essentially reduces to an effective one-dimensional Schr{\"o}dinger-like master equation that schematically goes like
\beq \frac{\dn^{2}R}{\dn r_{\ast}^{2}}+(\omega^{2}-V(r))R=0\;.\label{eq:1deomintro}\eeq
Here $R$ is a one-dimensional radial wavefunction, $r_{\ast}$ is a (radial) tortoise coordinate, $\omega$ is the QNM frequency, and $V(r)$ is an effective potential dependent on the spacetime geometry and probe field type. Low-lying QNMs are primarily determined by a peak in the potential outside of the horizon, and are therefore fairly insensitive to a black hole's interior. For asymptotic QNMs, by contrast, where the overtone number is large relative to other relevant scales, the potential $V(r)$ is dominated by its behavior near the origin $r=0$. Thus, though all QNMs are, by definition, vibrational modes exterior to the horizon, when highly-damped they are nonetheless sensitive to the interior structure of a black hole. This has made asymptotic QNMs preferential probes of black hole interiors \cite{Das:2004db, Ghosh:2005aq, Babb:2011ga,Lan:2022qbb,Konoplya:2022hll,Grozdanov:2026ktq,Grozdanov:2026lnc}. 

In holography, this connection is sharpened by the identification of bulk QNM frequencies with poles of retarded thermal correlation functions in the boundary theory. Earlier work showed that near-singularity Kasner exponents govern nonanalytic corrections to analytically continued thermal correlators~\cite{Frenkel:2020ysx}, motivating the search for corresponding signatures of the singularity geometry in the large-overtone QNM spectrum. This sensitivity to the deep interior is especially relevant when considering quantum corrections, which may remain perturbative in the exterior while becoming much more pronounced behind the horizon. Asymptotic QNMs therefore offer a natural means of diagnosing differences between quantum-corrected and classical black hole geometries.

\vspace{2mm}

\noindent \textbf{Quantum black holes.} In this article, by ``quantum'' black holes, we mean black hole solutions to semi-classical gravity, partially governed by the semi-classical Einstein equations,
\beq G_{ab}(g)+\Lambda g_{ab}=8\pi G_{N}\langle T_{ab}^{\text{mat}}\rangle\;.\label{eq:semieineq}\eeq
The left-hand side is the classical geometric part of the Einstein equations for a spacetime with metric $g_{ab}$ and cosmological constant $\Lambda$, while on the right-hand side $\langle T_{ab}^{\text{mat}}\rangle$ is the expectation value of the (renormalized)  quantum matter stress tensor in an appropriate quantum state. Semi-classical gravity should be viewed as an effective theory, invalid near the Planck scale when quantum gravitational effects become relevant.  Even above the Planck scale, 
since metric and matter fluctuations enter at the same order, it is generally inconsistent to treat the background geometry as purely classical~\cite{Ford:1982wu}. 
Still, in an appropriate regime of validity, uncovering quantum black hole solutions in semi-classical gravity remains an open problem. 

The technical challenge is due to backreaction: how quantum matter modifies the classical background geometry and vice versa. A straightforward approach is to treat backreaction perturbatively. Doing so provides an iterative procedure, whereby one first computes the renormalized quantum stress-tensor in a fixed classical background, and then perturbatively solves for quantum matter corrections to the geometry via  (\ref{eq:semieineq}). Typically, this procedure cannot be carried out beyond the leading quantum corrections, cf. \cite{York:1983zb,York:1984wp,Hochberg:1992rd,Hochberg:1992xt,Anderson:1994hh,Souradeep:1992ia,Steif:1993zv,Matschull:1998rv,Casals:2016ioo,Casals:2019jfo}. Further, this method only captures perturbative quantum effects. \emph{Exact} quantum black holes, solutions that incorporate non-perturbative quantum effects, are thus hard and rare to uncover. 

\vspace{2mm}

\noindent \textbf{Probing quantum black hole interiors: the need for non-perturbative solutions.} Probing the interior of any black hole requires an exact solution. To see why knowing only the leading perturbative corrections to a classical solution is insufficient, consider a simple example involving the four-dimensional Schwarzschild black hole. Suppose that the two leading corrections to the metric function can be written in the form 
\be 
f(r) = 1 - \frac{2 M }{r} + \frac{\alpha_1}{r^{n_1}} + \frac{\alpha_2}{r^{n_2}} + \cdots
\label{eq:blackfactintro}\ee
where $\alpha_1$ and $\alpha_2$ are dimensionful constants, and $n_1$ and $n_2$ are integers satisfying $n_2>n_1>1$. If these two corrections arise due to the same underlying physics, then it is natural to re-express $\alpha_i$ in terms of the length scale which governs that new physics. We denote this scale $\ell$. Let us therefore write $\alpha_i = \lambda_i \, \ell^{n_i}$ where $\lambda_i$ are dimensionless. Assume $M\gg\ell$ and dimensionless coefficients $\lambda_i$ of order unity. The corrections are then small compared with the Schwarzschild term for $r\gg\ell$. Their hierarchy, however, is lost when successive corrections become comparable in magnitude, at
\be 
r_c=\ell\left|\frac{\lambda_2}{\lambda_1}\right|^{1/(n_2-n_1)}
\sim\ell.
\ee
At this radius, the corrections are still parametrically smaller than the Schwarzschild term. The first correction would become comparable to $2M/r$ only at
\be 
r_1 = \ell \left(\frac{|\lambda_1| \ell}{2M}\right)^{1/(n_{1}-1)} \ll \ell \, .
\ee
But at this radius, the second correction term is already larger than the first! Thus, without an additional hierarchy among the $\lambda_i$, retaining only the leading correction does not provide a controlled description of the geometry in the region where it would modify the singularity. 

% By requiring $M \gg \ell$, it is possible to realize a separation of scales that makes the Schwarzschild $2 M/r$ behavior dominant over a parametrically large range of $r$,
% \be 
% r > \ell \left(\frac{|\lambda_1| \ell}{2M}\right)^{1/(n_{1}-1)} \, .
% \ee
% However, sufficiently deep in the black hole interior the corrections will become important and dominate over the Schwarzschild behavior. Generically, however, once the first such correction becomes relevant, so will the entire tower of corrections. To see this, examine the radial scale at which the two leading corrections written above become comparable, 
% \be 
% r_c = \ell \left(\frac{\lambda_2}{\lambda_1}\right)^{\frac{1}{n_2 - n_1}} \, .
% \ee
% At this radius, the corrections may still be small relative to the Schwarzschild term. However, for $r \sim r_c$

% Up to a numerical factor (which in general will be of order unity), this is the same scale $r \sim \ell$ at which the first correction becomes important. Hence, unless $\lambda_i$ are disparate in magnitude, a separation of scales between individual corrections controlled by~$\ell$ cannot be realized.

The asymptotic QNM spectra we study here will probe the full black hole geometry. Their frequencies are fixed by the \textit{exact} horizon structure, asymptotics, and ---crucially--- the singularity.  Hence, to understand the implications of quantum effects on the asymptotic QNM spectrum, we require complete non-perturbative quantum black hole solutions.

\vspace{2mm}

\noindent \textbf{Exact quantum black holes at the end-of-the-world brane.} Fortunately, the framework of braneworld holography \cite{deHaro:2000wj} grants us a context in which the semi-classical approximation can be consistently realized, and non-perturbative quantum black holes can be explicitly constructed. Briefly, 
in this context, a codimension-1 end-of-the-world (ETW) brane is embedded into a bulk $(d+1)$-dimensional anti-de Sitter (AdS) spacetime that is taken to have a dual description in terms of a $d$-dimensional conformal field theory (CFT) via the AdS/CFT correspondence. The brane boundary conditions are such that the brane geometry is dynamical; from the brane perspective, the theory is interpreted as an effective semi-classical theory of gravity with higher-derivative corrections incorporating backreaction effects due to the CFT living on the brane. This interpretation is valid in the  planar limit of the CFT at large central charge $c$. In this limit the bulk is described by Einstein gravity, while the semi-classical brane gravity theory is consistently understood as an asymptotic expansion in $cG_{N}\hbar$, with metric fluctuations suppressed relative to matter loops as $G_{N}\hbar\to0$.

Via this set-up, it was conjectured that quantum black holes can be exactly constructed by searching for \emph{classical} bulk black holes that localize on ETW branes \cite{Emparan:2002px}.  Notably, the problem of explicitly solving the semi-classical gravitational equations of motion is traded for uncovering braneworld black holes, solutions of the classical bulk gravitational equations  obeying braneworld boundary conditions. Such constructions explicitly exist in $(2+1)$ dimensions \cite{Emparan:1999wa,Emparan:1999fd,Emparan:2020znc,Emparan:2022ijy,Panella:2023lsi,Climent:2024nuj,Feng:2024uia,Climent:2024wol,Bhattacharya:2025tdn,Cao:2026jls} and, in principle, higher-dimensional spacetimes. See \cite{Panella:2024sor} for a modern review.
%\footnote{A non-holographic setting in which quantum black holes can be consistently and exactly uncovered are in  models of $(1+1)$-dimensional dilaton-gravity coupled to conformal matter, e.g., \cite{Christensen:1977jc,Callan:1992rs,Russo:1992ax,Bose:1995pz,Fabbri:1995bz,Fabbri:2005mw}.}

The power of braneworld holography is that it grants us access to full-fledged non-perturbative solutions through the bulk classical theory. In particular, in this work we will consider perturbations to charged and neutral $(2+1)$-dimensional black holes that have blackening factors of
the type (\ref{eq:blackfactintro}) where the number of corrections to the classical solution is finite. This is because the bulk classical gravity performs a resummation of the infinite tower of higher derivative terms resulting in a complete non-perturbative solution. 

\vspace{2mm}

\noindent \textbf{Main results.} In this article we compute the large-overtone limit of QNMs of exact charged and neutral quantum black holes in (2+1)-dimensional AdS and neutral black holes in (2+1)-dimensional flat space. (QNMs of quantum-corrected black holes have been previously explored in 
\cite{Babb:2011ga,Gong:2023ghh,Cartwright:2024iwc,Cartwright:2025fay}.) We do this using analytic methods originally developed in \cite{Motl:2002hd,Motl:2003cd} (see also \cite{Musiri:2003rs,Cardoso:2004up,Natario:2004jd}) where the radial coordinate is analytically continued to the complex plane. Doing so allows for the use of powerful techniques in complex analysis, namely, Stokes line matching and the monodromy theorem. This will allow us to match the behavior of solutions to the wave equation (\ref{eq:1deomintro}) near the curvature singularity, the horizon, and asymptotic infinity, leading to the QNM quantization condition. Specifically, the contributions to the QNM frequency from the leading order behavior of the effective potential are
\begin{equation}
\omega = 
\begin{cases}
\frac{\pi n}{\xi_{0}}+\frac{i\log 2}{4\xi_{0}}\;, \quad&\text{neutral}\;\;\text{AdS}_{3}\;,\\
\frac{\pi n}{\xi_{0}}\;,&\text{charged}\;\;\text{AdS}_{3}\;,\\
- i \kappa_h \left(n - \frac{1}{2} \right) \;,&\text{neutral}\;\;\text{flat}\;,
\end{cases}
\end{equation}
%\begin{align}
%\omega&=\frac{\pi n}{\xi_{0}}+\frac{i\log(2)}{4\xi_{0}}\;,\quad\, \text{neutral}\;\;\text{AdS}_{3}\;,\nonumber\\
%\omega&=\frac{\pi n}{\xi_{0}}\;, \qquad \qquad\;\;\;\;\;\;\,\text{charged}\;\;\text{AdS}_{3}\;,\\
%\omega&=\frac{i\kappa}{2}(1-2n)\;, \quad \;\;\;\;\;\,\text{neutral}\;\;\text{flat} \nonumber\;,
%\end{align}
where $n =1,2,\dots$ is the overtone number, $\xi_{0}$ is the value of the tortoise coordinate at the AdS$_{3}$ boundary, and $\kappa_h$ is the horizon surface gravity. These results are expected to provide good agreement when $n$ is large. Collectively, the frequencies all fit the generic form $\omega=(\text{offset})+n(\text{gap})$, where the gap is a function of the relaxation time.  Evidently the neutral AdS$_{3}$ and neutral flat cases have an offset while the charged AdS$_{3}$ does not, in contrast with its classical higher-dimensional counterpart. 
%Meanwhile, the asymptotic QNMs for the neutral flat black hole are purely imaginary, which, notably, cannot be recovered in any continuous limit from the AdS$_{3}$ system.
Moreover, the strength of quantum effects, encoded in the length scale $\ell$, is implicit in $\xi_{0}$ or $\kappa$. Notably, we find that, in the limit where this scale becomes large, i.e., when non-perturbative quantum backreaction effects are  relevant, the leading large-overtone spacings of the  brane and bulk black holes are equal. Thus, in this limit, the large overtone modes of the braneworld black hole are sensitive to the presence of higher dimensions.

The asymptotic spectra also allow us to address an inverse problem: what information about the black hole interior can be recovered from its QNMs? We address this by turning our attention to the subleading corrections to the asymptotic frequencies. We first show that, under an assumption on Stokes topology, a scaling exponent of the singularity is encoded in the offset of the asymptotic spectrum. We then compute further subleading corrections to the spectrum and show how the leading correction due to the angular momentum of the scalar field provides a natural way to read off the scaling of the singularity in a manner that does not rely on the Stokes topology. While the computation is technical, its essence is simple and based on a scaling argument that has considerably larger scope than the context in which we apply it here. A subleading term $\delta V\propto \xi^{q-2}$ in the near-singularity potential, with $q>0$ and $\xi$ the (complex) tortoise coordinate, enters the radial equation with strength $\omega^{-q}$. Since the leading asymptotic frequencies grow linearly with $n$, a nonvanishing first-order contribution therefore gives $\delta\omega\propto n^{-q}$. For metrics with a single blackening factor and near-singularity behavior $f(r) \sim r^{-s}$ with $s > 0$, we find that the leading corrections due to angular momentum $m$ scale as $m^2 \, n^{-s/(s+1)}$.  Consequently, comparing modes at the same overtone number and different angular quantum numbers gives
\begin{equation}
\Delta_n
\equiv \omega_{m_1,n}-\omega_{m_2,n}
\propto (m_1^2-m_2^2)\, n^{-s/(s+1)}.
\end{equation}
This subtraction cancels the common spacing, offset, and angular-momentum-independent corrections and therefore allows one to determine the scaling exponent $s$ by inverting the above. In the case where the singularity is spacelike, we show how this process can be used to determine Kasner exponents.

We compute the leading metric and angular momentum corrections analytically for neutral and charged quantum BTZ (qBTZ) black holes. The angular dependence also allows us to track transitions between different scaling regimes inside the black hole. For neutral qBTZ at weak backreaction, the interior contains an extended BTZ-like region before approaching a quantum-dominated Kasner singularity. The corresponding Kasner exponents change from $(p_t,p_\phi)=(0,1)$ to $(-1/3,2/3)$. Our numerical spectra exhibit the associated crossover in the effective exponent extracted from $\Delta_n$, providing a spectral diagnostic of the transition between ``Kasner eons''~\cite{Bueno:2024fzg}. For charged qBTZ, we similarly estimate the crossover between regimes controlled by the neutral and charge contributions to the metric. The charged asymptotic regime can be postponed to very high overtones when the charge is small. We also observe an oscillatory approach to the charged asymptotic spectrum, which complicates the extraction of subleading coefficients at finite overtone number. These oscillations may reflect a contribution associated with the inner horizon.

% \textcolor{blue}{
% \sout{Given that the asymptotic QNM frequencies of neutral and charged AdS black holes differ by an offset, we confirm that, under a general assumption about the Stokes topology, the offset is sensitive to the black hole singularity structure. In fact, from the offset to the leading order asymptotic QNM spectra, we show how to reconstruct the scaling exponent of the singularity (cf. (\ref{eq:qnm_quant_leading})). By further computing subleading corrections to the asymptotic QNMs, we provide a robust method, i.e., without relying on any assumption about the global Stokes topology, for extracting the scaling of
% the metric function in the vicinity of the singularity. Our findings exemplify the transition between ``Kasner eons'' \cite{Bueno:2024fzg}, the period in which non-perturbative quantum effects (captured by higher-derivative terms in the induced brane theory) become dominant on approach to a spacelike singularity. } }

\vspace{2mm}

\noindent \textbf{Outline.} The remainder of this article is organized as follows. In Section \ref{sec:3Dqbhs}, we review the family of quantum black holes arising in the context of braneworld holography. In Section \ref{sec:scalar-qbh-brane}, we formulate the equation of motion for a massless scalar perturbation on a three-dimensional quantum black hole and discuss its connection to the four-dimensional bulk problem in the tensionless limit. In Section \ref{sec:asympqnms}, we analytically compute the asymptotic QNMs of various quantum black holes ---neutral and charged qBTZ black holes, as well as flat neutral quantum black holes--- by continuing the radial coordinate into the complex plane and applying Stokes line and monodromy matching techniques.
Section \ref{sec:lookinqBH} is devoted to understanding to what extent the singularity structure can be reconstructed from an asymptotic QNM spectrum, and expands on \cite{Coviello:2026tol}. In Section \ref{sec:numericmethods}, we obtain the QNMs numerically, using pseudospectral and Leaver-inspired methods, and confirm that the numerical results agree with the analytical asymptotic predictions in the appropriate large-overtone regime.
Our results culminate in Section \ref{sec:eondetection}, where we use the asymptotic QNM spectra to detect transitions between Kasner eons inside the quantum BTZ  black hole.
Finally, Section \ref{sec:disc} contains  our concluding remarks and a brief outlook. Appendices are included to provide additional technical details. 

\paragraph{Note added.}  While this work was being completed, Refs.~\cite{Hartnoll:2026vhu,Xiao:2026pir} appeared with results overlapping with aspects of Section~\ref{sec:lookinqBH}. Hartnoll and Zhiboedov~\cite{Hartnoll:2026vhu} showed how near-singularity Kasner data can be extracted from the momentum dependence of subleading large-overtone QNM corrections, and how transitions between Kasner epochs leave signatures in the overtone dependence. Xiao, Nian and Li~\cite{Xiao:2026pir} studied how Kasner exponents can be reconstructed using distinct mass- and momentum-dependent spectral corrections, allowing ``temporal'' and spatial Kasner exponents to be determined independently. Our overlapping results were obtained independently. Our findings also show how momentum-dependent corrections can be used to extract scaling exponents of singularities. We apply our findings to spacelike Kasner singularities as well as timelike singularities.

\section{Three-dimensional quantum black holes: lightning review} \label{sec:3Dqbhs}

Here we very briefly review the holographic construction of quantum black holes via braneworld holography. All known exact constructions of three-dimensional quantum black holes are understood to correspond to specific asymptotically AdS$_{4}$ black holes that localize on an end-of-the-world brane \cite{Emparan:1999wa,Emparan:1999fd,Emparan:2002px}. 

\subsection{Set-up: bulk and brane}

There are two perspectives relevant for our analysis: the classical `bulk' gravitational theory, and the induced semi-classical brane perspective. We take the bulk to be characterized by Einstein-Maxwell-AdS$_{4}$ gravity, 
\beq I=\frac{1}{16\pi G_{4}}\int \dn^{4}x\sqrt{-\hat{g}}\left[\hat{R}+\frac{6}{\ell_{4}^{2}}-\frac{\ell_{\star}^{2}}{4}\hat{F}^{2}\right]\;, \quad \ell^{2}_{\star}=\frac{16\pi G_{4}}{g_{\star}^{2}}\;.\eeq
Here  $\hat{g}_{ab}$ denotes the bulk asymptotically AdS$_{4}$ metric with Ricci scalar $\hat{R}$, $G_{4}$ is the four-dimensional Newton's constant, $\hat{F}_{ab}=\partial_{a}A_{b}-\partial_{b}A_{a}$ is the bulk Maxwell field strength tensor, $\ell_{\star}$ is a coupling constant with dimensions of length, $g_{\star}$ is the dimensionless gauge coupling constant, and $\ell_{4}$ is the AdS$_{4}$ radius.
% For the theory to have a well-posed variational problem, we append the bulk action with a Gibbons-Hawking-York boundary term (suitable for Dirichlet boundary conditions), though this detail will be irrelevant for our purposes. 
We further minimally couple the bulk theory to a codimension-1 ETW brane $\mathcal{B}$ of tension $\tau$ characterized by
\beq I_{\text{brane}}=-\tau\int_{\mathcal{B}}\dn^{3}x\sqrt{-h}\;,\label{eq:genbraneact}\eeq
for induced brane metric $h_{ij}$.

The ETW brane excises a portion of the AdS$_{4}$ bulk, including its conformal boundary. Throughout, we shall work with a $\mathbb{Z}_{2}$-construction in which the space is surgically completed by introducing a second copy of the AdS$_{4}$ bulk plus brane geometry and suturing along the common $\mathcal{B}$. This leads to a jump discontinuity in the extrinsic curvature $K_{ij}$ across the brane which the Israel junction conditions~\cite{Israel:1966rt} relate to the brane tension
\beq \Delta K_{ij}-h_{ij}\Delta K=-8\pi G_{4}S_{ij}=8\pi G_{4}\tau h_{ij}\;.\label{eq:israeljuncconds}\eeq
Here $\Delta K_{ij}=K^{+}_{ij}-K_{ij}^{-}$ denotes the difference between the extrinsic curvatures on the  `$+$' and `$-$' sides of the brane,\footnote{Here we take $K_{ij}^{+}=-K_{ij}^{-}$ such that $\Delta K_{ij}=2K_{ij}$.} and $S_{ij}$ is the brane stress-tensor; for a purely tensional brane (\ref{eq:genbraneact}) one has $S_{ij}\equiv-\frac{2}{\sqrt{-h}}\frac{\delta I_{\text{brane}}}{\delta h^{ij}}=-\tau h_{ij}$.  
The bulk Maxwell field strength must also obey junction conditions \cite{Lemos:2021jtm}; however, we will not need the explicit form of these conditions here.

From the brane perspective, the relevant physics is captured by the Karch--Randall version of braneworld holography \cite{Karch:2000ct}. In this framework, the bulk geometry is entirely classical but integrating out the bulk region induces a three-dimensional theory of gravity coupled to a cutoff CFT$_3$. Schematically, the brane effective action takes the form
\begin{equation}
    I_{\rm eff}
    =
    I_{\rm Bgrav}[h_{ij},A_i]
    +
    I_{\rm CFT}[h_{ij},A_i],
    \label{eq:genactind-new}
\end{equation}
where $h_{ij}$ is the induced metric on the brane and $A_i$ is the pullback of the bulk Maxwell potential. The induced gravitational action $I_{\rm Bgrav}$ is intricate, consisting of the Einstein-Hilbert action supplemented by an infinite tower of higher-derivative corrections including non-minimal couplings to the gauge sector. We shall not need the explicit form of this induced theory, and therefore refer the reader to, e.g.,~\cite{Climent:2024nuj, Bueno:2023dpl} for additional details on the brane actions. The important point is that a classical bulk solution whose horizon intersects the brane gives an exact solution of this induced semi-classical theory. Finally, let us note that the three-dimensional couplings induced by the four-dimensional bulk theory are
\begin{equation}
    G_3
    =
    \frac{G_4}{2\ell_4},
    \label{eq:effGd-new}
    \qquad
    \frac{1}{L_3^2}
    =
    \frac{2}{\ell_4^2}
    \left(
        1-2\pi G_4\ell_4\tau
    \right) \,,  
    \qquad 
    \tilde \ell_\star^2 =
    \frac54\,\ell_\star^2,
    \qquad
    \tilde g_\star^2
    =
    \frac{2}{5}\,\frac{g_\star^2}{\ell_4}.
\end{equation}
The particular value of the brane tension for which $L_3 \to \infty$ shall be referred to below as the critical tension, 
\be
    \tau_{\rm crit} \equiv \frac{1}{2\pi G_4\ell_4} \, .
    \label{eq:critical_tension}
\ee

\vspace{2mm}

\noindent \textbf{AdS$_{4}$ C-metric.} To leverage braneworld holography to describe quantum black holes first requires AdS$_{4}$ black hole solutions that localize on the ETW brane. In general this is a difficult problem in classical gravity, but luckily exact solutions with this property are known. The relevant solution to the Einstein-Maxwell field equations is the  AdS$_{4}$ C-metric, which (in the absence of the ETW brane) may be interpreted as a black hole undergoing uniform acceleration~\cite{Griffiths:2006tk}. The metric and gauge field for a static and electrically charged AdS$_{4}$ C-metric are 
\beq
\begin{split}
 \dn s^2 &= \frac{\ell^2}{(\ell + xr)^2} \biggr[  -H(r) \dn t^2 + H^{-1}(r) \dn r^2+ r^2 \left(G^{-1}(x)\dn x^2 + G(x) \dn \phi^2  \right) \biggr] \ ,
 \\ 
 A&=A_{b}\dn x^{b}=-\frac{2}{\ell_{\star}}\frac{\ell q}{r}\dn t\;,\
 \end{split}
\label{eq:rotatingCmetqbtz}
\eeq
with field strength $\hat{F}=\dn A$.
Here $(t,r)$ are time and radial coordinates, and the ranges of the angular coordinates $(x, \phi)$ will be clarified momentarily. Notably, the boundary is located where the conformal factor in the metric (\ref{eq:rotatingCmetqbtz}) diverges, i.e., at $xr=-\ell$.

%The precise form of the metric functions $H(r)$ and $G(x)$ depend to the 
The metric functions will generically take the form
\beq
\begin{split}
    &H(r)= \frac{r^2}{\ell_3^2}+ \kappa -\frac{\mu \ell}{r}+\frac{q^{2}\ell^{2}}{r^{2}}  \ , \quad G(x)= 1-\kappa x^2-\mu x^3-q^{2}x^{4}\;,
\end{split}
\label{eq:metfuncsrotqbtz}\eeq
for dimensionless mass and charge parameters $\mu\geq0$ and $q$, respectively, and with the length scales $\ell_{3}$ and $\ell$ related to the AdS$_{4}$ length scale via
\beq \frac{1}{\ell_{4}^{2}}=\frac{1}{\ell_{3}^{2}}+\frac{1}{\ell^{2}}\;.\label{eq:AdS4lenghscale}\eeq
In the absence of a brane, the parameter $\ell$ is inversely proportional to the acceleration of the black hole. Finally, $\kappa=\pm1,0$ is a discrete parameter.  

Zeroes of $H(r)$ describe horizons. For $\mu\neq0$ and non-vanishing charge, the metric (\ref{eq:rotatingCmetqbtz}) has inner and outer black hole horizons, $r_{-}$ and $r_{+}$, respectively, satisfying $H(r_{\pm})=0$ and $0<r_{-}<r_{+}$. The geometry has an extremal limit when the inner and outer horizons coincide, $r_{-}=r_{+}$. Since the C-metric is accelerating, there can also be an acceleration horizon. When present, this is located at negative values of $r$. In this article we will consider scenarios where the acceleration horizon can be ignored.

% in general there will also be an acceleration horizon. Depending on the brane geometry, this horizon can be effectively removed. We will comment on this briefly below, though in this article we will consider scenarios where the acceleration horizon can be ignored.

On the other hand, zeroes $\{ x_i\}$ of $G(x)$ describe axes where $\phi$ degenerates. These are generically conical singularities. One of these conical singularities can be removed by imposing regularity on the azimuthal coordinate $\phi$ to ensure smoothness of the geometry along the axis of rotational symmetry, 
\beq \phi\sim \phi+\Delta\phi\;,\quad \Delta\phi=\frac{4\pi}{|G'(x_{i})|}=4\pi\biggr|-2\kappa x_{i}-3\mu x_{i}^{2}-4q^{2}x_{i}^{3}\biggr|^{-1}\;.\label{eq:azimuthid}\eeq
We denote the smallest positive root as $x=x_{1}$ and fix the periodicity such that the conical singularity at $x_{i}=x_{1}$ is removed, leaving conical singularities at $x_{i}\neq x_{1}$. The restricted region $0\leq x\leq x_{1}$ is therefore regular, free of conical singularities. 
%Clearly,  $x_{1}$ is a function of $\mu,q$ and $a$. 
It proves to be convenient to view $\mu$ as a `derived' parameter by solving $G(x_{1})=0$, 
\beq \mu=\frac{1}{x_{1}^{3}}\left[-q^{2}x_{1}^{4}-\kappa x_{1}^{2}+1\right]\;,\label{eq:muderiv}\eeq
which we take to be non-negative. 

\vspace{2mm}

\noindent \textbf{Umbilic surfaces and ETW branes.} A crucial geometric feature of the C-metric (\ref{eq:rotatingCmetqbtz}) is that the hypersurface $x=0$ is totally umbilic, i.e., the extrinsic curvature $K_{ij}$ is proportional to the induced metric at $x=0$; specifically $K_{ij}=-\ell^{-1}h_{ij}$. Positioning a brane $\mathcal{B}$ at $x=0$ guarantees the junction conditions (\ref{eq:israeljuncconds}) are satisfied, and the brane has the uniform tension
\beq \tau=\frac{1}{2\pi G_{4}\ell}\;.\label{eq:branetengen}\eeq
There are two limits worth emphasizing. First, in the infinite tension/acceleration limit, $\ell\to0$, the brane approaches the AdS$_{4}$ conformal boundary and $\ell_{4}\to\ell$. In this limit,
% , upon a double-Wick rotation, 
the geometry at the boundary is a (non-dynamical)  charged defect in AdS$_{3}$.\footnote{These geometries can be obtained as the boundary geometries of solutions obtained by performing a double Wick rotation of the  Schwarzschild/Reissner-Nordstr\"om-AdS black holes. For vanishing charge the boundary is the rotating BTZ black hole \cite{Hubeny:2009rc} (see Appendix B of \cite{Emparan:2020znc} for the appropriate double-Wick rotation), while for vanishing rotation one has a charged defect in AdS$_{3}$ \cite{Climent:2024nuj}.} 
In the opposite limit, $\ell\to\infty$, the brane becomes tensionless, and the C-metric reduces to the standard Reissner-Nordstr\"om-AdS solution. We will revisit this below.

Treating the $x=0$ hypersurface as an ETW brane, the region $x<0$ is cut off from the rest of the AdS$_{4}$ bulk spacetime.  The space can be surgically completed by introducing a second copy of the remaining $0\leq x\leq x_{1}$ region, glued along the common $x=0$ hypersurface. This procedure results in a $\mathbb{Z}_{2}$-symmetric double-sided braneworld \cite{Randall:1999vf,Karch:2000ct} free of conical singularities.

\subsection{Quantum black hole zoo}
\label{subsec:quantum-black-hole-zoo}

Thus far we have described the bulk AdS$_4$ C-metric and branes therein. We now consider the description of physics on the brane and reinterpret these geometries from the intrinsic three-dimensional point of view. The result is a small zoo of exact quantum black holes. The word ``quantum'' refers to the fact that, on the brane, these metrics solve an induced semi-classical theory of gravity coupled to a large-$c$ CFT, rather than the classical three-dimensional Einstein equations.

\vspace{2mm}
\noindent \textbf{Localized black holes on the brane.} Consider the induced geometry on the $x = 0$ brane, which reads
\beq 
\dn s^{2}|_{\mathcal{B}}=-H(r)\dn t^{2}+H^{-1}(r)\dn r^{2}+r^{2}\dn \phi^{2}\;,
\label{eq:naivemetrot}
\eeq
where the metric function $H(r)$ was given in Eq.~(\ref{eq:metfuncsrotqbtz}). The gauge field at $x=0$ is\footnote{Note, however, that the Maxwell gauge field does not localize on the brane in the same way as gravity --- see \cite{Climent:2024nuj} for an elaboration on this point. }
\beq 
A_{\mu}\dn x^{\mu}|_{\mathcal{B}}=-\frac{2}{\ell_{\star}}\frac{\ell q}{r}\dn t\;.
\eeq

Due to the identification (\ref{eq:azimuthid}), the metric (\ref{eq:naivemetrot}) is not yet cast in canonically normalized coordinates. For the azimuthal coordinate to have the standard $2\pi$-periodicity, introduce normalized coordinates $(\bar{t},\bar{r},\bar{\phi})$ 
\beq 
t=\eta \,  \bar{t}\;,\quad r=\frac{\bar{r}}{\eta}\;,\quad \phi=\eta\, \bar{\phi}\;,
\label{eq:cancoords}
\eeq
where $\eta\equiv \Delta\phi/(2\pi)$.
In these coordinates, the induced metric on the brane is 
\beq 
\begin{split}
\dn s^{2}|_{\mathcal{B}}&=-H(\bar{r})\dn \bar{t}^{2}+H^{-1}(\bar{r})\dn \bar{r}^{2}+\bar{r}^{2}\dn \bar{\phi}^{2}\;, \quad H(\bar{r})=\frac{\bar{r}^{2}}{\ell_{3}^{2}}+\kappa \eta^{2}-\frac{\ell}{\bar{r}}\mu\eta^{3}+\frac{\ell^{2}}{\bar{r}^{2}}q^{2}\eta^{4}\;,
\end{split}
\label{eq:cannormmet}\eeq
while the gauge potential becomes
\beq
\begin{split} 
\bar{A}_{b}\dn \bar{x}^{b}&=-\frac{2q\ell\eta^{2}}{\ell_{\star}\bar{r}}\dn \bar{t}\;.
\end{split}
\label{eq:Agaugbarred}\eeq

Going forward, each of the metrics we consider will have the form
\begin{equation}
    \dn s_{\mathcal B}^2
    =
    -f(r)\dn t^2
    +
    \frac{\dn r^2}{f(r)}
    +
    r^2\dn \phi^2,
    \qquad
    \phi\sim\phi+2\pi,
    \label{eq:brane-static-zoo}
\end{equation}
with different blackening factors $f(r)$. We also henceforth drop the bars on the canonical coordinates.

\vspace{2mm}

\noindent \textbf{Quantum BTZ black holes.} Consider the charged solution described above. We can write the brane metric in the more conventional form 
    \begin{equation}
    f_{\rm cqBTZ}(r)
    =
    \frac{r^2}{\ell_3^2}
    -
    8\mathcal G_3M
    -
    \frac{\ell W(M,Q)}{r}
    +
    \frac{\ell^2 P(M,Q)}{r^2},
    \label{eq:charged-qbtz-mass-form-zoo}
\end{equation}
where
\begin{equation}
    8\mathcal G_3M=-\kappa\eta^2,
    \qquad
    W(M,Q)=\mu\eta^3,
    \qquad
    P(M,Q)=q^2\eta^4 \, ,
    \label{eq:charged-qbtz-functions-zoo}
\end{equation}
and we note that $W(M, Q)$ and $P(M, Q)$ may in some cases require an additional branch specification~\cite{Emparan:2020znc, Climent:2024nuj}. Here $M$ is the mass, while $\mathcal G_3$ is the `renormalized' three-dimensional Newton constant obtained by incorporating the higher-derivative corrections from the brane,
\begin{equation}
        \mathcal G_3
    =
    \frac{G_4}{2\ell}
    =
    \frac{\ell_4}{\ell}G_3 \, .
\end{equation}
The electric charge is most easily computed from the AdS$_4$ bulk, 
\begin{equation}
    Q
    =
    \frac{2}{g_\star^2}
    \int \star \hat F
    =
    \frac{8\pi\eta qx_1\ell}{g_\star^2\ell_\star}\;,
    \label{eq:charged-qbtz-physical-charge-zoo}
\end{equation}
while the corresponding electric potential, in a gauge where $A_t$ vanishes at infinity, is
\begin{equation}
    \Phi
    =
    - A_t(r_+)
    =
    \frac{2q\ell\eta^2}{\ell_\star r_+}.
    \label{eq:charged-qbtz-potential-zoo}
\end{equation}

The description provided by braneworld holography is reliable when the Einstein-Hilbert term of the induced brane theory dominates over the higher-derivative terms. In practical terms, this means that the brane lies close to the would-be AdS boundary. In this limit, $\ell\sim \ell_4\ll \ell_3$, the effective AdS$_3$ radius obeys
\begin{equation}
    \frac{1}{L_3^2}
    =
    \frac{1}{\ell_3^2}
    \left[
        1
        +
        \frac{\ell^2}{4\ell_3^2}
        +
        O\!\left(
            \frac{\ell^4}{\ell_3^4}
        \right)
    \right].
    \label{eq:L3sqrel-new}
\end{equation}
Thus $L_3\simeq \ell_3$ to leading order. The central charge of the cutoff CFT$_3$ may be written as
\begin{equation}
    c_3
    =
    \frac{\ell_4^2}{G_4}
    =
    \frac{\ell}{2G_3\sqrt{1+\ell^2/\ell_3^2}},
\end{equation}
so that
\begin{equation}
    2c_3G_3
    =
    \ell_4
    =
    \frac{\ell}{\sqrt{1+\ell^2/\ell_3^2}}
    =
    \ell
    \left[
        1
        -
        \frac12\frac{\ell^2}{\ell_3^2}
        +
        O\!\left(
            \frac{\ell^4}{\ell_3^4}
        \right)
    \right].
    \label{eq:c3G3l-new}
\end{equation}
The semi-classical expansion on the brane is therefore an expansion in small but finite $c_3G_3\sim \ell$, with $c_3$ large. At the same time, the three-dimensional Planck length behaves as $L_P = \hbar G_3$. By taking $c_3$ to be large (which is already a necessary condition such that the four-dimensional bulk is approximately classical), it is therefore possible to realize a separation of scales $\ell \gg L_P$. In this sense, the braneworld semi-classical solutions are not only exact solutions of semi-classical equations but are also macroscopic. One need not in general worry about quantum gravitational fluctuations entering at the same order as the semi-classical effects and thereby spoiling the validity of the semi-classical approximation. 

The parameter $\ell$ controls the strength of the semi-classical backreaction. When this parameter vanishes, the quantum matter does not couple to gravity and the solution describes either a (charged) conical defect for $M < 0$, or a classical BTZ black hole for $M > 0$ immersed in a (charged) CFT plasma. In the case of the conical defects, the effect of the quantum backreaction is to `dress' the conical singularity with a horizon, thereby implementing a form of quantum cosmic censorship~\cite{Emparan:2002px,Frassino:2025buh}. For non-vanishing charge, the causal structure of the black hole is equivalent to that of the Reissner-Nordstr\"om-AdS black hole. When the charge vanishes, we have $W(M, Q) = W(M)$ and $P(M, Q) = 0$ and the metric reduces to the neutral qBTZ metric described in~\cite{Emparan:2020znc}. In this case, the causal structure is equivalent to that of the Schwarzschild-AdS black hole. 

\vspace{2mm}

\noindent \textbf{Flat quantum black holes.} By taking the induced cosmological constant to vanish, it is possible to realize  asymptotically flat quantum black holes on the brane. In this same limit, the brane tension approaches its critical value given in~\eqref{eq:critical_tension}, 
\begin{equation}
    \tau \to \tau_{\rm crit} \, \qquad \text{when} \qquad L_3 \to \infty \, .
\end{equation}
Importantly, this limit does not turn off the bulk cosmological constant and we have $\ell_4 = \ell$. 

In conventional form, the blackening factor of the induced brane metric reads
\begin{equation}
    f_{\rm flat}(r) = 1 - 8 G_3 M - 
    \frac{\ell W(M,Q)}{r}
    +
    \frac{\ell^2 P(M,Q)}{r^2} \, ,
    \label{eq:charged-qschw-mass-form-zoo}
\end{equation}
where 
\begin{equation}
    8 G_3M =1-\kappa\eta^2,
    \qquad
    W(M,Q)=\mu\eta^3,
    \qquad
    P(M,Q)=q^2\eta^4 \, .
    \label{eq:charged-qschw-functions-zoo}
\end{equation}
Note that the `renormalized' Newton constant goes over to the bare three-dimensional Newton constant in this limit. For non-vanishing charge, the metric describes an asymptotically flat quantum black hole with two horizons whenever $\mu^2\geq 4q^2$, located at (for $\kappa = +1$)
\begin{equation}
    r_\pm
    =
    \frac{\ell_4\eta}{2}
    \left(
        \mu
        \pm
        \sqrt{\mu^2-4q^2}
    \right)\;.
    \label{eq:charged-flat-horizons-zoo}
\end{equation}
The extremal limit occurs when the inner and outer radii coincide, or $\mu=2|q|$. The causal structure of the non-extremal solution is identical to that of the Reissner-Nordstr\"om black hole. When the charge vanishes, we have $P(M, Q) = 0$ and the solution describes a quantum Schwarzschild black hole. There is an event horizon and a spacelike singularity, with the causal structure equivalent to that of Schwarzschild.

\section{Scalar probes of quantum black holes on the brane}
\label{sec:scalar-qbh-brane}

Our ultimate goal is to probe a static quantum black hole and analyze the structure of asymptotic QNMs. To this end, we focus on massless scalar perturbations on a general static, circularly symmetric three-dimensional black hole with line element (\ref{eq:brane-static-zoo}).
% \begin{equation}
%     \dn s_{\mathcal B}^2
%     =
%     -f(r)\dn t^2+\frac{\dn r^2}{f(r)}+r^2\dn\phi^2,
%     \qquad
%     \phi\sim\phi+2\pi .
%     \label{eq:general-brane-metric}
% \end{equation} 
On this background we consider a minimally coupled massless scalar field $\psi$ on the brane, with equation of motion
\begin{equation}
    0=\Box_{3}\psi= -\frac{1}{f(r)}\partial_t^2\psi
    +\frac{1}{r}\partial_r\left(r f(r)\partial_r\psi\right)
    +\frac{1}{r^2}\partial_\phi^2\psi\;.
    \label{eq:brane-kg-general}
\end{equation}
Separating variables according to
\begin{equation}
    \psi(t,r,\phi)
    =
    e^{-i\omega t+i m\phi}
    \frac{R(r)}{\sqrt r},
    \qquad
    m\in\mathbb Z,
    \label{eq:brane-scalar-ansatz-general}
\end{equation}
for azimuthal number $m$, and introducing the tortoise coordinate 
\begin{equation}
   r_{*}=\int\frac{\dn r}{f(r)}\;,\quad  \frac{\dn r_*}{\dn r}
    =
    \frac{1}{f(r)}\;,
    \label{eq:brane-tortoise-general}
\end{equation}
the radial equation is brought to a Schr{\" o}dinger-like form
\begin{equation}
    \frac{\dn^2 R}{\dn r_*^2}
    +
    \left[
        \omega^2-V_{\mathcal B}(r)
    \right]R
    =
    0\;,
    \label{eq:brane-radial-schrodinger-general}
\end{equation}
with effective potential
\begin{equation}
    V_{\mathcal B}(r)
    =
    \frac{f(r)}{4r^2}
    \left[
        4m^2-f(r)+2r f'(r)
    \right].
    \label{eq:brane-potential-general}
\end{equation}
This is the master radial equation for a minimally coupled scalar propagating on a static quantum black hole geometry \eqref{eq:brane-static-zoo}.

\subsection{Tensionless limit of brane geometry}

We now compare the brane scalar problem with the corresponding four-dimensional bulk problem in the tensionless limit. The main purpose of this analysis is to show that scalars on these two geometries share parts of their spectra in the limit of asymptotic overtones. For the purposes of illustration we will focus on the  neutral static C-metric, the metric (\ref{eq:rotatingCmetqbtz}) for $q=0$. 
To obtain a finite four-dimensional black hole mass as $\ell\to\infty$, we scale
\begin{equation}
    \mu=\frac{2\widetilde{M}}{\ell},
    \label{eq:mu-scaling-tensionless-new}
\end{equation}
while holding $\widetilde M$ and $\ell_4$ fixed. Then 
\begin{equation}
    H(r)
    =
    \kappa-\frac{2\widetilde M}{r}
    +r^2\left(
        \frac{1}{\ell_4^2}
        -\frac{1}{\ell^2}
    \right)
    \longrightarrow
    F(r)
    \equiv
    \kappa-\frac{2\widetilde M}{r}
    +\frac{r^2}{\ell_4^2},
    \label{eq:H-to-F-new}
\end{equation}
and, for $\Omega\equiv 1+\frac{xr}{\ell}$,
\begin{equation}
    \Omega\to1,
    \qquad
    G(x)\to1-\kappa x^2.
    \label{eq:Omega-G-limit-new}
\end{equation}
The spherical Schwarzschild-AdS$_4$ limit corresponds to $\kappa=1$ with $x=\cos\theta$, such that
\begin{equation}
    \frac{\dn x^2}{1-x^2}+(1-x^2)\dn\phi^2
    =
    \dn\theta^2+\sin^2\theta\,\dn\phi^2.
    \label{eq:x-theta-new}
\end{equation}
Thus, in the tensionless limit, the bulk AdS$_{4}$ geometry becomes
\begin{equation}
    \dn s_4^2
    =
    -F(r)\dn t^2
    +\frac{\dn r^2}{F(r)}
    +r^2\left(
        \dn\theta^2+\sin^2\theta\,\dn\phi^2
    \right),
    \qquad
    F(r)
    =
    1-\frac{2\widetilde M}{r}
    +\frac{r^2}{\ell_4^2}.
    \label{eq:sads4-limit-new}
\end{equation}
Meanwhile, on the brane, using $\eta\to1$ in the same limit, the induced metric becomes
\begin{equation}
    \dn s_{\mathcal B,0}^2
    =
    -F(r)\dn t^2
    +\frac{\dn r^2}{F(r)}
    +r^2\dn\phi^2.
    \label{eq:tensionless-brane-metric-new}
\end{equation}
Evidently, the tensionless brane geometry is the equatorial, totally geodesic slice of the Schwarzschild-AdS$_4$ geometry (\ref{eq:sads4-limit-new}). 

We emphasize that, while the bulk and brane scalar equations are not identical, because the scalar fields live in different dimensions, they are governed by the same blackening factor $F(r)$. To see this, first consider a minimally coupled massless scalar in the Schwarzschild-AdS$_4$ background. The equation of motion is
\begin{equation}
    \Box_4\Psi=0.
    \label{eq:bulk-kg-sads-new}
\end{equation}
Using the standard separation ansatz,
\begin{equation}
    \Psi
    =
    e^{-i\omega t}
    Y_{lm}(\theta,\phi)
    \frac{R_4(r)}{r},
    \qquad
    \nabla_{S^2}^2Y_{lm}
    =
    -l(l+1)Y_{lm},
    \label{eq:bulk-sads-separation-new}
\end{equation}
where $l$ is the usual spherical harmonic index and $m$ is the azimuthal quantum number, and using the tortoise coordinate
\begin{equation}
    \frac{\dn r_*}{\dn r}
    =
    \frac{1}{F(r)},
    \label{eq:bulk-tortoise-new}
\end{equation}
the radial equation becomes
\begin{equation}
    \frac{\dn^2 R_4}{\dn r_*^2}
    +
    \left[
        \omega^2-V_4(r)
    \right]R_4
    =
    0,
    \label{eq:bulk-sads-radial-new}
\end{equation}
with potential
\begin{equation}
    V_4(r)
    =
    F(r)
    \left[
        \frac{l(l+1)}{r^2}
        +
        \frac{F'(r)}{r}
    \right].
    \label{eq:bulk-sads-potential-new}
\end{equation}
Specifically, for Schwarzschild-AdS$_4$
\begin{equation}
    V_4(r)
    =
    F(r)
    \left[
        \frac{l(l+1)}{r^2}
        +
        \frac{2\widetilde M}{r^3}
        +
        \frac{2}{\ell_4^2}
    \right].
    \label{eq:bulk-sads-potential-expanded-new}
\end{equation}

Meanwhile, on the tensionless brane, the scalar equation follows from the general brane result \eqref{eq:brane-radial-schrodinger-general} by setting $f(r)=F(r)$. Hence,
\begin{equation}
    \frac{\dn^2 R_{\mathcal B}}{\dn r_*^2}
    +
    \left[
        \omega^2
        -
        V_{\mathcal B,0}(r)
    \right]R_{\mathcal B}
    =
    0,
    \label{eq:brane-tensionless-radial-new}
\end{equation}
where the same tortoise coordinate \eqref{eq:bulk-tortoise-new} appears, and
\begin{equation}
    V_{\mathcal B,0}(r)
    =
    \frac{F(r)}{4r^2}
    \left[
        4m^2-F(r)+2rF'(r)
    \right].
    \label{eq:brane-tensionless-potential-new}
\end{equation}
For Schwarzschild-AdS$_4$,
\begin{equation}
    V_{\mathcal B,0}(r)
    =
    F(r)
    \left[
        \frac{m^2-\frac14}{r^2}
        +
        \frac{3\widetilde M}{2r^3}
        +
        \frac{3}{4\ell_4^2}
    \right].
    \label{eq:brane-tensionless-potential-expanded-new}
\end{equation}
Clearly, the two potentials \eqref{eq:bulk-sads-potential-expanded-new} and \eqref{eq:brane-tensionless-potential-expanded-new} are different. Therefore the low-lying bulk and brane quasinormal spectra are not expected to agree mode by mode. Nevertheless, in the large overtone limit, their spectra have the same spacing to leading order. This can be seen most directly from a monodromy analysis, as we will present momentarily.

\section{Asymptotic QNMs of quantum black holes}\label{sec:asympqnms}

In this section we analytically solve for the asymptotic QNMs of quantum BTZ black holes using the method of Stokes line matching. First we provide some general comments on the method, and then we apply the technique to neutral and non-extremal charged qBTZ black holes. Our results are confirmed by a numerical evaluation in Section \ref{sec:numericmethods}.  We then compute asymptotic QNMs for flat neutral quantum black holes adapting the method of monodromy matching.

\subsection{General comments} \label{ssec:gencoms}

\noindent\textbf{Setup of the problem.} We are interested in probing circularly symmetric and asymptotically AdS$_{3}$ spacetimes with metrics of the type (\ref{eq:brane-static-zoo}). Analyzing the propagation of said probes amounts to solving the Schr{\" o}dinger-like equation (\ref{eq:brane-radial-schrodinger-general}).
 %repeated here for convenience, 
% \begin{equation}
%     \frac{\dn^2 R}{\dn r_*^2}
%     +
%     \left[
%         \omega^2-V(r)
%     \right]R
%     =
%     0\;,
%     \label{eq:brane-radial-schrodinger-generalv2}
% \end{equation}
% for tortoise coordinate $r_{\ast}$ (\ref{eq:brane-tortoise-general}). 
%Since the effective potential $V(r)$ is short-range, 
The solutions are subject to boundary conditions
\beq 
\begin{split}
&R\to 0\;,\quad \;\;r \to \infty\;,\\
&R\propto e^{-i\omega r_{\ast}}\;,\quad \;\;r_{\ast}\to-\infty\;.
\end{split}
\label{eq:asympsolns}\eeq
The first condition refers to having a wave that is vanishing at the asymptotic AdS boundary $(r\to\infty)$, while the second has the wave purely ingoing at the horizon. In general, the QNM frequency $\omega$ is complex. In our conventions, where we have chosen the time dependence of the perturbation to be $e^{-i\omega t}$, damped modes have negative imaginary part, $\text{Im}(\omega)<0$; unstable, growing modes have $\text{Im}(\omega)>0$. We are interested in the high damping regime. For real $r_{\ast}$, it becomes operationally difficult to distinguish between exponentially vanishing and exponentially growing terms, and hence difficult to implement the boundary conditions.

The so-called ``monodromy method'' provides an analytic treatment of this problem that allows for the exact computation of asymptotic QNMs for spherically symmetric black holes. 
%are established analytical approaches to circumvent this technical problem and exactly compute asymptotic QNMs, collectively dubbed the ``monodromy method''. 
%The starting point is to analytically continue the radial coordinate $r$ to complex values. 
The essential simplifying aspect of the monodromy method is to analytically continue the radial coordinate and tortoise coordinate  to complex values; denote the complex tortoise coordinate as $\xi$. Then, traversing along the Stokes line, 
\beq \text{Im}(\omega \xi)=0\;,\label{eq:realstokeslinecond}\eeq
the asymptotic solutions (\ref{eq:asympsolns}) are purely oscillatory such that it is straightforward to impose the boundary conditions.
%\footnote{Note that for asymptotic QNMs in Schwarzschild or RN in asymptotically flat or de Sitter backgrounds, where $|\text{Im}(\omega)|\gg|\text{Re}(\omega)|$, the real Stokes line condition (\ref{eq:realstokeslinecond}) implies $\text{Re}(r_{\ast})=0$. For AdS-Schwarzschild and AdS-RN black holes, $|\text{Im}(\omega)|\sim |\text{Re}(\omega)|$, and thus one directly uses (\ref{eq:realstokeslinecond}).}
As we will see, analyzing the behavior of solutions to (\ref{eq:brane-radial-schrodinger-general}) as one traverses the Stokes line will lead to a determination of the asymptotic QNMs.

Historically, the monodromy method was initially developed in \cite{Motl:2002hd,Motl:2003cd} for asymptotically flat backgrounds. The general idea is the following. The first step is to locate two closed homotopic contours: a ``small'' contour that encloses only the pole at the horizon(s), 
%such that neither contour encloses the origin $r=0$ or `fictitious horizons' in the complex plane.
and a ``large'' contour taken to be close to spatial infinity $r=\infty$.
%while the other contour is close to the horizon. 
The radial function $R$ is then solved in each asymptotic region, evaluating the change in phase (the monodromy) along each contour. Since the two contours are homotopic, via the monodromy theorem,  the  monodromies of the small and large contours must be equal. This then determines the quantization condition on the QNM frequencies. Notably, the closed contours are constructed based on the topology of the Stokes lines. This method was confirmed to be equivalent to a complex WKB approximation in \cite{Andersson:2003fh}. 

Inspired by the monodromy method of \cite{Motl:2002hd,Motl:2003cd}, similar analytical techniques were  soon extended to asymptotically (A)dS spacetimes in \cite{Musiri:2003rs,Cardoso:2004up} (see \cite{Natario:2004jd} for a compendium). Rather than explicitly evaluating the monodromies of two homotopic contours, however, this approach relies on matching solutions obtained in particular regions of the Stokes line. In other words, this approach is a Stokes matching problem, despite also being referred to as the monodromy method in the literature. We adopt this approach here and refer to it as the Stokes matching method.

%The general idea of the monodromy method is then as follows. 

\vspace{2mm}

\noindent Let us now describe the general Stokes matching method in more detail before applying it to specific quantum black holes in later subsections.

\vspace{2mm}

\noindent \textbf{Complex tortoise coordinate behavior.}  Introduce the complex tortoise coordinate 
\begin{equation}
    \xi(r)
    =
    \int_0^r \frac{\dn \bar r}{f(\bar r)} \;,
    \label{eq:complex-tortoise-xi}
\end{equation}
for blackening factor $f(r)$ of the asymptotically AdS type. Since the integrand has poles at horizons along the real axis, our convention for evaluating the integral is to take the contour that goes around the poles from below.  The curvature singularity at $r=0$ maps to $\xi=0$. More precisely, in cases of interest to us, we will find that the tortoise coordinate near the origin $r=0$ in the complex-$r$ plane goes like 
\beq \xi\approx ar^{b}\;,\label{eq:tortnearorigingen}\eeq
for real coefficient $a$ and positive integer $b$ with $b\geq2$.

%We assume all horizons are nondegenerate. The integrand has poles at horizons that occur along the real axis, and our convention for the integral is that the contour goes around the poles from below. This convention reduces to the one in the previous section in the case where $1/f(r)$ can be written entirely as a partial fraction expansion. 

We  denote the location of the AdS boundary in tortoise coordinates via 
\begin{equation}
    \xi_0
    \equiv \lim_{r\to\infty}\xi(r)=
    \int_0^\infty \frac{\dn\bar{r}}{f(\bar{r})}\;.
    \label{eq:xi0-def}
\end{equation}
The integral converges because of the assumption of AdS asymptotics. Thus, we have near the AdS boundary, 
 \beq \xi-\xi_{0}\approx -\frac{L^{2}}{r}\;,\label{eq:tortcoordnearinfgen}\eeq
 for AdS length $L$.

%Physical horizons are defined as real roots of the blackening factor, i.e., $f(r_{+})=0$ for $r_{+}\in\mathbb{R}$, while unphysical horizons are complex roots, $r_{+}\in\mathbb{C}$.
Denote all roots of the blackening factor by $r_{k}$. We assume that all horizons are nondegenerate. Consequently, about any root, $f(r)\approx (r-r_{k})f'(r_{k})+\frac{1}{2!}(r-r_{k})^{2}f''(r_{k})+...$, such that, locally near the horizons $r_{k}$ 
\beq \xi(r)\approx \int_{0}^{r} \sum_{k}\frac{\dn\bar{r}}{(\bar{r}-r_{k})f'(r_{k})}=\sum_{k}\frac{1}{2\kappa_{k}}\log\left(1-\frac{r}{r_{k}}\right)\;,\label{eq:xilognearh}\eeq
 for horizon surface gravity $\kappa_{k}\equiv \frac{f'(r_{k})}{2}$. Thus, horizons are logarithmic singular points of $\xi(r)$, and the complex tortoise coordinate is multivalued around any horizon. In cases where $1/f(r)$ is a rational function (as is the case in this section), we can further simplify the result for $\xi_0$ as 
 \begin{equation}\label{eq:xi0_sum}
    \xi_0=\sum_{k}\frac{1}{2\kappa_k}\log\left(-\frac{1}{r_k}\right).
\end{equation} 
Note that because $\sum_k(2\kappa_k)^{-1}=0$ for the rational AdS metrics under consideration, any arbitrary reference length scale inserted into the logarithms cancels.

\vspace{2mm}

\noindent \textbf{Stokes line topology.} We will obtain the asymptotic QNMs by tracing along the Stokes line and matching solutions emanating from the origin toward complex infinity and vice versa, following \cite{Cardoso:2004up,Natario:2004jd}. Crucial to this analysis is the behavior of the Stokes line (\ref{eq:realstokeslinecond}) near the origin $r=0$ and at infinity in the complex $r$-plane.

To this end, first note that we will find the product $\omega \xi\to \omega \xi_{0}$ is real to leading order at large overtone. Consequently, the argument $\text{arg}(\omega\xi_{0})=0$ (mod $\pi$) leading to $\text{arg}(\omega)=-\text{arg}(\xi_{0})\equiv -\theta_{0}$ (mod $\pi$).\footnote{To see this, consider $\omega\equiv |\omega|e^{i\theta_{\omega}}$ and $\xi_{0}\equiv |\xi_{0}|e^{i\theta_{0}}$. Then, $\text{Im}(\omega\xi_{0})=0$ implies $\theta_{\omega}+\theta_{0}=k\pi$ for integer $k$.} Express now $r=\varrho e^{i\theta}$ for $\varrho\in\mathbb{R}^{+}$ and $\theta\in[0,2\pi]$. Near the origin the tortoise coordinate behaves as (\ref{eq:tortnearorigingen}).
%In what follows, near the origin $r=0$, the tortoise coordinate (\ref{eq:complex-tortoise-xi}) will behave like $\xi=ar^{b}$ for real coefficient $a$ and positive integer $b$ with $b\geq2$. 
Therefore, $\text{arg}(\xi)=\text{arg}(a) + b\theta$ leading to $\text{arg}(\omega\xi)=b\theta-\theta_{0}$. The Stokes line condition, $\text{Im}(\omega\xi)=0$, implies $\text{arg}(\omega\xi)=k\pi$ for integer $k$. Hence, locally near the origin, 
\beq\label{stokesline_condition} r=\varrho e^{i\theta}\;,\quad \text{with}\quad \theta=\frac{\theta_{0}-\text{arg}(a) + k\pi}{b}\;.\eeq
However, since $a$ is real, we will have $\text{arg}(a) =0$ or $\pi$. Hence, this can be absorbed into the definition of $k$. 
Further, since $\theta\sim \theta+2\pi$, increasing $k$ by $2b$ changes the angle by $2\pi$. The values $k=0,\ldots,2b-1$ therefore label the $2b$ distinct half-lines emanating from the origin, equally spaced, each separated by an angle $\pi/b$.
%This branching of the Stokes graph is of a different nature from the branching in the tortoise coordinate near the horizon described above.

Let us now consider the Stokes line near infinity, $r\to\infty$, where the tortoise coordinate behaves as (\ref{eq:tortcoordnearinfgen}). Going through a similar analysis as above, it is straightforward to show from the condition $\text{Im}(\omega(\xi-\xi_{0}))=0$ that $\theta=\text{arg}(\omega)+k\pi$ for integer $k$. Consequently, near complex infinity, there are only two distinct Stokes lines. %Evidently, moreover, near infinity $\xi$ has no monodromy. 

Finally, consider the Stokes condition near the singular points of $\xi(r)$, i.e., the horizons. For illustrative purposes, select one non-degenerate horizon $r_{+}$. The Stokes line condition is $\text{Im}\left(\frac{\omega}{2\kappa_{+}}\log(z)\right)=0$, for $z\equiv 1-\frac{r}{r_{+}}$. This implies $\frac{\omega}{2\kappa_{+}}\log(z)\equiv\mathfrak{r}\in\mathbb{R}$. Consequently, 
%writing $\frac{\omega}{2\kappa_{+}}\equiv (\alpha+i\beta)^{-1}$ for real $\alpha,\beta$, then $z=e^{(\alpha+i\beta)\mathfrak{r}}$. 
\beq \label{eq:stokes_spiral_horizon}
z=1-\frac{r}{r_{+}}=e^{\frac{2\kappa_{+}}{\omega}\mathfrak{r}}\;.\eeq
Except for when $\omega/2\kappa_{+}$ is purely real or  imaginary, curves approaching the singularity at $z=0$ (or, $r=r_{+}$) form spirals. In general, one expects the Stokes line to hit all of the `horizons'. 
%For intuition, please refer to the \figref{fig:stokesline} to see the shape of Stokes line in the neutral quantum BTZ.

\vspace{2mm}

\noindent \textbf{Stokes line matching.} So far we have provided a local analysis about special points of the Stokes lines. The next step of the method is to study the solutions to the radial equation (\ref{eq:brane-radial-schrodinger-general}) near the endpoints of the Stokes line, and match the solutions guided by matching the Stokes lines. As we demonstrate below, we will separately write the solution near the origin $r=0$ and the AdS boundary, and match by analytically continuing the solution near the origin. In particular, the Stokes line will provide for us a matching condition between the solutions near the origin and the asymptotic boundary, (\ref{eq:Rori}) and (\ref{eq:RinftyqBTZ1}), respectively.  A second matching condition is attained by  continuing the near-origin solution along another Stokes branch reaching the horizon. Together, the two matching conditions will yield the asymptotic QNMs. Note that in this approach, rather than explicitly computing a monodromy, we will directly impose the AdS boundary condition (\ref{eq:asympsolns}).

\subsection{Neutral qBTZ}

We begin with the uncharged quantum BTZ black hole. The Schr{\"o}dinger-like master equation (\ref{eq:brane-radial-schrodinger-general}) for the radial wavefunction $R(\xi)$ has effective potential (\ref{eq:brane-potential-general}),
\beq
\begin{split}
    V_{\mathrm{qBTZ}}(r)&=\frac{f(r)}{4 r^2}\left(4 m^2- f(r)+2 r f'(r)\right)\;,
\end{split}
\label{eq:effpotqBTZ}\eeq
for blackening factor (\ref{eq:charged-qbtz-mass-form-zoo}) with zero charge. Let us solve for $R(\xi)$ and study its behavior near the curvature singularity at $r=0$ and near the asymptotic AdS$_{3}$ boundary at $r\to\infty$. 

\vspace{2mm}

\noindent \textbf{Near curvature singularity.} Near $r=0$, the complex tortoise coordinate (\ref{eq:complex-tortoise-xi}) is 
\beq \xi\approx -\frac{1}{2\ell W(M)}r^{2}\;,\label{eq:xiqbtzneu}\eeq
such that $\xi=0$ at the singularity. Further, near $r=0$, the effective potential (\ref{eq:effpotqBTZ}) at leading order is
\beq V_{\text{qBTZ}}\approx -\frac{3}{4}\frac{\ell^{2}W(M)^{2}}{r^{4}}\approx-\frac{3}{16\xi^{2}}\;.\label{eq:Vqbtzneu}\eeq
The general solution for the radial master field near the origin is then given by 
\beq R_{(0)}(\xi)=A^{(0)}_{+}\sqrt{2\pi \omega \xi}J_{1/4}(\omega\xi)+A^{(0)}_{-}\sqrt{2\pi \omega \xi}J_{-1/4}(\omega\xi)\;,\label{eq:Rori}\eeq
where $J_\nu(z)$ is the Bessel function of the first kind and $A_{\pm}^{(0)}$ are complex integration constants.

\vspace{2mm}

\noindent \textbf{Near asymptotic boundary.} Near the boundary of AdS$_{3}$, the complex tortoise coordinate behaves as 
\beq \xi_{0}-\xi\approx \frac{\ell_{3}^{2}}{r} \;,\label{eq:comptortasy}\eeq
such that the effective potential (\ref{eq:effpotqBTZ}) at leading order near infinity goes like
\beq V_{\text{qBTZ}}\approx \frac{3r^{2}}{4\ell_{3}^{4}}\approx \frac{3}{4(\xi_{0}-\xi)^{2}}\;.\label{eq:vqbtzasyminf}\eeq
Consequently, in this regime, the radial master field has general solution
\beq R_{(\infty)}(\xi)=A^{(\infty)}_{+}\sqrt{2\pi \omega (\xi-\xi_{0})}J_{1}(\omega(\xi-\xi_{0}))+A^{(\infty)}_{-}\sqrt{2\pi \omega (\xi-\xi_{0})}Y_{1}(\omega(\xi-\xi_{0}))\;,\label{eq:RinftyqBTZ1}\eeq
for complex integration constants $A_{\pm}^{(\infty)}$ and where $Y_{\nu}(z)$ is the Bessel function of the second kind. At the boundary, $\xi=\xi_{0}$, we impose the boundary condition $R=0$. Using the asymptotic expansion for $Y_{1}(z)\approx -\frac{2}{\pi z}$ for $|z|\ll 1$, it follows that we must set $A_{-}^{(\infty)}=0$. 
% Thus, near the AdS boundary, using the asymptotic behavior $J_{1}(z)\approx \frac{z}{2}$ for $|z|\ll1$, the radial solution behaves as 
% \beq R_{(\infty)}(\xi)\approx \frac{A_{+}^{(\infty)}}{2}\sqrt{2\pi}(\omega(\xi-\xi_{0}))^{3/2}\;.\label{eq:asympsolnv1}\eeq

\vspace{2mm}

\noindent \textbf{Stokes lines and solution matching.} Let us now consider the Stokes line $\text{Im}(\omega \xi)=0$. As indicated by the numerics, asymptotically, $\omega \xi\to \omega \xi_{0}$ is real. Thus, the Stokes line $\text{Im}(\omega\xi)=0$ should extend out to asymptotic infinity. %Further, since $\omega\xi_{0}$ is real asymptotically, then the argument $\text{arg}(\omega\xi_{0})=0$ (mod $\pi$) such that $\text{arg}(\omega)=-\text{arg}(\xi_{0})$ (mod $\pi$).
%\footnote{To see this, consider $\omega\equiv |\omega|e^{i\theta_{\omega}}$ and $\xi_{0}\equiv |\xi_{0}|e^{i\theta_{0}}$. Then, $\text{Im}(\omega\xi_{0})=0$ implies $\theta_{\omega}+\theta_{0}=n\pi$ for integer $n$.}
Since the tortoise coordinate $\xi$ is multi-valued about the horizons, so too is $\text{Im}(\omega \xi)$, with a branch cut across each horizon. The idea, following \cite{Natario:2004jd}, is to choose one branch of $\xi(r)$ and follow the Stokes line along this branch; since the branch cuts are not defined uniquely, it is always possible to shift the branch cuts such that they do not intersect the Stokes line. 

\begin{figure}[t!]
    \centering
    \includegraphics[width=10cm]{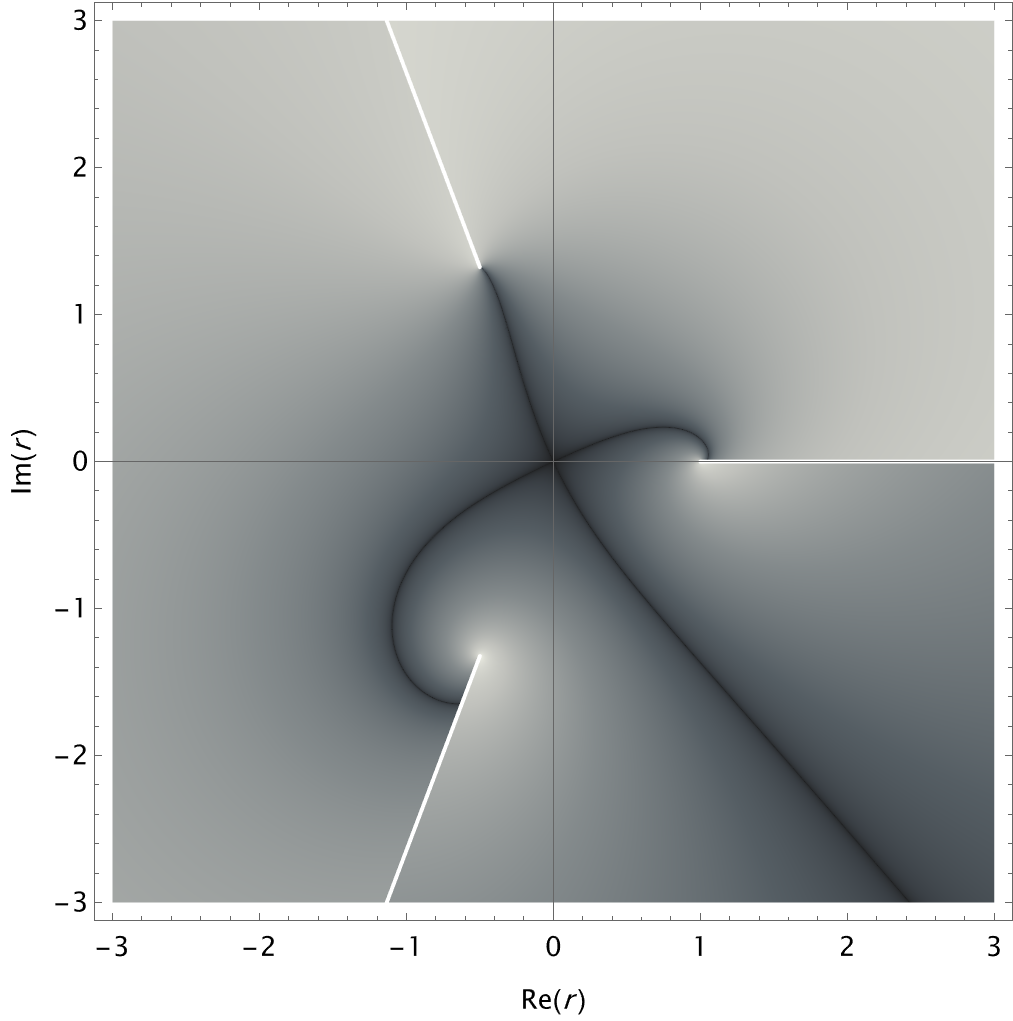}
    \caption{Stokes lines (shown in black), given by the condition $\im(\omega\xi)=0$, for the uncharged qBTZ black hole in the complex $r$-plane. The branch cuts, indicated in white, remove the spiral-like behavior near the (complex) horizons. Close to the origin, neighboring Stokes lines are spaced by an angle of $\pi/2$. }
    \label{fig:stokesline}
\end{figure}

To trace out the Stokes line, we are interested in its critical points, i.e., $\frac{\dn \xi}{\dn r}=f^{-1}(r)=0$. This occurs at the origin $r=0$, while singularities in $\frac{\dn \xi}{\dn r}$ arise at the horizons. Hence, we are interested in the near-origin behavior of a Stokes line extending out to infinity. The Stokes line will provide for us a matching condition between the solutions near the origin and the asymptotic boundary, (\ref{eq:Rori}) and (\ref{eq:RinftyqBTZ1}), respectively.  A second matching condition is attained by  continuing the near-origin solution along another Stokes branch reaching the horizon. For a numerical calculation of the Stokes lines refer to Figure~\ref{fig:stokesline}.

Consider first the near-origin solution (\ref{eq:Rori}). As we are interested in asymptotic QNMs, where $|\omega\xi|\to\infty$, i.e., when  $\lvert\omega\xi\rvert\gg1$,
% $1/|\omega|\ll |\xi|\ll 1$, 
%(despite our solution (\ref{eq:Rori}) arising from solving the master equation near the origin $\xi=0$).
%In this region, our solution (\ref{eq:Rori}) approximates as
we have
\begin{equation}
\begin{split}
    R_{(0)}(\xi)&\approx 2A_{+}^{(0)}\cos\left(\omega\xi-\frac{3\pi}{8}\right)+2A_{-}^{(0)}\cos\left(\omega\xi-\frac{\pi}{8}\right)\\
    &= e^{i\omega \xi}\left(A^{(0)}_+ e^{-3i\pi /8}+A^{(0)}_-e^{-i\pi /8}\right)+e^{-i\omega \xi}\left(A^{(0)}_+ e^{3i\pi /8}+A^{(0)}_-e^{i\pi /8}\right)\;,
    \end{split}
    \label{eq:match1}
\end{equation}
where we used the asymptotic expansion
\beq J_{\nu}(z)\sim \sqrt{\frac{2}{\pi z}}\cos\left(z-\frac{\nu \pi}{2}-\frac{\pi}{4}\right)\;,\quad |z|\gg1\;,\label{eq:Jnuasy1}\eeq
with $z\equiv \omega\xi$. 

Next, consider the asymptotic solution (\ref{eq:RinftyqBTZ1}) with $A_{-}^{(\infty)}=0$, as one follows the Stokes line emanating from the origin. From (\ref{eq:comptortasy}) we see $\omega(\xi-\xi_{0})\approx -\frac{\omega\ell_{3}^{2}}{r}$, such that on approach to the origin $\omega(\xi-\xi_{0})\ll-1$. In this regime, the radial solution (\ref{eq:RinftyqBTZ1}) goes like
\begin{equation}
\begin{split}
    R_{(\infty)}(\xi)&\approx2A_{+}^{(\infty)}\cos\left(\omega(\xi-\xi_{0})+\frac{3\pi}{4}\right)\\
    &=e^{i\omega \xi}\left(A^{(\infty)}_+ e^{-i \omega \xi_0}e^{ 3i \pi /4}\right)+e^{-i\omega \xi}\left(A^{(\infty)}_+ e^{i \omega \xi_0}e^{-3i\pi /4}\right)\;,
    \end{split}
    \label{eq:match2}\end{equation}
where we used the asymptotic expansion of the Bessel function
% \begin{equation}\label{eq:large_negative_bessel}
%     J_\nu(z)\approx \sqrt{\frac{2}{\pi z}} \cos\left(z+\frac{\nu \pi}{2}+\frac{\pi }{4}\right)\;, \qquad z\ll -1\;\;,
% \end{equation}
for $z\equiv \omega(\xi-\xi_{0})$. Matching the solutions \eqref{eq:match1} with \eqref{eq:match2} we find 
\begin{equation}
    \left(A^{(0)}_+ e^{-3\pi i/8}+A^{(0)}_-e^{-\pi i/8}\right)e^{i\omega \xi_0}e^{-3\pi i/4}=\left(A^{(0)}_+ e^{3\pi i/8}+A^{(0)}_-e^{\pi i/8}\right)e^{-i\omega \xi_0}e^{3\pi i/4}\;,
    \label{eq:condition1}
\end{equation}
where we divided out the coefficient $A_{+}^{(\infty)}$. Equation \eqref{eq:condition1} gives our first matching condition, relating the coefficients of the near-origin solution to the solution satisfying the boundary condition at infinity.

Let us now obtain our second matching condition. Consider again the solution near the curvature singularity (\ref{eq:Rori}). To move from the branch extending out to infinity to the branch which goes to the horizon we must rotate in $r$ by an angle $\pi/2$ counterclockwise; correspondingly, a rotation by $\pi$ counterclockwise in $\xi$ since $\xi\sim -r^2$ in this region. Carrying out this rotation we have 
\begin{equation}
    R_{(0)}(\xi)\approx A^{(0)}_+e^{3i\pi/4}\sqrt{2\pi\omega \xi}J_{1/4}(\omega \xi)+A^{(0)}_-e^{i\pi/4}\sqrt{2\pi\omega \xi}J_{-1/4}(\omega \xi)\;,
\end{equation}
where we used
\begin{equation}
    \sqrt{z e^{i \pi}}J_\nu(ze^{i\pi})=e^{i\pi(\frac{1}{2}+\nu)}\sqrt{z}J_\nu(z)\;.
\end{equation}
On this branch of the Stokes line, $\omega \xi$ picks up an additional factor of $-1$ relative to the branch extending to infinity. Thus, to propagate the solution away from the origin we apply the asymptotic  of the Bessel expansion %\eqref{eq:large_negative_bessel} 
to yield
\begin{equation}
     R(\xi)\sim(A^{(0)}_+ e^{3i\pi /8}+A^{(0)}_- e^{i\pi/8})e^{-i\omega \xi}+(A^{(0)}_+ e^{9i\pi/8}+A^{(0)}_-e^{3i\pi/8})e^{i\omega \xi}.
\end{equation}
The ingoing boundary condition at the horizon demands $R(\xi)\sim e^{-i\omega \xi}$ as $\xi\to-\infty$, thus restricting
\begin{equation}
\label{eq:condition2ne}
    A^{(0)}_+ e^{9i\pi/8}+A^{(0)}_- e^{3i\pi/8}=0\;.
\end{equation}
This provides for us our second matching condition. 

Together, the matching conditions \eqref{eq:condition1} and \eqref{eq:condition2ne}  form a homogeneous linear system for $A^{(0)}_\pm$. A nontrivial solution exists only if
% Combined, the matching conditions \eqref{eq:condition1} and \eqref{eq:condition2ne} constrain the coefficients $A^{(0)}_\pm$ %are constrained by the conditions \eqref{eq:condition1} and \eqref{eq:condition2ne}
% to satisfy the following condition for a nontrivial solution of the linear system
\begin{equation}
    \begin{vmatrix}
e^{ 9i\pi/8} & e^{ 3i\pi/8} \\
e^{i(\omega \xi_0 - 9\pi/8)} - e^{-i(\omega \xi_0 - 9\pi/8)} & e^{i(\omega \xi_0 - 7\pi/8)} - e^{-i(\omega \xi_0 - 7\pi/8)}
\end{vmatrix}
= 0\;.
\end{equation}
Solving, we obtain
\begin{equation}\label{eq:analytic_omega1}
    \omega=\frac{\pi n}{\xi_0} +i\frac{ \log 2}{4 \xi_0}\;,
\end{equation}
for integer $n$ (valid only for large $n$).  
The dependence on the backreaction parameter $\ell$ enters implicitly through $\xi_0$, fixed by the roots of the metric function $f$.\footnote{So far we have computed the QNM frequencies where $\text{Re}(\omega)>0$. To attain QNMs with $\text{Re}(\omega)<0$, we make a different choice in branch cuts. Specifically, set $\tilde{\xi}_0=\xi_R-i\xi_I$, with $\xi_R, \xi_I>0$. The associated Stokes line has its branch extending to spatial infinity in the first quadrant, akin to a reflection in the line $\im(r)=0$ of the line in Figure \ref{fig:stokesline}. Then, since $\tilde{\omega}=-\omega_R-i\omega_I$, with $\omega_R, \omega_I>0$, the sign of the argument of the Bessel function is modified in different regions, necessitating the use of different expansions from those used above. Repeating the matching calculation then yields the other branch of QNMs, $\tilde\omega=-\bar\omega =-\frac{\pi n}{\tilde{\xi}_0} +i\frac{ \log 2}{4 \tilde{\xi}_0}$.}

% For $\ell=100$, $x_0=0.671033 + i\, 0.800889 $, and the analytic asymptotic frequency from \eqref{eq:analytic_omega1} is \begin{equation} 
%     \omega(n) = (1.93102-2.30471 i)n + 0.127125 + 0.106513 i,
% \end{equation}
% while the corresponding numerical fit to the modes in \cref{tab:qnm_100_new}, with $n\geq 20$ gives
% \begin{equation}
%     \omega(n) = (1.931018 - 2.304718i)n + 0.127272 + 0.106521i.
% \end{equation}
% The analytical result is in excellent agreement with the numerical result.

%Consequently, we find that the Stokes line is as shown in Figure \ref{fig:stokesline}, which we have determined by numerically solving $\im(\omega x)=0$....

\subsubsection*{Comparing brane and bulk spectra}

We now have enough to analytically show that in the large overtone limit the bulk and brane QNM spectra have the same spacing at leading order. To wit, both the bulk and brane radial equations can be written as
\begin{equation}
    \frac{\dn^2 R_i}{\dn \xi^2}
    +
    \left[
        \omega^2-V_i(r)
    \right]R_i=0\;,
    \qquad
    i\in\{4,\mathcal B\}\;,
    \label{eq:common-monodromy-radial}
\end{equation}
with the same complex tortoise coordinate $\xi$ (\ref{eq:complex-tortoise-xi}), using the blackening factor $F(r)$ introduced in (\ref{eq:H-to-F-new}). The distinction between the two problems is entirely in the effective potential $V_{i}$. Near the AdS boundary, $r\to\infty$, 
\begin{equation}
    F(r)\simeq \frac{r^2}{\ell_4^2},
    \qquad
    \xi_0-\xi\simeq \frac{\ell_4^2}{r}\;.
    \label{eq:boundary-tortoise-asymptotic}
\end{equation}
The bulk (\ref{eq:bulk-sads-potential-expanded-new}) and brane (\ref{eq:brane-tensionless-potential-expanded-new}) potentials thus behave as
\begin{equation}
    V_4
    \simeq
    \frac{2}{(\xi_0-\xi)^2}\;,
    \qquad
    V_{\mathcal B,0}
    \simeq
    \frac{3}{4(\xi_0-\xi)^2}\;,
    \label{eq:boundary-potentials-monodromy}
\end{equation}
becoming singular as $\xi\to\xi_{0}$. 
More compactly, we write
\begin{equation}
    V_i
    \simeq
    \frac{\nu_{\infty,i}^2-\frac14}{(\xi_0-\xi)^2}\;,
    \label{eq:boundary-bessel-index-def}
\end{equation}
with
\begin{equation}
    \nu_{\infty,4}=\frac32\;,
    \qquad
    \nu_{\infty,\mathcal B}=1\;.
    \label{eq:boundary-bessel-indices}
\end{equation}
The different values of $\nu_{\infty,i}$ encode the different AdS falloffs of the four-dimensional bulk scalar and the three-dimensional brane scalar.

Similarly, near the curvature singularity, $r=0$, we have (for the neutral black hole)
\begin{equation}
    F(r)\simeq -\frac{2\widetilde M}{r}\;,
    \qquad
    \xi\simeq -\frac{r^2}{4\widetilde M}\;,
    \label{eq:origin-tortoise-asymptotic}
\end{equation}
and the potentials become
\begin{equation}
    V_4
    \simeq
    -\frac{1}{4\xi^2}\;,
    \qquad
    V_{\mathcal B,0}
    \simeq
    -\frac{3}{16\xi^2}\;,
    \label{eq:origin-potentials-monodromy}
\end{equation}
diverging at the curvature singularity.
Equivalently,
\begin{equation}
    V_i
    \simeq
    \frac{\nu_{0,i}^2-\frac14}{\xi^2}\;,
    \label{eq:origin-bessel-index-def}
\end{equation}
with
\begin{equation}
    \nu_{0,4}=0\;,
    \qquad
    \nu_{0,\mathcal B}=\frac14\;.
    \label{eq:origin-bessel-indices}
\end{equation}

Near either singular endpoint, the radial equation (\ref{eq:common-monodromy-radial}) reduces to a Bessel equation. For example, near the curvature singularity $\xi=0$,
\begin{equation}
    \frac{\dn^2 R_i}{\dn \xi^2}
    +
    \left[
        \omega^2
        -
        \frac{\nu_{0,i}^2-\frac14}{\xi^2}
    \right]R_i
    \simeq 0\;,
    \label{eq:origin-bessel-equation}
\end{equation}
whose local solutions are\footnote{For cases with integer $\nu$, one should replace $J_{-\nu}(z)$ with $Y_\nu(z)$.}
\begin{equation}
    R_i
    \sim
    \sqrt{\omega \xi}\,
    J_{\pm\nu_{0,i}}(\omega \xi)\;.
    \label{eq:origin-bessel-solutions}
\end{equation}
Likewise, near the AdS boundary, the local solutions are proportional to Bessel functions $\sqrt{\omega(\xi-\xi_{0})} J_{\pm \nu_{\infty,i}}(\omega(\xi-\xi_{0}))$. 
%with index $\nu_{\infty,i}$ and argument $\omega(\xi_0-\xi)$.
Analytic continuation acts on the local Bessel basis through frequency-independent connection matrices. For noninteger orders these can be represented by phase factors in the $J_{\pm nu}(z)$ basis; for integer orders, one uses a $J_{\nu}(z)$, $Y_\nu(z)$ basis, and these will mix.  Thus, values $\nu_{0,i}$ and $\nu_{\infty,i}$ enter the Stokes matching calculation only through frequency-independent phase factors.

Since the bulk and brane radial equations share the same blackening factor $F(r)$, they also share the same horizon surface gravity and have the same logarithmic structure of the tortoise coordinate. Therefore the only difference between the bulk and brane problems is in the Bessel indices,
\begin{equation}
    (\nu_{0,4},\nu_{\infty,4})
    =
    \left(0,\frac32\right)\;,
    \qquad
    (\nu_{0,\mathcal B},\nu_{\infty,\mathcal B})
    =
    \left(\frac14,1\right)\;.
    \label{eq:bessel-index-comparison}
\end{equation}
These indices change the constant phase appearing in the quantization condition, but not the coefficient multiplying the frequency.

The matching takes the schematic form
\begin{equation}
    e^{2i\omega \xi_0}
    =
    \mathcal C_i\;,
    \qquad
    i\in\{4,\mathcal B\}\;,
    \label{eq:monodromy-quantization-exponential}
\end{equation}
where constants $\mathcal C_i$ are independent of $\omega$ at leading order in the large-$|\omega|$ expansion. The constants $\mathcal C_i$ depend on the Bessel indices, and hence on the spacetime dimension, but the exponential factor is governed by the same complex tortoise length $\xi_0$ in both problems. Equivalently, for large overtone number $n$,
\begin{equation}
    \omega_n^{(i)}\xi_0+\varphi_i
    =
    \pi n+ \text{ subleading}\;,
    \qquad
    n\gg1,
    \qquad
    i\in\{4,\mathcal B\}\;,
    \label{eq:monodromy-quantization-phase}
\end{equation}
for some frequency-independent complex offset $\varphi_i$.
It follows that
\begin{equation}
    \lim_{n\to\infty}
    \left(
        \omega_{n+1}^{(4)}
        -
        \omega_n^{(4)}
    \right)
    =
    \lim_{n\to\infty}
    \left(
        \omega_{n+1}^{(\mathcal B)}
        -
        \omega_n^{(\mathcal B)}
    \right)
    =
    \frac{\pi}{\xi_0}.
    \label{eq:monodromy-spacing-equivalence}
\end{equation}
This is the precise sense in which the tensionless brane spectrum relates to the bulk Schwarzschild-AdS$_4$ spectrum: there is an equivalence of the asymptotic mode spacings. The potentials, Bessel indices, and phase offsets are different, so the low-lying modes and the additive constants in the large-$n$ expansion need not agree. However, the singular endpoints, horizon monodromy, and tortoise coordinate are the same, and hence the leading large-overtone spacing is the same. In this sense, the large overtone modes of the braneworld black hole probe the existence of higher dimensions.

\subsection{Charged qBTZ}

\begin{figure}
    \centering
    \includegraphics[width=10cm]{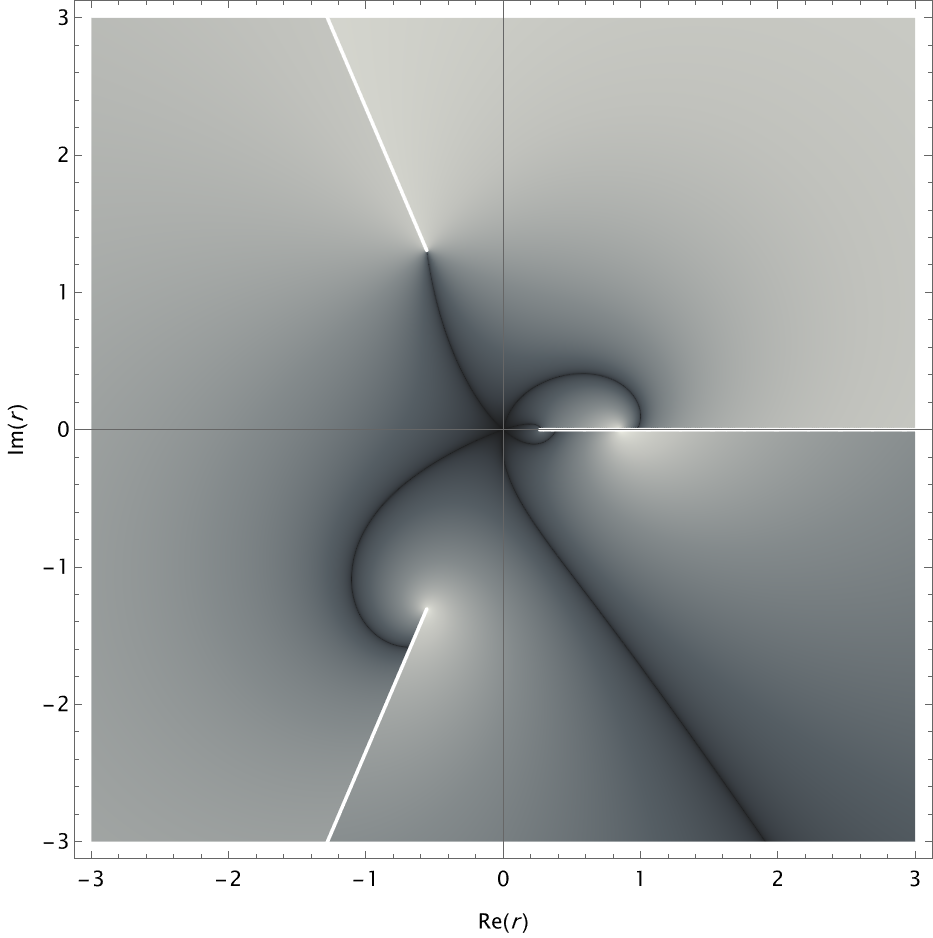}
    \caption{The Stokes lines (in black), defined by $\im(\omega\xi)=0$, for the charged qBTZ black hole computed numerically and depicted in the complex $r$-plane. Branch cuts are indicated in white, and they dampen the spiral behavior near the (complex) horizons. Near the origin, adjacent Stokes lines are separated by an angle of $\pi/3$. Compared to the uncharged case, the presence of an inner horizon introduces an additional region where the Stokes lines converge.}
    \label{fig:Stokes-charged}
\end{figure}

Let us now determine the asymptotic quasinormal modes for the charged qBTZ black hole, with blackening factor (\ref{eq:charged-qbtz-mass-form-zoo}). As in the neutral case, the Schr{\"o}dinger-like master equation (\ref{eq:brane-radial-schrodinger-general}) for the radial wavefunction $R(\xi)$ has effective potential (\ref{eq:effpotqBTZ}). We will again solve for $R(\xi)$ and study the Stokes behavior near the curvature singularity at $r=0$ and near the asymptotic AdS$_{3}$ boundary at $r\to\infty$, following the algorithm carried out above for the neutral system. For a numerical computation of the Stokes lines in this case refer to Figure~\ref{fig:Stokes-charged}.
%We begin with the non-extremal charged qBTZ, such that there are no degenerate horizons. 

%\subsubsection{Non-extremal qBTZ}

\vspace{2mm}

\noindent \textbf{Master field solutions.} Near $r=0$, the complex tortoise coordinate (\ref{eq:complex-tortoise-xi}) behaves as 
\begin{equation}
    \xi\approx \frac{r^3}{3a}\;,
\end{equation}
for $a\equiv \ell^{2} q^{2}\eta^{4}=\ell^{2} P(M,Q)$. The effective potential goes like
\begin{equation}
    V_{\mathrm{cqBTZ}}\sim-\frac{5 }{36 \xi^2}\;,
    %= \frac{(\frac{2}{3})^2-1}{4x^2}.
\end{equation}
from which the general solution of the Schr\"odinger-like equation in this region is
\begin{equation}\label{eq:solclose0}
    R_{(0)}(\xi)\sim A^{(0)}_+ \sqrt{2\pi\omega \xi}J_{1/3}(\omega \xi) + A^{(0)}_- \sqrt{2\pi\omega \xi}J_{-1/3}(\omega \xi)\;, 
\end{equation}
for complex coefficients $A_{\pm}^{(0)}$. 

Toward the AdS boundary, $r\to\infty$, the charged black hole behaves in the same way as the neutral solution, with the same limiting behavior of the tortoise coordinate (\ref{eq:comptortasy}), effective potential (\ref{eq:vqbtzasyminf}), and radial master field solution (\ref{eq:RinftyqBTZ1}).

\vspace{2mm}

\noindent \textbf{Stokes lines and solution matching.} We are interested in the Stokes line
$\text{Im}(\omega \xi)=0$ (see Figure \ref{fig:Stokes-charged}).
As with the neutral case, we begin with the near-origin solution (\ref{eq:solclose0}), considering the branch with ${\rm Re}(\omega \xi) > 0$ as $|\omega \xi| \to \infty$. Using the asymptotic expansion  (\ref{eq:Jnuasy1}) we find
%we can expand this solution as
\begin{equation}
    R_{(0)}(\xi)\approx e^{i\omega \xi}\left(A^{(0)}_+ e^{-5i\pi/12}+A^{(0)}_-e^{-i\pi/12}\right)+e^{-i\omega \xi}\left(A^{(0)}_+ e^{5i\pi/12}+A^{(0)}_-e^{i\pi/12}\right)\;.
    \label{eq:origToinfCharged}
\end{equation}
Again, our analysis at asymptotic infinity is unchanged with respect to the neutral system.  
Hence, imposing the boundary condition at infinity, $A_{-}^{(\infty)}=0$ on solution (\ref{eq:RinftyqBTZ1}), propagating the solution along the branch of the Stokes line away from infinity and matching with \eqref{eq:origToinfCharged} gives our first matching condition
\begin{equation}
    \left(A^{(0)}_+ e^{-5i\pi/12}+A^{(0)}_-e^{-i\pi/12}\right)e^{i\omega \xi_0}e^{-3i\pi/4}=\left(A^{(0)}_+ e^{5i\pi/12}+A^{(0)}_-e^{i\pi/12}\right)e^{-i\omega \xi_0}e^{3i\pi/4}\;,
\end{equation}
where $\xi_0$ is the boundary value of the tortoise coordinate.

Now we want to continue the solution from the origin to the horizon branch. First note that locally near the origin, the Stokes line condition $\text{Im}(\omega\xi)=0$ amounts to  $\text{Im}(\omega r^3)=0$, from which we obtain that the Stokes lines emerging from the origin are therefore separated by angles of $\pi/3$.\footnote{To see this, note that the condition $\text{Im}(\omega r^3)=0$ gives $r = \varrho e^{(i\theta_0+ip\pi)/3}$, for $\varrho\in\mathbb{R}$ and integer $p$.} Hence, adjacent Stokes lines are separated by an angle of $\pi$ in $\xi$.  
 With our prescription for handling the branch cuts, due to the multi-valuedness of $\xi$ near the horizons, the line that goes to the horizon  
 is rotated by $\pi$ in the complex $\xi$-plane.
After this rotation, the near-origin solution (\ref{eq:solclose0}) becomes
\begin{equation}
    R_{(0)}(\xi)\approx A^{(0)}_+e^{5i\pi/6}\sqrt{2\pi\omega \xi}J_{1/3}(\omega \xi)+A^{(0)}_-e^{i\pi/6}\sqrt{2\pi\omega \xi}J_{-1/3}(\omega \xi)\;.
\end{equation} 
Propagating this solution away from the origin, we again use the asymptotic expansion for the Bessel function 
%\eqref{eq:large_negative_bessel}
such that 
\begin{equation}
     R_{(0)}(\xi)\approx (A^{(0)}_+ e^{5i\pi/12}+A^{(0)}_- e^{i\pi/12})e^{-i\omega \xi}+(A^{(0)}_+ e^{5i\pi/4}+A^{(0)}_-e^{i\pi/4})e^{i\omega \xi}\;.
\end{equation}
The ingoing boundary condition imposes $R(\xi)\sim e^{-i\omega \xi}$ as $\xi\to-\infty$, setting
\begin{equation}
\label{eq:condition2}
    A^{(0)}_+ e^{i5\pi/4}+A^{(0)}_- e^{i\pi/4}=0\;.
\end{equation}

Combining the above two conditions, we must solve
\begin{equation}
    \begin{vmatrix}
e^{ 5i\pi/4} & e^{i \pi/4} \\
e^{i(\omega \xi_0 - 7\pi/6)} - e^{-i(\omega \xi_0 - 7\pi/6)} & e^{i(\omega \xi_0 - 5\pi/6)} - e^{-i(\omega \xi_0 - 5\pi/6)}
\end{vmatrix}
= 0\;,
\end{equation}
for $\omega$. It is easy to verify  
\begin{equation}
    \omega= \frac{ \pi n}{\xi_0}\;.
\label{eq:asymqnmwchar}\end{equation}
Again, the dependence on $\ell$ enters through $\xi_0$. 
%Our finding matches the numerics (see, for example, Table \ref{tab:tensionlessCharged}, considering that in this case $\xi_0=0.106537 + 0.209245 I$).

\subsection{Flat quantum black holes}

So far we have focused on the quantum BTZ family of black holes, having AdS$_{3}$ asymptotics. Let us now analytically compute the asymptotic QNMs of flat quantum black holes. Due to a change in the asymptotic structure of the black hole geometry, there are a few broad differences worth pointing out. We focus on the neutral case.

\subsubsection*{General comments}

We are interested in probing circularly symmetric and asymptotically Mink$_{3}$ spacetimes with metrics of the form (\ref{eq:brane-static-zoo}) with blackening factor~\eqref{eq:charged-qschw-mass-form-zoo}. Mode propagation follows from solving the Schr{\" o}dinger-like equation (\ref{eq:brane-radial-schrodinger-general}). The first essential difference with the prior analysis for AdS black holes is the choice of boundary conditions. For flat black holes we impose
\beq 
\begin{split}
&R\propto e^{+i\omega r_{\ast}}\;,\quad \;\;r_{\ast}\to+\infty\;,\\
&R\propto e^{-i\omega r_{\ast}}\;,\quad \;\;r_{\ast}\to-\infty\;.
\end{split}
\label{eq:asympsolnsflat}\eeq
The second condition is the usual choice to have purely ingoing waves at the horizon, while the first condition, in contrast with the AdS case (\ref{eq:asympsolns}), has the wave be purely outgoing at infinity. 

Another crucial difference from the analysis of AdS black holes is that the asymptotic QNMs obey 
\beq |\text{Im}(\omega)|\gg |\text{Re}(\omega)|\;,\eeq
with $\text{Im}(\omega)\to-\infty$. Thus, $\omega$ is large and approximately purely imaginary, and, consequently, the Stokes line condition (\ref{eq:realstokeslinecond}), $\text{Im}(\omega\xi)=0$, implies $\text{Re}(\xi)=0$ for complex tortoise coordinate $\xi$. 
%Hence, rather than $\omega\xi$ being asymptotically real for complex tortoise coordinate $\xi$, now $\omega\xi\in\mathbb{R}$
Writing $r=\rho e^{i\theta}$, we then conclude that, near the origin $r=0$, where the tortoise coordinate behaves as (\ref{eq:tortnearorigingen}), we find
\beq 0=\text{Re}(\xi)=a\rho^{b}\cos(b\theta)\;.\eeq
Solving gives $\theta=\frac{\pi}{2b}+\frac{k\pi}{b}$, for $k=0,1,...,2b-1$. Again, in a neighborhood of the origin in the complex $r$-plane, the Stokes line produces $2b$ half-lines equally spaced by an angle $\pi/b$.  

\vspace{2mm}

\noindent Let us now proceed with the analysis of computing the asymptotic QNMs for the neutral flat quantum black hole. Due to the asymptotic nature of the spacetime, we can and will employ the monodromy method, instead of the Stokes matching procedure. 

% \subsubsection*{(Un)charged quantum black holes}

% To streamline the presentation, we will evaluate the neutral and (non-extremal) charged flat quantum black holes simultaneously, emphasizing their differences when they arise.

\vspace{2mm}

\noindent \textbf{Near asymptotic infinity.} We begin by focusing on solutions near infinity, $r\to\infty$. Since the potential (\ref{eq:brane-potential-general}) vanishes at infinity, the solutions to (\ref{eq:brane-radial-schrodinger-general}) near infinity go like
\beq R_{(\infty)}(\xi)=A_{+}^{(\infty)}e^{i\omega \xi}+A_{-}^{(\infty)}e^{-i\omega \xi}\;.\label{eq:nearasyminfflat}\eeq
To ensure the boundary condition at spatial infinity, we impose $A_{-}^{(\infty)}=0$. 

\vspace{2mm}

\noindent \textbf{Near the origin.} Near the origin $r=0$, the tortoise coordinate behaves as (\ref{eq:xiqbtzneu})
%\beq \xi=-\frac{r^{2}}{2F\ell}\;,%\quad \xi_{\text{charged}}=\frac{r^{3}}{3\ell^{2}P}\;,\eeq
and the potential goes like (\ref{eq:Vqbtzneu}),
%\beq V_{\text{uncharged}}=-\frac{3}{16\xi^{2}}\;,
%\quad V_{\text{charged}}=-\frac{5}{36\xi^{2}}\;,\eeq
as for neutral qBTZ. Consequently, near the origin, the wavefunction has the generic solution (\ref{eq:Rori}).
%while the charged system has solution (\ref{eq:solclose0}).
From the asymptotic expansion for the Bessel function (\ref{eq:Jnuasy1}), the solution near the origin has the form
\beq R_{(0)}(\xi)\approx \left(A_{+}^{(0)}e^{-i\alpha_{+}}+A^{(0)}_{-}e^{-i\alpha_{-}}\right)e^{i\omega\xi}+\left(A_{+}^{(0)}e^{i\alpha_{+}}+A_{-}^{(0)}e^{i\alpha_{-}}\right)e^{-i\omega\xi}\;,\label{eq:nearoriflat}\eeq
for $\alpha_{\pm}=\frac{\pi}{4}\left(1\pm\frac{1}{2}\right),$
%\;,\quad \alpha_{\pm}^{\text{charged}}=\frac{\pi}{4}\left(1\pm \frac{2}{3}\right)\;,\eeq
and the complex coefficients $A^{(0)}_{\pm}$.
%are in general different between neutral and charged cases.

\vspace{2mm}

\noindent \textbf{Near the horizon.} 
%Away from extremality, denote the inner and outer horizons of the charged black hole by $r_{+}>r_{-}$. 
Near the horizon the complex tortoise coordinate behaves as (\ref{eq:xilognearh}) and the potential is vanishing, $V\approx 0$. Hence, the solution near the horizon goes like
\beq R_{(\text{hor})}(\xi)\approx A_{+}^{(\text{hor})}e^{i\omega\xi}+A_{-}^{(\text{hor})}e^{-i\omega \xi}\;.\label{eq:Rhorflat}\eeq
Imposing the horizon boundary condition implies setting $A_{+}^{(\text{hor})}=0$. 

\vspace{2mm}

\begin{figure}
    \centering
    \includegraphics[width=10cm]{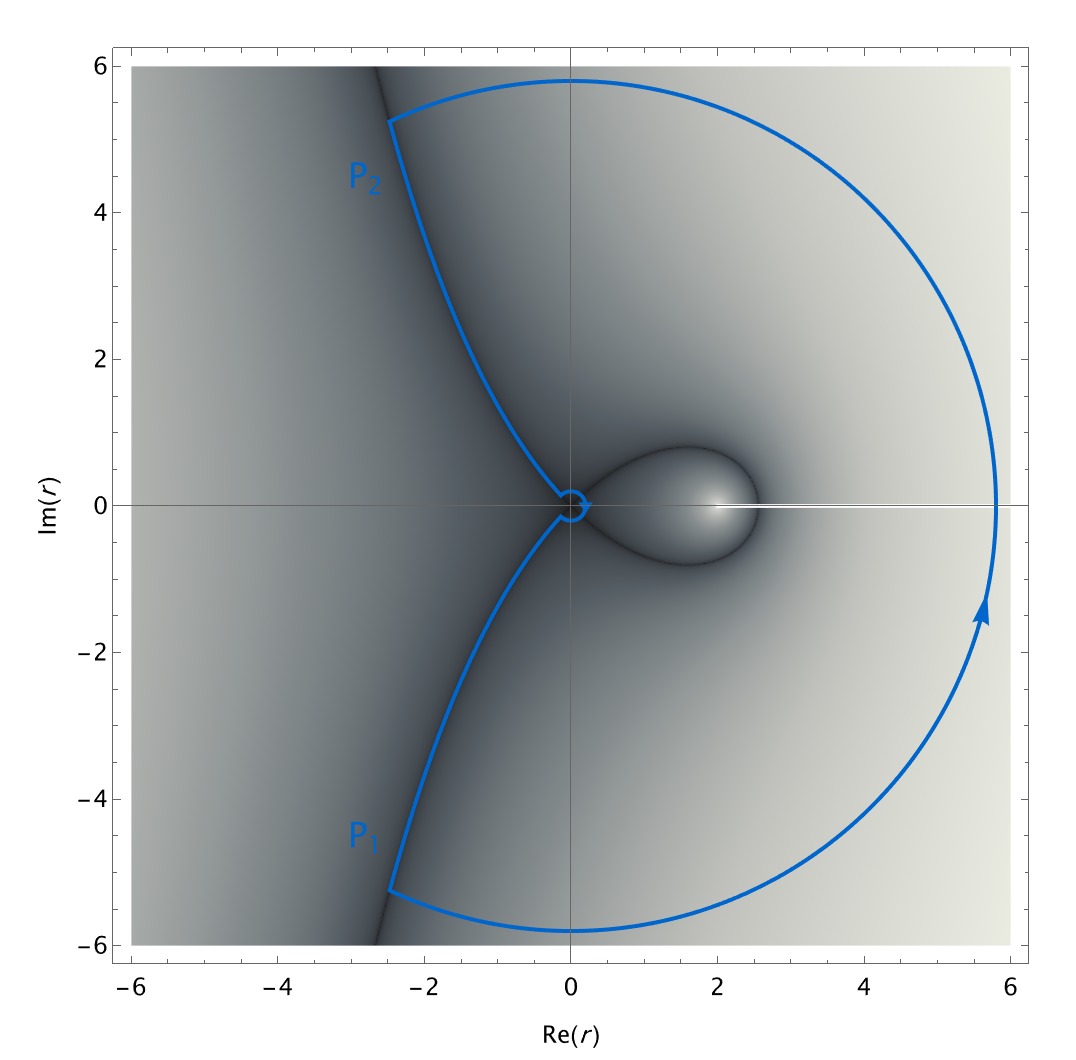}
    \caption{Numerically computed Stokes lines (black), defined by $\text{Re}(\xi)=0$, for the quantum corrected three-dimensional asymptotically flat black hole. The branch cut and horizon are indicated by the white line and white point, respectively. Because the Stokes lines do not terminate at the horizon, the QNM spectrum cannot be obtained by simply following Stokes lines, requiring a monodromy matching technique instead. The contour followed to evaluate this monodromy is represented in blue.}
    \label{fig:flatstokesline}
\end{figure}

\noindent \textbf{Stokes line and monodromy matching.} We must now match the solutions along the Stokes line, $\text{Im}(\omega\xi)=0$ (or, equivalently, $\text{Re}(\xi)=0$) which has the shape of Figure~\ref{fig:flatstokesline}. In this way, neither of the exponentials $e^{\pm i\omega\xi}$ will dominate over the other. Note that the only singular point of the curve $\text{Re}(\xi)=0$ occurs at the origin $r=0$ (as follows by considering the zeros of $\dn\xi/\dn r$). Recall that two of the branches of the Stokes line emanating from the origin are unbounded. As in the AdS case, around the horizon the complex tortoise coordinate is multivalued, thus having branch points. Meanwhile, near asymptotic infinity, the tortoise coordinate behaves as $\xi\propto r$.

We will consider two closed contours: (i) a large contour that does not encircle the origin (but approaches it), partially traces along two unbounded branches of the Stokes line and encircles the horizon, and (ii) a smaller contour that encircles the black hole horizon. We will compute the monodromy that follows from traversing each contour (see Figure \ref{fig:flatstokesline}).

\vspace{1mm}

\noindent \emph{Large contour.} We match the solution near the origin (\ref{eq:nearoriflat}) to the solution near asymptotic infinity (\ref{eq:nearasyminfflat}), say at a point $P_{1}\in\mathbb{C}$, where we impose $A_{-}^{(\infty)}=0$,
such that
\beq A_{+}^{(0)}e^{i\alpha_{+}}+A_{-}^{(0)}e^{i\alpha_{-}}=0\;.\label{eq:flatcond1}\eeq
We will use the contour prescription in \cite{Natario:2004jd}. The contour connects the two unbounded branches through a point $P_1$ on one branch and a point $P_2$ on the other.  
% Broadly, the contour connects the two unbounded branches of the Stokes line, connected by points $P_{1}$, and a second point at the opposite end, denoted by point $P_{2}$. 

Connecting the unbounded branches amounts to rotating from $P_{1}$ to $P_{2}$; the complex tortoise coordinate is rotated by an angle of $3\pi$, i.e., $\xi\to e^{3\pi i}\xi$. Carrying out this rotation on the solution (\ref{eq:Rori}) gives\footnote{Here we used $\sqrt{2\pi e^{3\pi i}\omega\xi}J_{\pm\frac{j}{2}}(e^{3\pi i}\omega\xi)=e^{\frac{3\pi i}{2}(1\pm j)}\sqrt{2\pi\omega\xi}J_{\pm\frac{j}{2}}(\omega\xi)\approx 2e^{6i\alpha_{\pm}}\cos(\omega\xi-\alpha_{\pm})$,
where the first equality follows from $J_{\nu}(z)=z^{\nu}w(z)$ for an even holomorphic function $w(z)$, and the second is a consequence of using the asymptotic expansion (\ref{eq:Jnuasy1}) with $z=|\omega\xi|$ and $\nu=\pm j/2=\pm\frac{1}{4}$.}
\beq R_{(0)}(\xi)\approx \left(A_{+}^{(0)}e^{7i\alpha_{+}}+A_{-}^{(0)}e^{7i\alpha_{-}}\right)e^{i\omega\xi}+\left(A_{+}^{(0)}e^{5i\alpha_{+}}+A_{-}^{(0)}e^{5i\alpha_{-}}\right)e^{-i\omega\xi}\;.\label{eq:R0flatrot}\eeq
We will close the contour at infinity, $\xi\propto r\to\infty$. Together with $\text{Im}(\omega)\ll0$, we have that the term proportional to $e^{-i\omega\xi}$ is exponentially small, and thus we consider instead the coefficient of the $e^{i\omega\xi}$ term. 

The monodromy of $R_{(0)}$ as we traverse the large contour follows from comparing solutions (\ref{eq:nearoriflat}) and (\ref{eq:R0flatrot}). For this, we need to know the monodromy of $e^{i\omega\xi}$ as it encircles the horizon.  To wit, consider the monodromy around a generic horizon. For simplicity, select a specific complex horizon $r_{k}$ and consider a closed contour $\gamma\subset\mathbb{C}$ that only encircles $r=r_{k}$ with either a clockwise or an anticlockwise orientation. The monodromy $\mathcal{M}$ of $\xi$ is equal to $\mathcal{M}(\xi)=-\frac{2\pi i}{2\kappa_{k}}$ for clockwise $\gamma$ and $\mathcal{M}(\xi)=+\frac{2\pi i}{2\kappa_{k}}$ for counterclockwise $\gamma$.\footnote{To see this, parametrize a small circle around $r=r_{k}$ for $r,r_{k}\in\mathbb{C}$ via $r=r_{k}+\epsilon e^{-i\theta}$ for $|\epsilon|\ll |r_{k}|$ and $\theta\in (0,2\pi)$. The argument of the logarithm (\ref{eq:xilognearh}) is then $1-\frac{r}{r_{k}}\approx -\frac{\epsilon}{r_{k}}e^{-i\theta}$. Then for $\theta\sim \theta+2\pi$, it follows $\log\left(-\frac{\epsilon}{r_{k}}e^{-i\theta}\right)\sim\log\left(-\frac{\epsilon}{r_{k}}e^{-i\theta}\right)-2\pi i$. Hence, traversing around $\gamma$ gives $\xi\sim \xi\pm\frac{i\pi}{\kappa_{k}}$.} Here $\kappa_{k}$ denotes the surface gravity associated with horizon $r_{k}$. The associated monodromy of the asymptotic waves is 
\beq \mathcal{M}(e^{\pm i\omega \xi})=e^{\pm \pi\omega/\kappa_{k}}\;.\eeq
Consequently, the monodromy of the large contour, traversing from point $P_{1}$ to $P_{2}$, is given by the ratio 
\beq \mathcal{M}[R_{(0)}]=\frac{\left(A_{+}^{(0)}e^{7i\alpha_{+}}+A_{-}^{(0)}e^{7i\alpha_{-}}\right)e^{\pi\omega/\kappa_h}}{\left(A_{+}^{(0)}e^{-i\alpha_{+}}+A^{(0)}_{-}e^{-i\alpha_{-}}\right)}\;,\eeq
where $\kappa_h$ is the surface gravity for the physical horizon of the flat black hole.

%in (\ref{eq:R0flatrot}).  

%Meanwhile, for the charged (non-extremal) case, the contour is such that connecting the unbounded branches amounts the complex tortoise coordinate is rotated by an angle of $2\pi$. 

\vspace{1mm}

\noindent \emph{Small contour.} Recall the near-horizon solution is (\ref{eq:Rhorflat}), such that, $ R_{(\text{hor})}(\xi)\approx A_{-}^{(\text{hor})}e^{-i\omega\xi}$, upon imposing the horizon boundary condition. Since the contour is small and encircles only the horizon, no other Stokes branches are crossed, and the monodromy is 
% Since the contour is small, encircling only the horizon and there is no hopping across other Stokes branches, the monodromy about this contour is  
\beq \mathcal{M}[R_{(\text{hor})}]=e^{-\pi \omega/\kappa_h}\;.\eeq

\vspace{1mm}

\noindent \emph{Monodromy matching.} We now match monodromies, $\mathcal{M}[R_{(0)}]=\mathcal{M}[R_{(\text{hor})}]$, yielding
\beq \left(A_{+}^{(0)}e^{7i\alpha_{+}}+A_{-}^{(0)}e^{7i\alpha_{-}}\right)e^{\pi\omega/\kappa_h}-\left(A_{+}^{(0)}e^{-i\alpha_{+}}+A^{(0)}_{-}e^{-i\alpha_{-}}\right)e^{-\pi \omega/\kappa_h}=0\;.\eeq
Together with the boundary condition (\ref{eq:flatcond1}), we obtain a homogeneous linear system for the coefficients $A^{(0)}_{+}$ and $A_{-}^{(0)}$, which admits a nontrivial solution only if
 \begin{equation}
    \begin{vmatrix}
e^{i\alpha_{+}} & e^{i\alpha_{-}} \\
e^{7i\alpha_{+}}e^{\pi\omega/\kappa_h}-e^{-i\alpha_{+}}e^{-\pi\omega/\kappa_h} & \;\;\; e^{7i\alpha_{-}}e^{\pi\omega/\kappa_h}-e^{-i\alpha_{-}}e^{-\pi\omega/\kappa_h}
\end{vmatrix}
= 0\;.
\end{equation}
Solving yields the frequency 
\beq \label{eq:flat-monodromy}\omega=-i\kappa_h \left(n - \frac{1}{2} \right)\;,\eeq
for $n = 1, 2, \dots$, which we identify with the overtone number. The monodromy calculation establishes this expression at leading asymptotic order for large overtone number and fixed $m$. For $m=0$, Appendix~\ref{app:zeroReFlat} independently shows this result is exact for every $n = 1, 2, \dots$. Hence, we see that, unlike in the case of the quantum BTZ solutions (\ref{eq:analytic_omega1}) and (\ref{eq:asymqnmwchar}), here the leading asymptotic QNM frequency is purely imaginary, and exactly imaginary for $m = 0$.

\section{Looking inside a quantum black hole}\label{sec:lookinqBH}

In the previous section,  we used analytic methods to extract asymptotic QNM spectra of a variety of quantum black holes. The methods relied crucially on the behavior of the analytically continued metric function in the vicinity of $r = 0$, and hence on the structure of the singularity. This had tangible consequences, such as the equivalence of the large overtone spacing in the quantum black holes and the higher-dimensional black holes with which they share certain properties.  It is therefore natural to ask the inverse question: given an asymptotic QNM spectrum, to what extent can one reconstruct the singularity structure?  

A primary obstacle is the dependence of the calculation on  the global Stokes topology of the metric, for example in the choice of contour. Here we approach this problem by making several simplifying assumptions. We begin by asking what can be inferred if the Stokes topology is assumed to be Schwarzschild-AdS-like (in a sense we shall clarify below), finding that features of the singularity are encoded in the offset of the QNMs. We then compute the subleading corrections to the asymptotic QNMs and find that the same features are encoded in the overtone dependence of the angular-momentum-dependent terms in the potential. We argue that this latter result is independent of the Stokes topology. 

For concreteness, in this section, expanding on \cite{Coviello:2026tol}, we work in three spacetime dimensions, though the main ideas clearly generalize to higher dimensions. We assume a metric that is characterized by a single blackening factor $f(r)$ which at large radius exhibits AdS asymptotics, and at small radius exhibits a power-law singularity,
\be  \label{eq:metric_f_ansatz}
f(r) \sim -a r^{-s} \left(1 + c r^p + \cdots  \right) \qquad \text{as } \quad r \to 0 \, , 
\ee
with $s > 0$ and $0< p < s+1$ . The singularity is spacelike for $a > 0$ and timelike for $a < 0$. Throughout we shall assume a massless test scalar on this geometry. The problem reduces to a Schr\"odinger-type problem (\ref{eq:brane-radial-schrodinger-general}).
% one of the form
% \be 
% \frac{{\rm d}^2 R}{{\rm d}\xi^2} + \left(\omega^2 - V(\xi) \right) R = 0 \, .
% \ee
Defining the coordinate $z = \omega \xi$ (and assuming $\omega \neq 0$) the problem can be recast in the form
\be 
\frac{{\rm d}^2 R}{{\rm d}z^2} + \left(1 - U(z) \right) R = 0 \, , \qquad \text{where} \qquad U(z) \equiv \frac{1}{\omega^2} V\left(\frac{z}{\omega}\right) \, .
\label{eq:schroprob1Dz}\ee
In the limit $|\omega| \to \infty$, the $\omega^{-2}$ factor drives the potential $U(z) \to 0$ at all points $\xi$ except those where $V(\xi)$ exhibits a compensating divergence. Hence, away from such regions, the large-$|\omega|$ solution is well-approximated by
\be 
R(z) \sim C_+ e^{i \omega \xi} + C_- e^{-i \omega \xi} \, .
\ee
The nontrivial information characterizing the QNM spectrum is thus localized near the singular points of the potential. The strategy is to solve the equation in the neighborhood of such points and then transport the resulting solutions, along with the appropriate boundary conditions, along a carefully chosen contour in the complex-$r$ plane. The contour is chosen to lie along Stokes rays defined by the condition ${\rm Im}(\omega \xi) = 0$. Along such rays the solution is purely oscillatory, with exponential growth and decay absent. Matching the solutions between these different regions then leads to the asymptotic QNM quantization condition.  Subleading corrections to the spectrum can then be obtained by applying perturbation theory to the subleading terms in the potential near one of its singular points.

\subsection{Leading asymptotic modes}

Here we shall compute the leading-order behavior of the QNMs at large overtone number. Our analysis follows the presentation in Section \ref{sec:asympqnms}. Note that in each regime studied below we separately use the coordinate $z$. Differences in these coordinates on different parts of the contour must be separately checked in the matching. For this reason, we always translate back into the local $\omega \xi$ plane wave basis when matching. 

% Throughout we will make use of the complexified tortoise coordinate $\xi$ defined by
% \be 
% \xi(r) = \int_0^r \frac{{\rm d}{\bar r}}{f(\bar{r})} \, .
% \ee
% We assume all horizons are nondegenerate. The integrand has poles at horizons that occur along the real axis, and our convention for the integral is that the contour goes around the poles from below. This convention reduces to the one in the previous section in the case where $1/f(r)$ can be written entirely as a partial fraction expansion. Our results will be phrased in terms of $\xi_0$ given by
% \be 
% \xi_0 \equiv \xi(\infty) \, .
% \ee
% The integral converges because of the assumption of AdS asymptotics. 

\paragraph{Large-$r$ solution.} We begin with the solution behavior at large radius, where the complex tortoise coordinate goes like (\ref{eq:comptortasy}) and the effective potential $V_{(\infty)}$ behaves as (\ref{eq:vqbtzasyminf}). 
% At large radius, the potential behaves as 
% \be 
% V_{(\infty)}  = \frac{3 r^2}{4 \ell_3^4} + O(1) \, .
% \ee
% In the same limit, the tortoise coordinate evaluates to
% \be 
% \xi - \xi_0 = - \frac{\ell_3^2}{r} + O(r^{-3}) \, .
% \ee
Let $z = \omega (\xi_0 - \xi)$. To leading order, the master equation (\ref{eq:schroprob1Dz}) reduces to 
\be 
\frac{\dn^{2} R}{\dn z^{2}} + \left(1 - \frac{3}{4 z^2}\right)R = 0 \, . 
\ee
Hence, the general solution can be written as
\be 
R_{(\infty)}(z) = A_1 \sqrt{2 \pi  z} \, J_1(z) + A_2 \sqrt{2 \pi z} \, Y_1(z) \, .
\ee
The requirement of normalizable boundary conditions at $r \to \infty$ requires us to set $A_2 = 0$. Expanding the resulting solution at large argument allows us to connect the result to the plane wave basis,
\be 
R_{(\infty)}(z) \sim A_1 e^{i \left(\frac{3 \pi}{4} - \omega \xi_0 \right)} e^{i \omega \xi} +  A_1 e^{-i \left(\frac{3 \pi}{4} - \omega \xi_0 \right)} e^{-i \omega \xi} \, .
\ee

 \paragraph{Near horizon solution.} We are assuming here a nondegenerate event horizon along the positive real axis at $r = r_h$. In the vicinity of the horizon, the tortoise coordinate behaves logarithmically, $\xi\sim\log(r-r_{h})$, cf. (\ref{eq:xilognearh}),
 and the potential vanishes linearly.
%at the horizon
% \be 
% V_h = f'(r_h) \left[\frac{m^2}{r_h^2} + \frac{f'(r_h)}{2 r_h} \right] (r-r_h) + O \left((r-r_h)^2 \right) \, .
% \ee
% In the vicinity of the horizon, the tortoise coordinate behaves logarithmically,
% \be
% \xi = \frac{1}{f'(r_h)} \log (r - r_h) + O(r-r_h) \, .
% \ee
Hence the local solution exhibits plane wave behavior
\be 
R_{(h)}(\xi) \sim B_1 e^{i \omega \xi} + B_2 e^{-i \omega \xi } \, .
\ee
The QNM boundary condition of purely ingoing waves at the horizon has us set $B_1 = 0$. Since the connection between the tortoise coordinate and the radius is logarithmic, subleading corrections to the plane wave solutions are exponentially suppressed near the horizon.

\paragraph{Near origin solution.} Near the origin, the potential behaves as
\be 
V_{(0)} = - \frac{(1+2s) a^2}{4} r^{-2(1+s)} + O\left(r^{-2-2s + {\rm min}(p, s)} \right) \, ,
\ee
while the tortoise coordinate behaves like 
\be \label{eq:tort_gen_origin}
\xi = -\frac{r^{s+1}}{a (s+1)} + O(r^{s+p+1}) \, .
\ee
For $z = \omega \xi$, the leading-order differential equation then takes the form
\be 
\frac{\dn^{2}R}{\dn z^{2}} + \left(1 + \frac{\frac{1}{4}-\nu^2}{z^2} \right) R = 0 \,, \qquad \text{where} \qquad \nu = \frac{s}{2 (s+1)} \, . 
\ee
The solution can then be expressed in terms of Bessel functions,
\be \label{eq:R0_gen_sol}
R_{(0)}(z) = C_1 \sqrt{2 \pi  z} J_\nu (z) + C_2 \sqrt{2 \pi  z} J_{-\nu }(z) \, .
\ee
Given our assumption that $s > 0$, the index of the Bessel function will never be an integer. Hence $\sqrt{2 \pi z} J_{\pm \nu}(z)$ provides an adequate basis. At large argument, the solution can be expanded in the plane wave basis,
\be 
R_{(0)}(z)  \sim \left(C_1 e^{-i \alpha_+} + C_2 e^{-i \alpha_-}\right) e^{i \omega \xi} + \left(C_1 e^{i \alpha_+} + C_2 e^{i \alpha_-}\right) e^{-i \omega \xi } \,, 
\ee
where we defined
\be 
\alpha_\pm = \frac{\pi}{4} \pm \frac{\pi \nu}{2} \, .
\ee
We emphasize that no boundary condition is imposed on the local solution near $r = 0$. 

\paragraph{Stokes rays and topology.} Above we have constructed local solutions in three regions: near infinity, near the horizon, and near the origin. The objective is now to obtain the QNM quantization condition by matching these solutions and their boundary conditions in the regime where the solutions are well-approximated by plane waves. However, in general such a matching is difficult because the solutions mix growing and decaying exponential terms. To circumvent this difficulty, we will perform the matching along Stokes rays where ${\rm Im}(\omega \xi) = 0$.\footnote{In the WKB literature, these are usually called \textit{anti}-Stokes lines. We will call them Stokes lines in keeping with the black hole literature.}

Near infinity, the Stokes ray will have ${\rm Im}(\omega \xi_0) = 0$ for large overtones. Given our contour prescription for the tortoise coordinate, the value of $\xi_0$ is fixed. As a matter of convention, on the Stokes ray connected to infinity we take ${\rm Re}(\omega \xi) > 0$ to leading order at large overtone. We then have 
\be 
{\rm Arg}(\omega) = - {\rm Arg}(\xi_0) \, .
\ee
Given  the relationship between $\xi$ and $r$, along with the branch choice ${\rm Re}(\omega \xi) > 0$, there is a single Stokes ray that extends to the AdS boundary. Near a horizon (be it real or complex) the Stokes line takes the form of a logarithmic spiral, just as we saw in the analysis of the neutral and charged qBTZ black holes in \eqref{eq:stokes_spiral_horizon}.  

% Near a horizon (be it real or complex), we have $\xi \propto {\rm log}(r-r_h) $. Let $r - r_h = \rho e^{i \vartheta}$. Then a Stokes line near the horizon satisfies
% \be 
% \vartheta = - \frac{{\rm Im}(\omega/f'(r_h))}{{\rm Re}(\omega/f'(r_h))} \, \log \rho + \, \text{const}. \, .
% \ee
% The Stokes line therefore spirals into the horizon in the complex radial plane. 

Finally, near the origin the tortoise coordinate admits the local expansion~\eqref{eq:tort_gen_origin}. Solving for the Stokes lines gives
\be \label{eq:stokes_sep_gen}
{\rm Arg}(r) = \frac{\pi k}{s+1} + \frac{{\rm Arg(a)}}{s+1} - \frac{{\rm Arg}(\omega)}{s+1} \, , \qquad k = 0, 1, 2, \dots \, .
\ee
A number of Stokes rays then emerge from $r = 0$ in the complex plane, separated by $\pi/(s+1)$. If $s$ is an integer, then there are $2(s+1)$ total rays emanating from there. However, if $s$ is non-integer, there will be a branch cut at the origin in general. 

We have therefore understood the \textit{local} geometry of the Stokes lines. However, we need to understand how the different Stokes lines connect. This requires global information about the metric function. Given the metric function $f(r)$, the Stokes rays can be obtained by solving a differential equation. Let $\sigma = \omega \xi \in \mathbb{R}$ be the coordinate along the Stokes ray. Then we have
\be 
\sigma \equiv \omega \xi = 
\omega \int^r \frac{{\rm d} \bar{r}}{f(\bar{r})} \qquad \Longrightarrow \qquad \frac{{\rm d} r}{{\rm d}\sigma} = \frac{f(r)}{\omega} \, .
\ee
In solving this differential equation, the only detail of $\omega$ that matters is its complex phase. The magnitude $|\omega|$ can be absorbed into the definition of $\sigma$. Hence, given our assumptions, without loss of generality, we can determine the Stokes lines by solving the equation
\be 
\frac{{\rm d} r}{{\rm d}\sigma} = \frac{f(r)}{e^{-i {\rm Arg}(\xi_0)}} \, . 
\ee
In practice, the equation is solved beginning near $r = 0$ (e.g.~taking $|r| \sim 10^{-5}$), choosing a value of $k$ in \eqref{eq:stokes_sep_gen} to set ${\rm Arg}(r)$, and integrating.

Given an explicit choice for the metric function $f(r)$, the above procedure allows for the construction of the Stokes lines. Here, we wish to perform the computation for a more general metric function. We therefore \textit{assume} that the Stokes topology is Schwarzschild-AdS-like. What we mean by this is the following. The relevant contour connecting infinity to the horizon has the structure: follow the Stokes line from infinity to $r = 0$, rotate counterclockwise by $\pi/(s+1)$ in the complex radius plane, and then follow the corresponding Stokes line to the horizon. This contour is known to be the correct one for neutral and charged qBTZ black holes. Its generality beyond that is, as we have said, an assumption.

\paragraph{Solution matching and QNM quantization.}

Given the assumption for the contour described above, we need to transform the near-origin solution \eqref{eq:R0_gen_sol} by a $\pi/(s+1)$ rotation in the complex-$r$ plane. This amounts to transforming $z \to z e^{i \pi}$. We then have
\be 
R_{(0)}^{(II)}(z) = C_1 e^{i 2 \alpha_+} \sqrt{2 \pi z} J_\nu (z) + C_2 e^{i 2 \alpha_-} \sqrt{2 \pi z} J_{-\nu} (z) \, .
\ee
We then expand the solution at large argument, expressed in the plane wave basis. The only subtlety here is that one must use the relationship between $z$ and $\xi$ that held on the infinity-to-origin portion of the contour, which involves the additional factor of $e^{i \pi}$. Expanding at large argument we then have:
\be \label{eq:gen_hor_R_cont}
R_{(0)}^{(II)}(z) \sim \left(C_1 e^{3 i \alpha_+} + C_2 e^{3 i \alpha_-} \right) e^{i \omega \xi} + \left( C_1 e^{i \alpha_+} + C_2 e^{ i \alpha_-} \right) e^{-i \omega \xi} \, .
\ee

Matching the solutions on the infinity-to-origin portion of the contour gives
\be \label{eq:infin-orgin-match}
\frac{C_2}{C_1} = - \frac{\sin \left(\omega \xi_0 - \frac{\pi \nu}{2} \right)}{\sin \left(\omega \xi_0 + \frac{\pi \nu}{2} \right)} \, .
\ee
Then, using this result in~\eqref{eq:gen_hor_R_cont} and setting the outgoing contribution to vanish yields the asymptotic QNM quantization condition,
\be \label{eq:qnm_quant_leading}
\omega \xi_0 = n \pi + \frac{i}{2} \log \left[2 \cos \frac{\pi s}{2(1+s)} \right]  \, .
\ee
As a simple check, one can verify that for $s = 1$ and $s=2$ this reduces to the qBTZ and charged qBTZ results, respectively. 

We therefore see that, if the global Stokes topology is consistent with our assumption, the scaling exponent of the singularity can be extracted from the offset in the asymptotic QNMs. For example, for the neutral and charged qBTZ black holes, the offset is a direct signal of the singularity.

\subsection{Subleading corrections}

Above we have obtained the leading-order QNMs at large overtone number. It is possible to go further than this and obtain subleading corrections by treating deviations from the exact Bessel problems that govern the near-singularity and near-boundary problems as perturbations~\cite{Musiri:2003bv, Cardoso:2003vt, Musiri:2005ev}. These corrections offer a particularly transparent way, largely independent of the Stokes topology, to read off the behavior of the singularity. Assumptions on the Stokes topology enter only insofar as the method needs that the black hole singularity appears in the dominant Stokes topology of the problem, along with the assumption that the coefficient of the leading $m^2$-dependent term does not vanish. 

\paragraph{Scaling of corrections with overtone number.} Let us start by illustrating how corrections to the potential near one of its poles lead to particular scalings of the QNM frequencies with the overtone number $n$. Suppose that near a second-order pole the potential has a subleading term of the form
\be 
V(\xi) = \frac{\nu^2 - 1/4}{\xi^2} + b \xi^{q-2} \, ,
\ee
where $q$ should not be confused with the charge parameter in the charged qBTZ metric.  Substituting $z = \omega \xi$, the corresponding radial equation is
\be 
\frac{\dn^{2}R}{\dn z^{2}} + \left[1 + \frac{1/4-\nu^2}{z^2}  - b \omega^{-q} z^{q-2} + \cdots \right] R = 0 \, .
\ee
Because $\omega \sim n$ at large overtone, provided $q > 0$ we have $\lambda \equiv -b \omega^{-q} \sim n^{-q}$ which serves as a parameter to perform perturbation theory in. Because $\lambda \sim n^{-q}$ at large overtone, it will therefore be the case that the correction to the QNM spectrum, expanded in $\lambda$, will have the leading behavior 
\be 
\delta \omega \propto n^{-q} \, .
\ee
This scaling is a consequence of the second-order pole of the potential near the origin and the assumption that the leading order QNMs scale linearly with $n$ at large overtone number. This does not make any specific assumptions about the topology of Stokes lines. 

\paragraph{Subleading corrections to the potential.} Having established the overtone scaling of subleading corrections to the potential, let us now proceed to identify the subleading corrections to the potential for our metric ansatz~\eqref{eq:metric_f_ansatz} and those arising from the angular momentum terms.  Given the near-origin behavior of~\eqref{eq:metric_f_ansatz}, we can first identify the correction to the tortoise coordinate,
\be 
\xi = - \frac{r^{s+1}}{a(s+1)} \left[1 - \frac{c(s+1)}{s+p+1} r^p + \cdots  \right] \, .
\ee
For notational convenience, let us define
\be 
\rho(\xi) \equiv [-a (s+1) \xi]_\Gamma^{1/(s+1)} \, ,
\ee
with the branch chosen such that $\rho/r \to 1$ on the origin-infinity Stokes ray. The subscript $\Gamma$ serves as a reminder that, because the objects inside the brackets are complex, one must take care with the corresponding phase in evaluating these quantities. Inverting this expression to obtain $r(\rho)$ gives
\be 
r = \rho \left[1 + \frac{c}{s+p+1} \rho^p + \cdots \right] \, .
\ee

Recall that for a massless scalar in three dimensions, the effective potential has the form~(\ref{eq:brane-potential-general}). 
% For a massless scalar in three dimensions, we recall the form of the effective potential,
% \be 
% V(r) = f(r) \left[\frac{m^2}{r^2} + \frac{f'(r)}{2r} - \frac{f(r)}{4 r^2} \right] \, .
% \ee
We are interested in seeing how the leading corrections to the metric and the angular momentum $m$ contributions modify the near-origin form of the potential. Expanding the potential near $r = 0$ and using~\eqref{eq:metric_f_ansatz} we write $V = V_f + V_m$ with
\begin{align}
    V_f(\xi(r)) &= a^2 r^{-2(s+1)} \left[-\frac{1+2s}{4} + \frac{c(p-2s-1)}{2} r^p + \cdots \right] \,, 
    \\
    V_m(\xi(r)) &= -a m^2 r^{-s - 2} \left[ 1 + c r^p + \cdots \right] \, .
\end{align}
After substituting for the tortoise coordinate (and using $\rho$ from above) we obtain
\begin{align}
    V_f(\xi) &= \frac{1}{\xi^2} \left[- \frac{1+2s}{4(s+1)^2} + \frac{cp(p-s)}{2 (s+1)^2 (s+p+1)} \rho^p + \cdots \right] \,,
    \\
    V_m(\xi) &= - \frac{m^2}{a(s+1)^2} \frac{\rho^s}{\xi^2} + \cdots  \, .
\end{align}
The leading term in $V_f(\xi)$ is simply the $1/\xi^2$ pole that makes the near-origin solution a Bessel problem. The next-to-leading term incorporates both the leading correction to the tortoise coordinate and the leading correction arising from the metric function. In $V_m(\xi)$ we include only the first correction incorporating angular momentum. In practice, the leading metric and angular-momentum corrections need not be consecutive terms in the asymptotic expansion of the potential --- additional contributions may occur at intermediate orders. Our aim here, however, is to isolate the leading correction associated with each of these two sources. 

Upon setting $z = \omega \xi$ and retaining the first metric and first angular momentum corrections, the differential equation near the origin becomes
\be 
\frac{\dn^{2} R}{\dn z^{2}} + \left[1 + \frac{1/4 -\nu^2}{z^2} + \lambda_f z^{q_f - 2} + \cdots + \lambda_m z^{q_m-2} + \cdots \right] R  = 0 \,,
\ee
where we defined 
\be \label{eq:q_lambda_metric}
q_f = \frac{p}{s+1} \, , \qquad \lambda_f =\frac{c p (s-p)}{2(s+1)^2(s+p+1)} \left[-\frac{a(s+1)}{\omega}\right]_\Gamma^{p/(s+1)} \, ,
\ee
which controls the leading correction due to the metric, and 
\be \label{eq:q_lambda_angmom}
q_m = \frac{s}{s+1} = 2 \nu \, , \quad \lambda_m = \frac{m^2}{a (s+1)^2}\left[-\frac{a(s+1)}{\omega}\right]_\Gamma^{s/(s+1)} \, ,
\ee
which controls the leading correction due to the momentum. Note that we have here once again given a subscript $\Gamma$ to the terms with the fractional powers as a reminder to take care with tracking the phases of these contributions. 

\paragraph{Near-origin solution for a general correction.}

Let us consider the modified near-origin equation
\be 
\frac{\dn^{2}R}{\dn z^{2}} + \left[1 + \frac{1/4 - \nu^2}{z^2} + \lambda z^{q-2} \right]R = 0
\ee
where $\lambda$ is a complex constant which we assume to have small modulus and $q$ is a real power. The leading order corrections arising from the metric and from the angular momentum terms both can be put into this form. We now solve for the leading correction to the QNM quantization condition. 

Let us identify $\mathcal{H}_0 = \partial_z^2 + 1 + (1/4-\nu^2)/z^2$ as the leading-order differential operator. Substituting $R = R^{(0)} + \lambda R^{(1)}$ and collecting powers of $\lambda$ gives
\be 
\mathcal{H}_0 R^{(0)} = 0 \, , \qquad \mathcal{H}_0 R^{(1)} = -z^{q-2} R^{(0)} \, .
\ee
Thus, the zeroth-order solution acts as a source for the correction. We can solve this problem using variation of parameters. To this end, let $R_\pm(z)$ be the two independent solutions of $\mathcal{H}_0 R^{(0)} = 0$. For the problem at hand we have $R_\pm (z) = \sqrt{2 \pi z} \, J_{\pm \nu}(z)$ which have Wronskian ${\cal W}[R_+, R_-] = -4 \sin (\pi\nu)$. We then write $R^{(1)} = c_1(z) R_+ + c_2(z) R_-$ and impose $c_1' R_+ + c_2' R_- = 0$ and find that the equation reduces to
\be \label{eq:subleading_corr_eq}
c_1' R_+' + c_2' R_-' = -  z^{q-2} R^{(0)} \, .
\ee
To solve this equation, it is useful to first introduce the integrals
\be 
H_{\sigma \tau} (z; q) = \int_{+\infty}^z {\rm d}t\, t^{q-1} J_{\sigma \nu}(t) J_{\tau \nu}(t) 
\ee
where $\sigma, \tau \in \{\pm 1\}$. These integrals will be convergent at the lower limit provided $q < 1$. This will always be the case for us here given the values for $q$ for the metric and angular momentum corrections.\footnote{Convergence at the upper limit requires care, and in some cases requires isolating the finite part of the integrals. We discuss this in Appendix~\ref{sec:convergence_suffering}.}We can then solve \eqref{eq:subleading_corr_eq} and obtain the solution with subleading correction as
\begin{align}
R(z) =&\, \left[C_1  - \frac{\lambda \pi}{2 \sin (\pi \nu)} \left(C_1 H_{+-}(z;q) + C_2 H_{--}(z;q) \right)\right] R_+(z) 
\nonumber 
\\
&+ \left[C_2 + \frac{\lambda \pi}{2 \sin (\pi \nu)} \left(C_1 H_{++}(z;q) + C_2 H_{+-}(z;q)\right) \right]R_-(z) + O(\lambda^2) \, . \label{eq:Rsol_subleading_pert_gen}
\end{align} 
We have chosen the integration constants such that $c_{1,2}(\infty) = 0$, ensuring that the large-$z$ limit of the solution agrees with~\eqref{eq:R0_gen_sol}. As a consequence, the matching to the large-$r$ solution is unaltered, leading to~\eqref{eq:infin-orgin-match}.

\paragraph{Modified matching conditions.}  As before, we match the solutions at infinity and the horizon. We do this by following a Stokes ray from infinity to $r = 0$, performing a counterclockwise rotation in the complex $z$-plane by $\pi$, and then following a Stokes ray to the horizon. For the leading-order solution, this amounts to analytic continuation of the Bessel functions with the only subtlety arising from the rotation near $r = 0$, which gives $R_\pm \to e^{2 i \alpha_\pm} R_\pm$. The perturbed solution is more subtle because we must track the integrals along the contour so that we can identify the accumulated phase difference. To be clear about this, let us call the specified contour $\Gamma$ and define
\be \label{eq:Delta_gen_contour}
\Delta_{\sigma \tau}(q) = \int_\Gamma {\rm d}t \, t^{q-1} J_{\sigma \nu}(t) J_{\tau \nu}(t) \, . 
\ee
We relegate the convergence analysis of this integral to Appendix~\ref{sec:convergence_suffering}, and present the final result:
\be 
\Delta_{\sigma \tau}(q) = \left(e^{i \pi [q + (\sigma + \tau)\nu]} - 1\right) I_{\sigma \tau} \, , \quad I_{\sigma \tau} = \int_0^\infty {\rm d}t \, t^{q-1} J_{\sigma \nu}(t) J_{\tau \nu}(t) \, ,
\ee 
where $I_{\sigma \tau}$ is a Weber-Schafheitlin integral which can be written as~\cite{spWeber} 
\be \label{eq:Weber-Schafheitlin}
\int_0^\infty {\rm d}t\,t^{q-1}J_\alpha(t)J_\beta(t)
=\frac{2^{q-1}\Gamma(1-q)\Gamma((\alpha+\beta+q)/2)}
{\Gamma(1+(\beta-\alpha-q)/2)\Gamma(1+(\alpha+\beta-q)/2)
\Gamma(1+(\alpha-\beta-q)/2)} \, ,
\ee
and convergence requires $2\nu < q < 1$. More specifically, $I_{++}$ requires $q + 2 \nu > 0$, $I_{+-}$ and $I_{-+}$ require $q > 0$, and $I_{--}$ requires $q - 2 \nu >0$. This means that the case $q = 2 \nu$, which is realized by the angular momentum term, requires a bit of separate care. Moreover, when $0 < q < 2\nu$ the combined $--$ integrals along the full contour can be naturally defined as the finite part of the above. We address convergence in the different required cases in Appendix~\ref{sec:convergence_suffering}. In the present case we have
\be 
I_{++} = \frac{2^{q-1} \Gamma (1-q) \Gamma(\nu +q/2)}{\Gamma^2(1-q/2) \Gamma(1+ \nu - q/2)} \, ,
\ee
while the remaining two integrals can be expressed as ratios
\be 
\frac{I_{-+}}{I_{++}}=\frac{I_{+-}}{I_{++}} = \frac{\sin \left[\pi (q/2 +\nu) \right]}{\sin (\pi q/2)} \, , \qquad \frac{I_{--}}{I_{++}} = \frac{\sin \left[\pi(q/2 + \nu) \right]}{\sin \left[\pi (q/2 - \nu) \right]} \, .
\ee
Here we have used the identity $\Gamma(x) \Gamma(1-x) = \pi/ \sin (\pi x)$ in simplifying the results. Putting this together we have
\be \label{eq:Delta_vals_horizon}
\Delta_{+-} = e^{-i\pi \nu} \Delta_{++} \,, \quad \Delta_{--} = e^{-2 i \pi \nu} \Delta_{++} \,, \quad \Delta_{++} = 2 i e^{i \pi(q/2+\nu)} \sin \left[\pi(q/2 + \nu)\right] I_{++} \, .
\ee
In the case $q = 2 \nu$, the only potentially problematic term is $\Delta_{--}$, which should then be understood as a limit. Specifically, the part of the contour where the solution is rotated from the infinity-origin curve to the origin-horizon curve makes a non-vanishing contribution and we have (see Appendix~\ref{sec:convergence_suffering})
\be 
\Delta_{--}(2\nu) = \frac{i \pi 2^{2\nu}}{\Gamma^2(1-\nu)} \, .
\ee

The behavior of $R(z)$ near the horizon is then given by substituting $H_{\sigma \tau} \to \Delta_{\sigma \tau}$ in solution~\eqref{eq:Rsol_subleading_pert_gen} using the results of ~\eqref{eq:Delta_vals_horizon}. We then find in the large-$z$ expansion the following plane wave asymptotics,
\be 
R^{(II)}(z) \sim B^{(II)}_+ e^{i \omega \xi} + B^{(II)}_- e^{-i \omega \xi} 
\ee
with
\begin{align}
B^{(II)}_- &= C_1 e^{i \alpha_+} + C_2 e^{i \alpha_-}  +O(\lambda^2)\,,
\\
B^{(II)}_+ &=  C_1 e^{3 i \alpha_+} + C_2 e^{3 i \alpha_-} + \pi \lambda \Delta_{+-} \left(C_1 e^{i \alpha_+} + C_2 e^{i \alpha_-} \right) + O(\lambda^2) \, .
\end{align}
The QNM boundary conditions then require $B_+^{(II)} = 0$, together with the matching~\eqref{eq:infin-orgin-match}. 

To solve for the corrected QNMs, define for convenience $X = \omega \xi_0$ and let
\be 
\gamma(X) \equiv - \frac{\sin \left(X - \pi \nu/2 \right)}{\sin \left(X + \pi \nu/2 \right)}  = \frac{C_2}{C_1} \, .
\ee
The ingoing boundary condition at the horizon then becomes
\be 
\gamma(X) + e^{3 i \pi \nu} - i \pi \lambda \Delta_{++} \left[ e^{i \pi \nu} + \gamma(X)\right] = 0 + O(\lambda^2)
\ee
% \be 
% \eta(X) + \frac{\pi \lambda \Delta_{++}}{2 \sin(\pi \nu)} \left( 1 + e^{-i \pi \nu} \eta(X) \right)^2  = - e^{3 i \pi \nu} + O(\lambda^2)
% \ee
To solve this, let us write $X = X_0 + \delta X$ and linearize to obtain the perturbative correction,
\be 
\delta X = \frac{\pi \lambda \Delta_{++}}{2 \left( 1 + e^{2 i \pi \nu}\right)}\, .
\ee
Fully explicitly, this is
\be 
\delta \omega \,\xi_0  = \frac{i \pi^2 e^{i \pi q/2} \Gamma(1-q)}{2^{2-q} \cos(\pi\nu) \Gamma^2(1-q/2) \Gamma(1+\nu -q/2) \Gamma(1-\nu-q/2)} \lambda \, ,
\ee
which is valid for $0 < q < 1$.  Here $\lambda$ itself depends on $\omega$. However, in the leading-order correction, we can substitute $\omega \to  n \pi/\xi_0$. This formula constitutes the main result of our calculations in this section. 

The result can be specialized to the metric and angular momentum corrections by substituting the values of $\lambda$ and $q$ as given in Eqs.~\eqref{eq:q_lambda_metric} and \eqref{eq:q_lambda_angmom}. More explicitly, for the metric correction we have
\be 
\delta \omega_f = \frac{\pi c p (s-p) \Delta_{++}}{4 (s+1)^2 (s+p+1) \xi_0 (1 + e^{2 i \pi \nu})}  \left[-\frac{a(s+1) \xi_0}{\pi}\right]_\Gamma^{p/(s+1)} n^{-p/(s+1)} \, ,
\ee
and for the angular momentum one 
\be \label{eq:ang_mom_correction_gen}
\delta \omega_m = \frac{i m^2 \pi^2 e^{i \pi \nu}}{2^{2-2\nu} a (s+1)^2 \xi_0 \cos(\pi \nu) \Gamma^2(1-\nu)} \left[-\frac{a (s+1) \xi_0}{\pi}\right]_\Gamma^{2 \nu} \, n^{-2 \nu } \, , \quad \nu = \frac{s}{2 (s+1)} \, .
\ee
We again stress that these corrections need not appear consecutively in the large overtone expansion of the QNMs.

\paragraph{Example: neutral qBTZ. } For the neutral qBTZ spacetime we have $s = 1$ and $p = s$. Hence the leading correction coming from the metric vanishes. We therefore have the QNMs with subleading corrections given by
\be \label{eq:neut-qbtz-subleading-analytical}
\xi_0 \omega_{m,n} = n \pi + \frac{i}{4} \log 2 + \frac{e^{i \pi/4} \Gamma^2(1/4)}{8 \sqrt{2 \pi }} m^2 \sqrt{\frac{\xi_0}{\ell W(M)}} \, n^{-1/2} + \cdots \, .
\ee

\paragraph{Example: charged qBTZ. } For the charged qBTZ black hole we have $s = 2$ and $p = 1$. The leading-order correction to the spectrum reads
\be  \label{eq:charged-qbtz-subleading-analytical}
\xi_0 \omega_{m,n} = n \pi  - \frac{\Gamma(1/3) W(M, Q)}{48 \ell P(M, Q)} \left( \frac{3 \ell^2 P(M, Q) \xi_0}{\pi} \right)^{1/3} \, n^{-1/3} + O(n^{-2/3})  \, ,
\ee
while the angular momentum contribution reads
\be \label{eq:charged-qbtz-subleading-analytical-ang-mom}
\xi_0 \, \delta \omega_{m, n} = \frac{i \Gamma(1/3)^2 m^2}{ 2^{1/3} \, 12 \, \ell^2 P(M, Q)} \left( \frac{3 \ell^2 P(M, Q) \xi_0}{\pi} \right)^{2/3} \, n^{-2/3}\, .
\ee
Here we have written the angular momentum contribution separately, because there is a further subleading correction from the metric that enters at the same order, which we have not computed the coefficient of in detail. 

\subsection{Extracting singularity data: a recipe}

Let us now return to the question posed at the beginning of this section: given a QNM spectrum, can one reverse engineer any information about the singularity? We have seen from the leading QNM spectra \eqref{eq:qnm_quant_leading} that the scaling exponent of the singularity is encoded in the offset of the large overtone QNMs. However, this result was not satisfactory because computing the offset requires an assumption on the topology of the Stokes lines. We then computed various subleading corrections to the large overtone asymptotics. The coefficients of these subleading corrections also depend sensitively on the Stokes topology and choice of contour. However, remarkably, their scaling with $n$ provides a means to identify the singularity scaling that is independent of the Stokes topology. 

This is most clear for corrections proportional to the square of angular momentum, for which we found 
\be 
\delta \omega_{m,n} \propto m^2 n^{-2 \nu} \, , \quad \nu = \frac{s}{2 (s+1)}
\ee
where the precise proportionality constant depends on the behavior of the Stokes lines. We calculated this explicitly for a Schwarzschild-AdS-like Stokes topology in Eq.~\eqref{eq:ang_mom_correction_gen}. Given a QNM spectrum, including large overtones, for any two angular quantum numbers $m_1^2 \neq m_2^2$ we have
\be \label{eq:Delta_two_mode}
\Delta_n \equiv \omega_{m_1,n} - \omega_{m_2, n} \propto (m_1^2-m_2^2) \,  n^{-2\nu}  \, .
\ee
By fitting the resulting data, it is possible to extract the value of $\nu$, and hence the scaling of the metric function in the vicinity of the singularity. 

\paragraph{Generalization to two blackening factors.} Throughout, we have assumed that the metric under consideration is described by a single blackening factor $f(r)$. This is the case most relevant to us here. However, it is not the most general case. In general, a static and circularly symmetric spacetime in three dimensions will be described by two blackening factors when written in terms of the `circumferential radius' $r$,
\be 
{\rm d}s^2 = -f(r) {\rm d}t^2 + \frac{{\rm d}r^2}{g(r)} + r^2 {\rm d}\phi^2 \, .
\label{eq:ansattwoblack}\ee
Assuming the spacetime is asymptotically AdS, our scaling argument for the large overtone quasinormal modes goes through with minimal modification. 

Suppose that the near-singularity metric is described by two independent scaling exponents,
\be 
f(r)  \sim  -a r^{-s_1}  \, , \quad g(r) \sim - b r^{-s_2}\,,  \qquad s_1, s_2 > 0 \, .
\ee
We assume that $a b > 0$ so that the geometry is Lorentzian. The tortoise coordinate is then defined as $\xi(r) = \int^r {\rm d}{\bar r}/\sqrt{fg}$ which near the singularity behaves like
\be 
\xi \propto r^{(s_1 + s_2+2)/2} \, .
\ee
As a result, the angular momentum contribution to the potential scales like
\be 
V_m(\xi) = \frac{m^2 f(r)}{r^2} \propto m^2 \xi^{-2(s_1+2)/(s_1+s_2+2)} \, .
\ee
Thus, assuming $\omega \sim n$ at large overtone, the same scaling argument gives
\be 
\delta \omega_{m,n} \propto m^2 n^{-q_m} \, , \quad \text{where} \quad q_m = \frac{2 s_2}{s_1+s_2+2} \, .
\ee
For $0 < q_m < 1$, this will be the leading angular momentum correction within the perturbative regime. Setting $s_1 = s_2 = s$ recovers the scaling $n^{-s/(s+1)}$, in agreement with our analysis. 

A key difference introduced by two blackening factors is that the angular momentum corrections cannot disentangle both scaling exponents individually. This could perhaps be done if the underlying theory and metric function were known, thereby allowing one to compute the coefficients of various corrections, which are sensitive to the Stokes topology.  One possible means to disentangle the scaling exponents would be by examining several different probe fields, e.g. a charged scalar, or including mass terms.\footnote{Ideas along these lines have been pursued in~\cite{Xiao:2026pir}, which appeared when this article was in preparation.} Another possibility is to use additional input from the theory, as we shall discuss below. 

\paragraph{Spacelike singularities: Kasner exponents.} Recall that near the spacelike singularity of a generic $D$-dimensional black hole the geometry has a universal Kasner-form
\beq ds^{2}=-d\tau^{2}+\sum_{i=1}^{D-1}\tau^{2p_{i}(x)}(e^{i})^{2}\;,\label{eq:Kasgeomgen}\eeq
when spatial curvature is negligible. 
Here $\tau$ is an ultralocal proper time coordinate, $e^{i}$ are co-frame one-forms, and, in vacuum Einstein gravity, the Kasner exponents satisfy $\sum_{i=1}^{D-1}p_{i}=\sum_{i=1}^{D-1}p_{i}^{2}=1$. The gravitational dynamics features ultra-locality and oscillatory, chaotic dynamics \cite{Belinsky:1970ew}; the full evolution consists of a sequence of Kasner epochs and eras connected
by transitions when the exponents change \cite{Belinski:2017fas} but continue to satisfy the additive constraints.

% Quantum effects, when manifested as higher-derivative corrections to general relativity, garner new transitions captured by Kasner \emph{eons}, periods of time (longer than epochs and eras) dominated by semiclassical/quantum gravitational physics \cite{Bueno:2024fzg}. Deep in the interior, quantum effects are expected to become dominant, which should correspond to a transition from a classical (Einsteinian) eon to a ``quantum eon''. 

Consider again our ansatz with two blackening factors (\ref{eq:ansattwoblack}).
In the case where $a, b > 0$ the singularity is spacelike. In the approach to such a singularity $r$ is the timelike coordinate. We can cast the local metric into a Kasner form by introducing the local proper time coordinate,
\be 
{\rm d} \tau = \frac{{\rm d} r}{\sqrt{-g(r)}} \,, \quad \Rightarrow \quad \tau \propto r^{(s_2 + 2)/2} \, .
\ee
After constant rescalings of the spatial coordinates, the near-singularity metric becomes
\be 
{\rm d} s^2 = - {\rm d}\tau^2 + \tau^{2p_t} {\rm d}t^2 + \tau^{2 p_\phi} {\rm d}\phi^2 \, ,
\label{eq:kasmet3d}\ee
with Kasner exponents
\be 
p_t = -\frac{s_1}{s_2+2} \,, \qquad p_\phi = \frac{2}{s_2+2} \, .
\label{eq:kasexp3d}\ee
In the case where there is a single blackening factor and the two scaling exponents $s_i$ are equal, it becomes possible to read off from the asymptotic QNM spectrum the Kasner exponents of the singularity. Meanwhile, in the case of two blackening factors, it is possible to read off the following specific combination of Kasner exponents,\footnote{As this article was in preparation, this result was independently uncovered in \cite{Hartnoll:2026vhu}.}
\be 
q_m = \frac{2(1-p_\phi)}{1 - p_t} \, .
\ee
To obtain both Kasner exponents, additional information is needed. For example, in certain theories of gravity the sum of the Kasner exponents always has a fixed, solution-independent value~\cite{Bueno:2024qhh}. Therefore, if one has prior knowledge of the theory and can safely assume the singularity is spacelike,  both Kasner exponents can be disentangled.  In our case of interest, however, the Kasner exponents arise from a higher-dimensional bulk spacetime, and hence this logic cannot be so straightforwardly applied. 

\section{Numerical results}
\label{sec:numericmethods}

In this section we numerically determine the asymptotic QNMs of neutral and charged circularly symmetric quantum-corrected black holes, finding precise agreement with our analytical computations in Sections~\ref{sec:asympqnms} and~\ref{sec:lookinqBH}. Our results are based on two independently implemented numerical schemes. First, we utilize a pseudospectral method to obtain on the order of fifty overtones. Second, for the AdS black holes, we employ a determinant-based Leaver method, which allows us to probe much higher overtones. The implementation of the latter method was performed using ChatGPT, while the pseudospectral method was implemented by the human authors. We discuss both methods, and benchmark their results, in Appendix~\ref{app:numerical_methods}.

\subsection{Numerical validation of leading order analytical results}  
We compute the QNM spectra with the pseudospectral method for uncharged qBTZ, charged qBTZ and neutral, asymptotically flat quantum black holes across various values of the azimuthal quantum number $m$, black hole mass $\widetilde M$, and charge $\widetilde Q$, while fixing $\ell_4=1$ for the asymptotically AdS quantum black holes. Here, $\widetilde M=\mu \ell/2$ and $\widetilde Q=q \ell$ are rescaled quantities defined such that the four-dimensional black hole mass and charge remain finite in the limit $\ell\to\infty$. The frequencies  are calculated and crosschecked using different numbers of collocation points as a convergence check. The numerical and analytical values for the asymptotic QNMs agree across all cases, validating our analytical results. 

\begin{figure}[!ht]
\centering
\includegraphics[width = \linewidth]{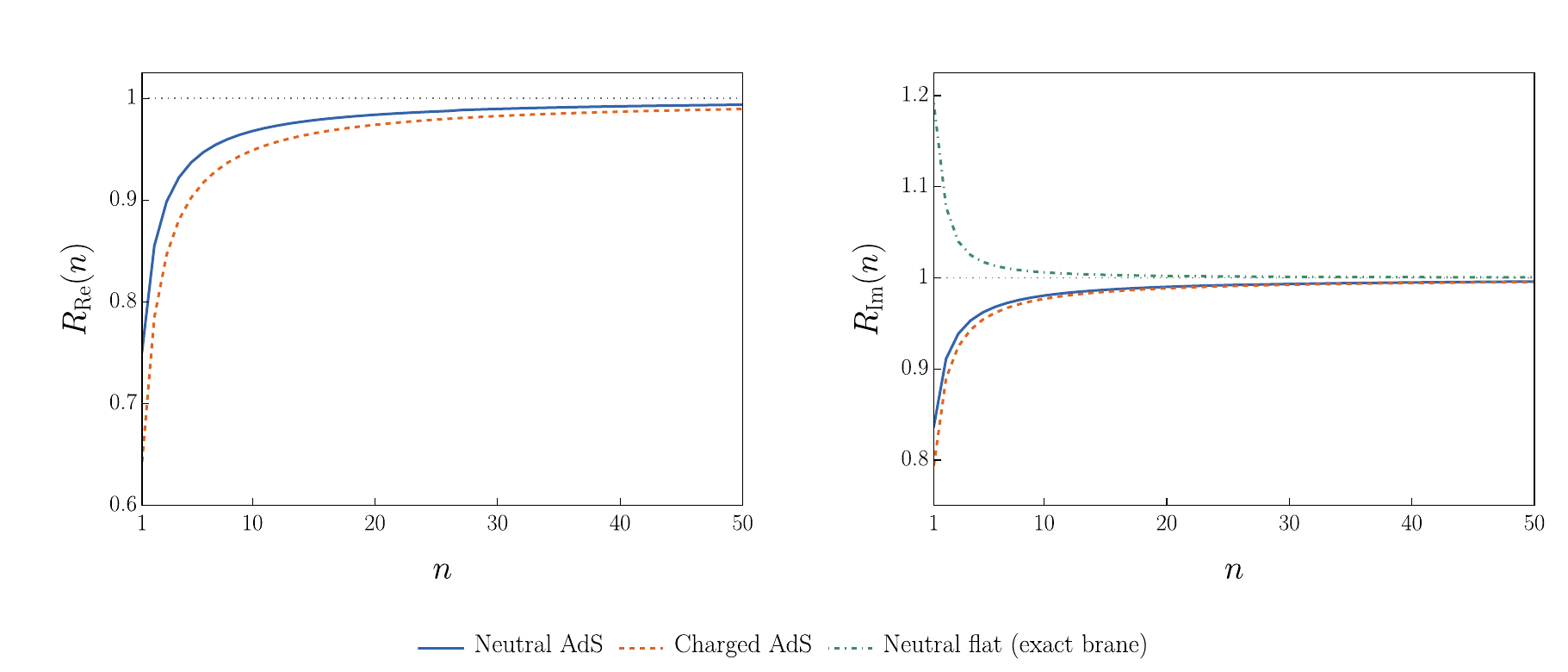}
\caption{Ratios between the real parts (left) and imaginary parts (right) of the QNMs of the quantum corrected black holes in the tensionless limit and those of the corresponding bulk black holes. For the uncharged AdS and asymptotically flat cases we set $\widetilde M=1$ and the azimuthal quantum number $m=0$. For the charged AdS case, we instead fix the event horizon radius to $r_h=5$ and choose the charge $\widetilde Q/\widetilde Q_{\text{extremal}}=2/3$. The modes of the bulk counterparts are computed with angular quantum numbers $s=l=0$.  All spectra are obtained numerically with the pseudospectral method, apart from the  flat quantum corrected black hole, for which we use the analytical result derived in Appendix~\ref{app:zeroReFlat}. In the flat case shown here, the
real part of the QNMs vanishes and it is therefore not included in the left panel; Appendix~\ref{app:zeroReFlat} also explains this behavior. As the overtone number $n$ increases, the ratios tend to 1, showing that the brane and bulk spectra agree increasingly well in the asymptotic regime.}
\label{fig:ratio-data-compare}
\end{figure}

We first compare the QNMs of the bulk and brane black holes in the tensionless case. In Figure~\ref{fig:ratio-data-compare} we show the ratios $R_{\rm Re}(n)$ and $R_{\rm Im}(n)$, defined respectively as the
ratios between the real and imaginary parts of the QNMs of the quantum corrected black holes in the tensionless limit and those of the corresponding bulk black holes.  We take $m=0$ throughout. For the asymptotically flat case, we use the analytical expression for the three-dimensional quantum black hole QNMs derived in Appendix~\ref{app:zeroReFlat}, while for all the other cases we use the numerical QNM spectra obtained with the pseudospectral method. In all cases, as the overtone number $n$ increases, the ratios approach unity, showing that the tensionless quantum corrected spectra converge at leading order to the corresponding bulk spectra in the asymptotic regime. 

In \Cref{tab:uncharged_summary,tab:summary-charge,tab:summary-flat}, we compare the analytical prediction for the offset and gap derived in Section~\ref{sec:asympqnms} with fits to the numerical QNMs obtained via the pseudospectral method, for several values of $\ell$. Regarding the charged qBTZ case, we find that the analytical approximation improves with increasing charge, as the neglected subleading terms in the effective potential become progressively smaller. Near the origin, at the next-to-leading order, the potential behaves as:
\begin{equation}
    V_{\rm uncharged} \approx-\frac{3}{16 \xi^2}+\frac{i m^2 }{4 \sqrt{\widetilde{M}} \xi^{3/2}}, \qquad  V_{\rm charged} \approx-\frac{5}{36 \xi^2}+\frac{4 {\widetilde M}}{12\times 3^{2/3} \widetilde{Q}^{4/3} \xi^{5/3}}.
\end{equation}
These expressions show that, in the uncharged case, the accuracy of the leading approximation improves as the mass increases. In the charged case, the subleading correction depends on both  the mass and charge. At fixed mass, increasing the charge suppresses this correction, improving the agreement with the analytical prediction. This behavior is also confirmed numerically, as shown in Table~\ref{tab:changingcharge}. Overall, the agreement between the fitted numerical spectra and the analytical expressions confirms that the pseudospectral computation validates the leading order analytical approximation to the asymptotic QNMs derived in Section~\ref{sec:asympqnms}.

\begin{table}[!htb]
\centering
\small
\begingroup
\setlength{\tabcolsep}{7pt}
\renewcommand{\arraystretch}{1.15}
\begin{tabular}{ccccc}
\toprule
$\ell$ &  $\pi/\xi_0$ (analytic) & $\omega^{\text{gap}}$ (fit)  & $i \log 2/(4\xi_0)$ (analytic) & $\omega^{\text{offset}}$ (fit)   \\
\midrule
$\infty$  & $1.969104-2.350189\,i$ & $1.969100-2.350189\,i$ & $0.129634+0.108614\,i$ & $0.129787+0.108625\,i$\\
%1000  & $1.965178-2.345505\,i$ & $1.965175-2.345505\,i$& $0.129375+0.108397i$ & $0.129527+0.108406i$ \\
500  & $1.961280-2.340851\,i$ & $1.961276-2.340853\,i$ & $0.129119+0.108182\,i$ & $0.129292+0.108213\,i$\\
%200  & $1.949744-2.327076\,i$ & $1.949741-2.327076\,i$ & $0.128359+0.107546i$ & $0.128508+0.107556i$\\
100  & $1.931021-2.304706\,i$ & $1.931018-2.304707\,i$ & $0.127125+0.106513\,i$ & $0.127267+0.106520 \,i$\\
50  & $1.895329-2.262016\,i$ & $1.895324-2.262017\,i$ & $0.124770+0.104544\,i$ & $0.124912+0.104555\,i$\\
\hline
\end{tabular}
\endgroup
\caption{Test of the analytical prediction (\ref{eq:analytic_omega1}) for the neutral quantum BTZ black hole at several values of the backreaction parameter $\ell$, with $\kappa=1$, $\widetilde M=1$, $\ell_4=1$, $m=0$. The complex tortoise length $\xi_0$ is computed from (\ref{eq:xi0_sum}) using the roots of $f$, independently of the QNM calculation. The fitted slope $\omega^{\text{gap}}$ is obtained from a linear fit $\omega(n)=\omega^{\text{offset}}+\omega^{\text{gap}}n$ to the numerical spectrum over $n\geq20$. Note that $\pi/\xi_0$ depends on $\ell$, so the asymptotic spacing agrees with the bulk value only in the tensionless limit.}
\label{tab:uncharged_summary}
\end{table}

\begin{table}[!htb]
\centering
\begingroup
\setlength{\tabcolsep}{7pt}
\renewcommand{\arraystretch}{1.15}
\begin{tabular}{ccc}
\toprule
$\ell$ &  $\pi/\xi_0$ (analytic) & $\omega^{\text{gap}}$ (fit)  \\
\midrule
$\infty$ &  $1.891687-2.352511\,i$ & $1.886244-2.352321\,i$\\
500 &  $1.887913- 2.347815 \,i$& $1.878502-2.341922\,i$\\
100 &  $1.855082-2.306954\,i$ & $1.850034-2.307479\,i$ \\
50 &  $1.820736 - 2.264140\,i$ & $1.815259 - 2.262940\,i$ \\
\hline
\end{tabular}
\endgroup
\caption{Test of the analytical prediction (\ref{eq:asymqnmwchar}) for the charged quantum BTZ black hole at several values of the backreaction parameter $\ell$, with $\kappa=1$, $\widetilde M=1$, $\ell_4=1$, $m=0$, $\widetilde Q/\widetilde Q_{\text{extremal}}=0.3$. The complex tortoise length $\xi_0$ is computed from (\ref{eq:xi0_sum}) using the roots of $f$, independently of the QNM calculation. The fitted slope $\omega^{\text{gap}}$ is obtained from a linear fit $\omega(n)=\omega^{\text{offset}}+\omega^{\text{gap}}n$ to the numerical spectrum over $n\geq20$. Note that $\pi/\xi_0$ depends on $\ell$, so the asymptotic spacing agrees with the bulk value only in the tensionless limit.}
\label{tab:summary-charge}
\end{table}

\begin{table}[!htb]
\centering
\renewcommand{\arraystretch}{1.15}
\begin{tabular}{@{}lccc@{}}
\toprule
Charge $\widetilde Q$ & $1.5$ & $3$ & $4$ \\
\midrule
Relative error $(\%)$ & $0.267$ & $0.038$ & $0.028$ \\
\bottomrule
\end{tabular}
% \begin{tabular}{cc}
% \toprule
% $q$ & $\left|\frac{\omega^{\rm gap}_{\rm fit}-\omega^{\rm gap}_{\rm ana}}{\omega_{\rm ana}^{\rm gap}}\right|$  \\
% \midrule
% 1.5 & $0.267\, \%$  \\
% 3 &  $0.038\, \%$\\
% 4 & $0.028\, \%$\\
% \hline
% \end{tabular}
\caption{Comparison between the analytical prediction \eqref{eq:asymqnmwchar} and numerical results for the charged quantum BTZ black hole in the tensionless limit, with $\kappa=1$, $\widetilde M=10$, $\ell_4=1$, $m=0$, varying the charge of the black hole. The numerical frequency gap $\omega^{\rm gap}_{\rm fit}$ is extracted from a linear fit to the spectrum for $n\geq20$. We report the (percentage) relative error $\left|(\omega^{\rm gap}_{\rm fit}-\omega^{\rm gap}_{\rm ana})/\omega_{\rm ana}^{\rm gap}\right|$ across several values of the  charge $\widetilde Q$. Note that the extremal charge is $\widetilde Q_{\text{extremal}}\approx4.78$. The analytical approximation becomes increasingly accurate as the charge increases. }
\label{tab:changingcharge}
\end{table}

\begin{table}[!htb]
\centering
\renewcommand{\arraystretch}{1.15}
\begin{tabular}{ccccc}
\toprule
$\ell$ &  $-i\kappa_h $ (gap analytic) & $\omega^{\text{gap}}$ (fit) & $i \kappa_h/2$ (offset analytic) & $\omega^{\text{offset}}$ (fit)  \\
\midrule
$\infty$  & $-0.25\,i$ & $-0.250005\,i$ & $0.125\,i$ & $0.125313\,i$\\
500 & $-0.248759 \,i$ & $ -0.249033 \,i$ & $0.124379\, i$ & $0.123204 \,i$ \\
100 & $ -0.245180\,i$ & $-0.245249\,i$ & $0.122590\, i $& $0.124902\,i$\\
50 & $-0.240691\,i$ & $-0.240700\,i$ & $0.120346\,i$ & $0.120744\,i$ \\
\hline
\end{tabular}
\caption{Test of the analytical prediction (\ref{eq:flat-monodromy}) for the flat  quantum corrected black hole at several values of the backreaction parameter $\ell$, with $\kappa=1$, $\widetilde M=1$, $m=0$. Here $\kappa_h$ is the surface gravity at the horizon. The fitted parameters are obtained from a linear fit $\omega(n)=\omega^{\text{offset}}+\omega^{\text{gap}}n$ to the numerical spectrum over $n\geq20$. }
\label{tab:summary-flat}
\end{table}

\subsection{Subleading corrections to the QNM spectrum}

In this section, we consider the subleading corrections to the asymptotic QNM spectrum. Accurately extracting these corrections requires computing a large number of modes, and we therefore make use of our determinant-based solver. In addition, because these calculations require very high working precision, we fix the solution parameters $\{\kappa, x_1, \ell_3, q\}$ and treat the physical parameters as derived quantities. This allows the solution parameters that enter our numerics to be specified exactly---for example, $x_1=1$. By contrast, fixing physical parameters such as $Q$ or $M$ would require solving cubic or quartic equations to determine the quantities entering the calculation. Doing so would substantially increase the required numerical precision without materially affecting our conclusions.

To extract the offset and subleading corrections, we fit our numerically determined frequencies at fixed $m$ to 
\be 
 \omega_{m,n}^{\mathrm{fit}}
 = \widehat{s}_m n+\widehat{b}_m
   +\sum_{j=1}^{K}c_{m,j}n^{-j\alpha},
 \label{eq:fit_ansatz}
\ee
where $\alpha=1/2$ for neutral qBTZ and $\alpha=1/3$ for charged qBTZ. These powers are chosen based on the expectations provided by the analytics. 

For the high-overtone comparisons of offsets and coefficients of the corrections, our principal fits use $K=3$ and $500\leq n\leq1000$. We perform an unweighted linear least-squares fit to the complex frequencies, leaving the spacing, offset and all correction amplitudes free. Each set of constant $m$ modes is fitted independently in the same way. To compare with the analytical expansion of $\xi_0\,\omega_{m,n}$, we define 
\be  
\widehat{\mathcal B}_m=\xi_0\widehat{b}_m, \qquad \text{and} \qquad \widehat{\mathcal C}_j(m)=\xi_0 c_{m,j}\,.
\ee

To extract the leading angular momentum dependent coefficient, we form the two combinations
\begin{align}
 \widehat{\mathcal A}_{(1,2)}
 &=\frac{\widehat{\mathcal C}_{j_*}(2)
              -\widehat{\mathcal C}_{j_*}(1)}{3},
 \label{eq:angular-two}\\
 \widehat{\mathcal A}_{4}
 &=-\frac{49}{36}\widehat{\mathcal C}_{j_*}(0)
   +\frac32\widehat{\mathcal C}_{j_*}(1)
   -\frac3{20}\widehat{\mathcal C}_{j_*}(2)
   +\frac1{90}\widehat{\mathcal C}_{j_*}(3),
 \label{eq:angular-combinations}
\end{align}
where $j_*=1$ for neutral qBTZ and $j_*=2$ for charged qBTZ. Both combinations are designed to cancel contributions independent of $m$ and preserve the coefficient of $m^2$. The four-point combination additionally cancels terms proportional to $m^4$ and $m^6$ in an angular expansion, although it does not remove subleading overtone corrections proportional to $m^2$. This subtraction is particularly useful in the charged case. That is because the full $n^{-2/3}$ coefficient contains an $m$-independent contribution which we have not computed. This term cancels in the subtractions. We can therefore compare the remaining angular coefficient directly with the analytical prediction without assuming a value for that contribution.

\paragraph{Extraction of the offset.} We first compare the fitted offset term with the analytical prediction
\be
 \xi_0\omega_{m,n}=\pi n+\mathcal B_{\rm off} + \cdots ,
 \qquad
 \mathcal B_{\rm off}=
 \begin{cases}
  \dfrac{i\log 2}{4},&\text{neutral},\\[4pt]
  0,&\text{charged}.
 \end{cases}
\ee
Table~\ref{tab:offset_fits} reports the real and imaginary parts of $\widehat{\mathcal B}_0-\mathcal B_{\rm off}$, obtained from the $m=0$ spectra. Agreement with the analytical prediction corresponds to both quantities vanishing.

\begin{table}[ht]
    \centering
    \small
    \begin{tabular}{llrrr}
    \toprule
    Case & $\ell$ & Fit Window & ${\rm Re} \left[\widehat{\mathcal B}_0- \mathcal{B}_{\rm off}\right]$ & ${\rm Im} \left[\widehat{\mathcal B}_0 - \mathcal{B}_{\rm off}\right]$
    \\
    \midrule 
    Neutral & $10^{-2}$ & 500--1000 & $-5.29 \times 10^{-8}$ & $2.63 \times 10^{-7}$
    \\
    Neutral & 1 & 500--1000 & $2.56 \times 10^{-9}$ & $1.01 \times 10^{-9}$
    \\
    Neutral & 100 & 500--1000 & $4.02 \times 10^{-9}$ & $-2.89 \times 10^{-10}$
    \\
    Neutral & 1000 & 500--1000 & $4.05 \times 10^{-9}$ & $-3.54 \times 10^{-10}$ 
    \\
    \midrule
    Charged & 250 & 500--1000 & $-5.12 \times 10^{-5}$ & $1.66 \times 10^{-4}$
    \\
    Charged & 500 & 500--1000 & $-1.20 \times 10^{-5}$ & $3.03 \times 10^{-5}$
    \\
    Charged & 1000 & 500--1000 & $-2.85 \times 10^{-6}$& $5.58 \times 10^{-6}$
    \\
    \bottomrule
    \end{tabular}
    \caption{Comparison of numerically fitted offset with the analytical prediction. Displayed fits use $K = 3$ and fit all terms in the ansatz, including the spacing $\hat{s}_m$. Agreement with the analytical prediction would imply vanishing of the terms displayed in the last two columns. Charged cases at smaller $\ell$ are omitted because the available spectra still probe the crossover from neutral-like behavior to the charged asymptotic regime. In these cases, the extracted offset depends strongly on the fitting window and the number of correction terms retained, preventing a reliable comparison with the asymptotic prediction. All cases use $m=0$, $\kappa = -1$, $x_1=\ell_3=1$ with $q = 0$ (neutral) or $q = 7/100$ (charged). Here $\widehat{\mathcal B}_m = \xi_0 \widehat{b}_m$ in the notation of Eq.~\eqref{eq:fit_ansatz}.}
    \label{tab:offset_fits}
\end{table}

The fitted offsets agree closely with the analytical predictions in the neutral case and in the charged case at sufficiently large $\ell$. For the cases displayed, the deviations from the predicted values are smaller than the variation observed when adjusting the fitting windows and adjusting the value of $K$, controlling the number of terms used in the fit. 

Charged cases at smaller $\ell$ are omitted because the available overtones remain sensitive to the crossover from neutral-like behavior to the charged asymptotic regime. In these cases, the extracted offset depends strongly on the fitting window and the number of correction terms $K$. This does not reflect a problem with the asymptotic analysis; it simply means that for smaller values of $\ell$ and the chosen parameters 1000 overtones are not sufficient for the results to enter the asymptotic regime.

\paragraph{Matching subleading overtone-dependent corrections: neutral case.} Next we consider the agreement between our analytical calculation and the numerical results for the subleading overtone-dependent corrections for the neutral qBTZ black hole. Our analytical result for this was presented in Eq.~\eqref{eq:neut-qbtz-subleading-analytical}. 

\begin{table}[!ht]
    \centering
    \small
    \begin{tabular}{rrr}
    \toprule
    $\ell$ & $\mathrm{Re}\,\widehat{\mathcal C}_1(0)$ &
    $\mathrm{Im}\,\widehat{\mathcal C}_1(0)$ \\\midrule
    
    $10^{-2}$ & $3.99\times 10^{-6}$ & $-1.68\times 10^{-5}$ \\
    
    $1$ & $-1.63\times 10^{-7}$ & $-6.68\times 10^{-8}$ \\
    
    $100$ & $-2.60\times 10^{-7}$ & $1.84\times 10^{-8}$ \\
    
    $1000$ & $-2.62\times 10^{-7}$ & $2.26\times 10^{-8}$ \\
    \bottomrule
    \end{tabular}
    \caption{Neutral $m=0$ correction at order $n^{-1/2}$, whose analytical
value is zero. Fits use $K=3$ and $500\leq n\leq1000$ with all
coefficients free. Here $\widehat{\mathcal C}_1(0) = \xi_0 c_{0, 1}$ in the notation of Eq.~\eqref{eq:fit_ansatz}.}
    \label{tab:neutral-metric-correction}
\end{table}

\begin{table}[!ht]\centering\footnotesize
\setlength{\tabcolsep}{4pt}\renewcommand{\arraystretch}{1.18}
\begin{tabular}{rlrr}\toprule
$\ell$ & Method & Analytical & Fitted \\ \midrule
$10^{-2}$ & Two-point & $0.5103127+23.01793\,i$ & $0.1398836+11.33276\,i$ \\
$$ & Four-point & $$ & $0.4321595+22.91431\,i$ \\
\addlinespace[3pt]
$1$ & Two-point & $0.3416804+1.697352\,i$ & $0.3192715+1.665915\,i$ \\
$$ & Four-point & $$ & $0.3445281+1.69882\,i$ \\
\addlinespace[3pt]
$100$ & Two-point & $0.02149479+0.08134548\,i$ & $0.0214946+0.08134339\,i$ \\
$$ & Four-point & $$ & $0.02149312+0.08134447\,i$ \\
\addlinespace[3pt]
$1000$ & Two-point & $0.004685904+0.01754069\,i$ & $0.004685661+0.01754045\,i$ \\
$$ & Four-point & $$ & $0.004685575+0.0175405\,i$ \\
\bottomrule\end{tabular}
\caption{Angular coefficient in the neutral case extracted from the fitted model using Eqs.~\eqref{eq:angular-two}--\eqref{eq:angular-combinations}. The fits use $K=3$ and $500\leq n\leq1000$, with the spacing, offset and all correction amplitudes free.  }
\label{tab:neutral-metric-correction-ang-mom}
\end{table}

At $m=0$ the analytical $n^{-1/2}$ coefficient vanishes. Table~\ref{tab:neutral-metric-correction} shows the results obtained by fitting our $m = 0$ numerical data to the model provided by Eq.~\eqref{eq:fit_ansatz}. The results are consistent with the analytical prediction.  The $m^2$ contribution is non-trivial in the neutral case. Table~\ref{tab:neutral-metric-correction-ang-mom} shows the result of our fitting for several different values of $\ell$ and using the two- and four-point combinations to isolate the $m$-dependence.  We see that the agreement between the fit to the numerical data and the analytical prediction improves as $\ell$ increases, and is generally better for the four-point combination than for the two-point combination. In the case of small $\ell$, higher
overtones would be needed to establish the asymptotic
angular coefficient more reliably.

\begin{figure}[!htbp]
    \centering
    \includegraphics[width = \textwidth]
        {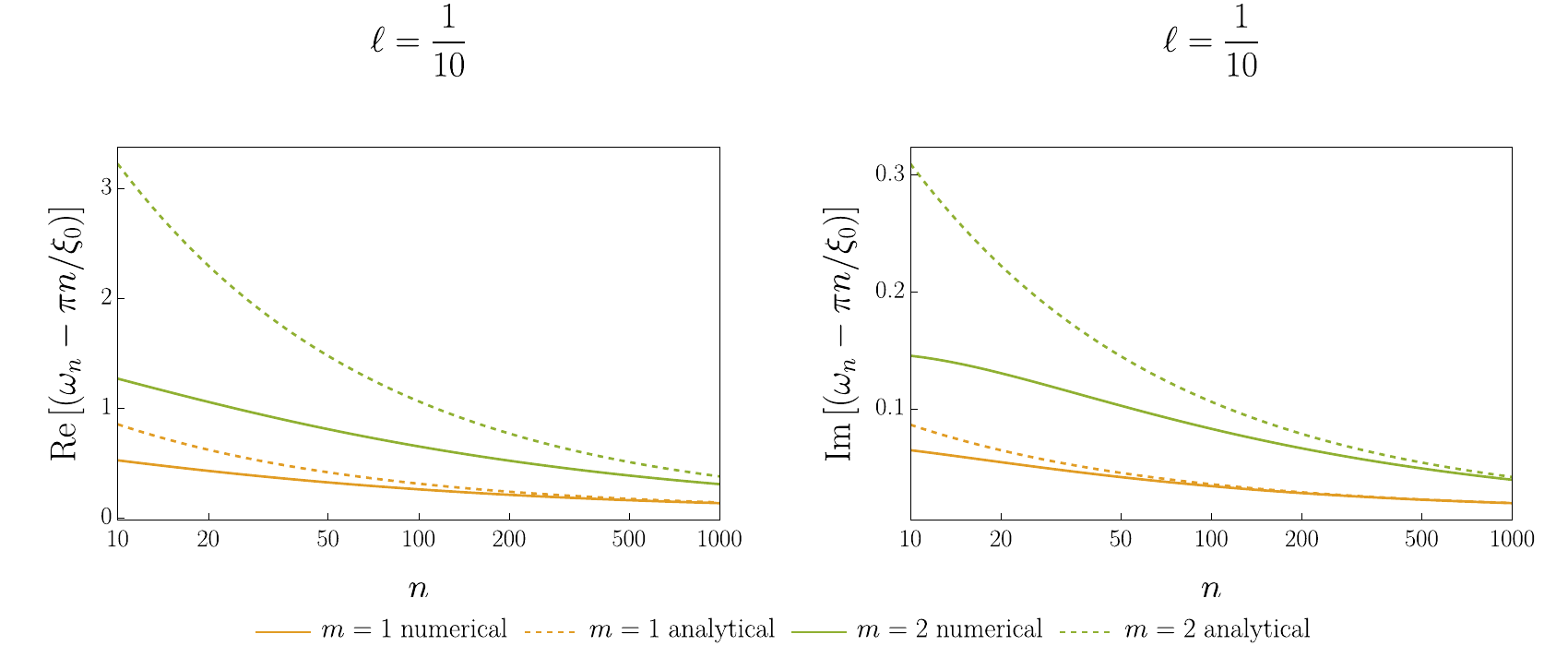}
    \includegraphics[width=\textwidth]
        {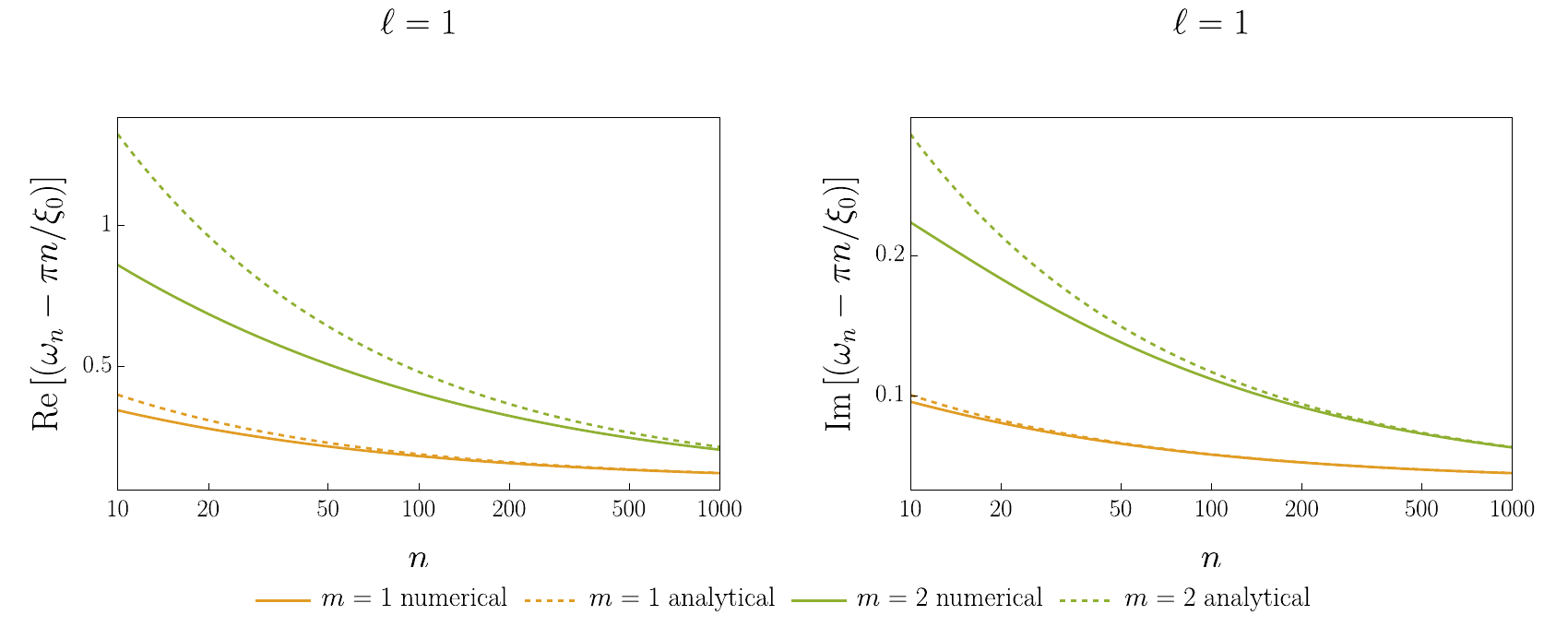}
    \includegraphics[width=\textwidth]
        {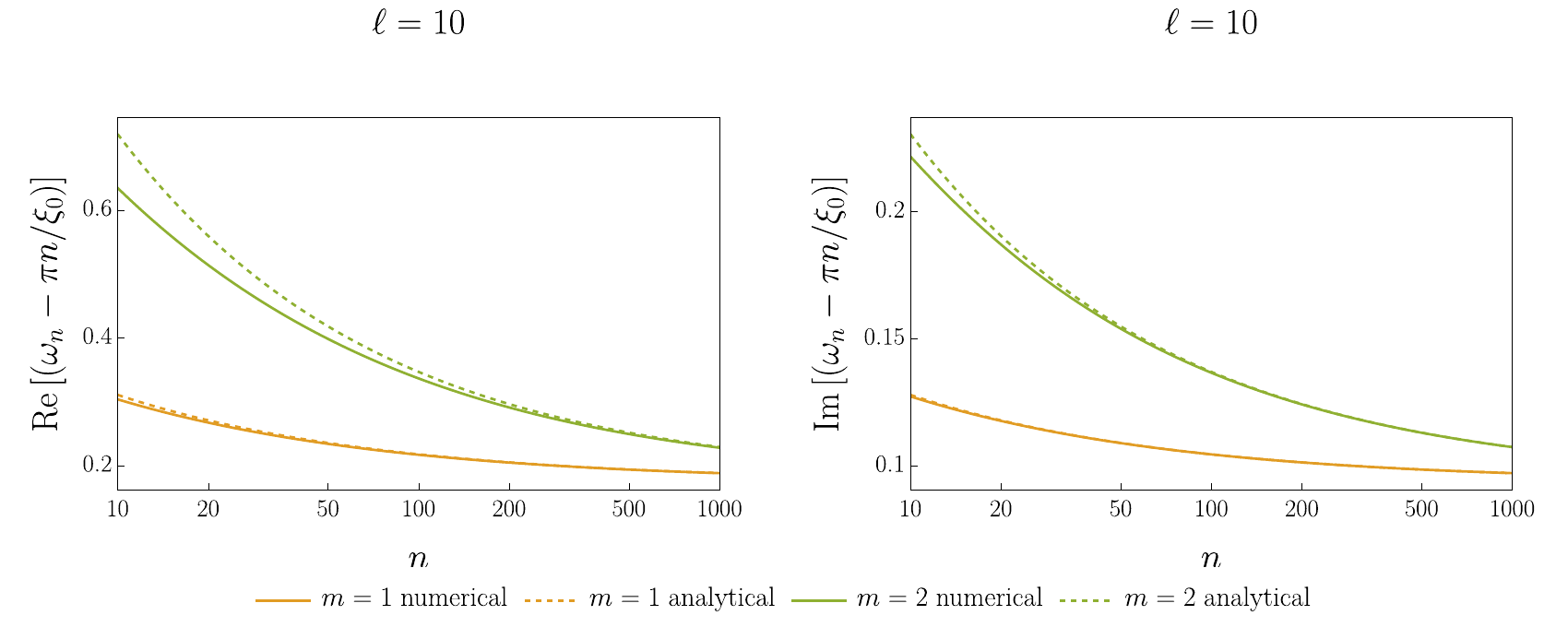}
    \caption{Comparison of numerically computed highly damped QNM
frequencies with the analytically predicted corrections. The left
and right columns show the real and imaginary parts, respectively,
of the frequency after subtracting the leading linear term
$\pi n/\xi_0$. Each row corresponds to a different value of $\ell$,
and each panel shows two values of the angular number $m$.
The numerical results generally converge more rapidly to the
analytical predictions for larger $\ell$ and smaller $m$.
The parameters are $\kappa = -1$, $x_1 = 1$, $\ell_3 = 1$, and $q = 0$.}
    \label{fig:compare_numerical_analytical_neutral}
\end{figure}

Figure~\ref{fig:compare_numerical_analytical_neutral} provides a complementary comparison of the frequencies with the analytical predictions. We subtract the leading linear term $\pi n/\xi_0$ to display the offset and overtone-dependent corrections. The agreement between the numerical QNMs and the analytical approximation improves systematically as the overtone number increases. At fixed overtone number, convergence to the asymptotic approximation is faster for larger values of $\ell$ and smaller values of $m$. In all cases we have explored, the real and imaginary parts of the QNMs approach the analytical correction monotonically from below.

\paragraph{Matching subleading overtone-dependent corrections: charged case.} We now consider the subleading corrections for the charged qBTZ black hole. The leading correction, given in
Eq.~\eqref{eq:charged-qbtz-subleading-analytical}, is proportional to $n^{-1/3}$ and is independent of angular momentum. The leading angular-momentum-dependent contribution enters at order $n^{-2/3}$, as given in \eqref{eq:charged-qbtz-subleading-analytical-ang-mom}.
We again employ the fitting ansatz \eqref{eq:fit_ansatz}, now with $\alpha=1/3$.

\begin{table}[!ht]\centering\footnotesize
\setlength{\tabcolsep}{5pt}\renewcommand{\arraystretch}{1.18}
\begin{tabular}{rrr}\toprule
$\ell$ & Analytical & Fitted $\widehat{\mathcal C}_1(0)$ \\\midrule
$100$ & $-0.4541244-0.1709298\,i$ & $-0.4407284-0.1036947\,i$ \\
$250$ & $-0.302062-0.1158559\,i$ & $-0.3007547-0.1199475\,i$ \\
$500$ & $-0.2213566-0.08707117\,i$ & $-0.2210442-0.08782438\,i$ \\
$1000$ & $-0.1615903-0.06593653\,i$ & $-0.1615149-0.06607593\,i$ \\
\bottomrule\end{tabular}
\caption{Leading charged correction, multiplying $n^{-1/3}$. The fit uses $m=0$.}
\label{tab:charged-leading}\end{table}

Table~\ref{tab:charged-leading} compares the fitted coefficient $\widehat{\mathcal C}_1(0)$ with the analytical prediction. Agreement improves as $\ell$ increases: the relative difference between the complex coefficients decreases from approximately $1.3\%$ at $\ell=250$ to $0.34\%$ at $\ell=500$ and $0.091\%$ at $\ell=1000$. The discrepancy is larger at $\ell=100$, where the spectrum exhibits an oscillatory approach to the asymptotic regime. In this case, the extracted coefficient also depends strongly on the fitting window and the number of correction terms retained, indicating that the available data do not yet provide a stable determination of the asymptotic coefficient.

\begin{figure}
    \centering
    \includegraphics[width=\linewidth]{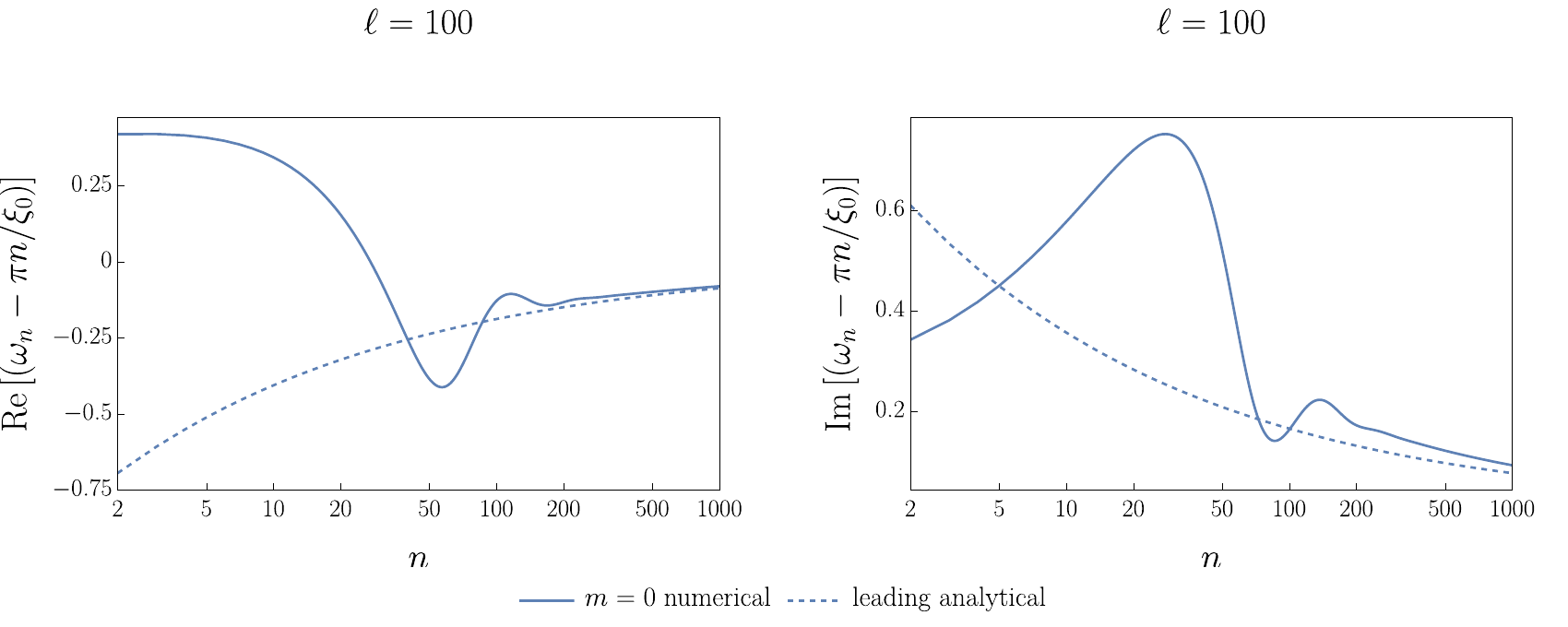}
    \includegraphics[width=\linewidth]{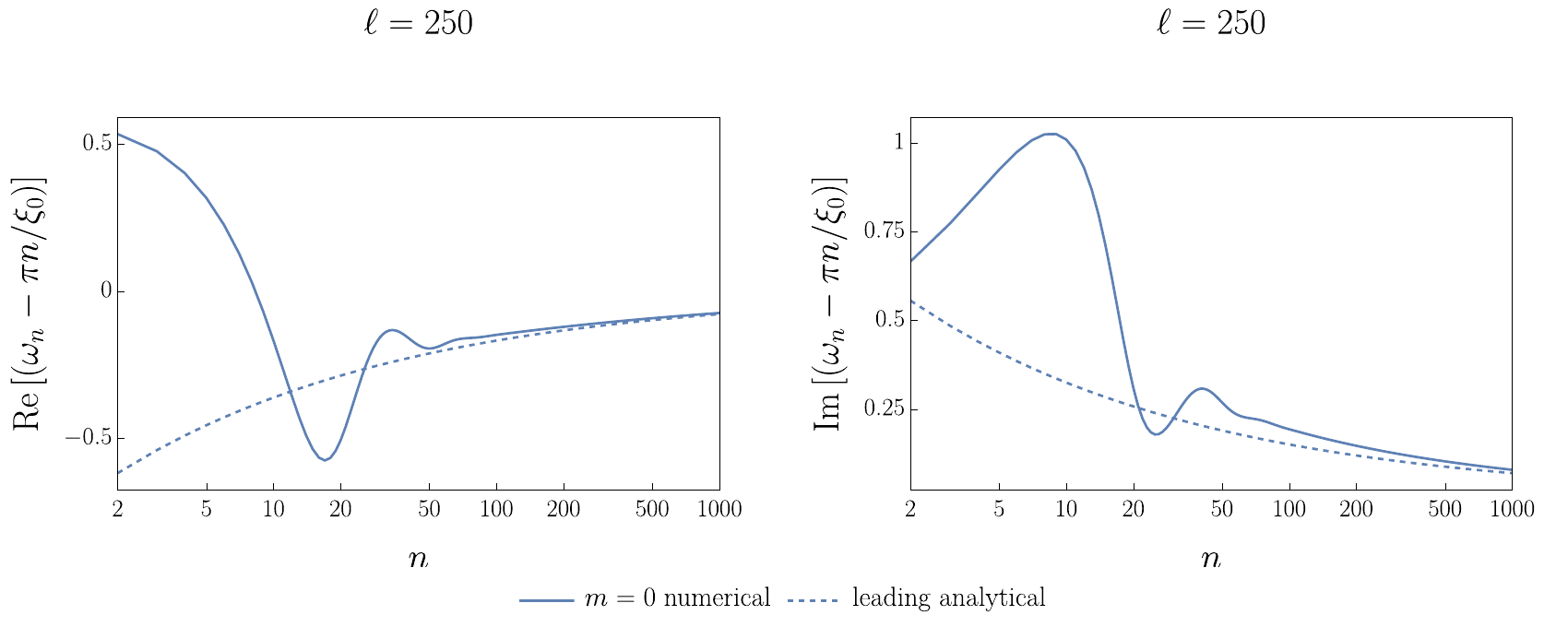}
    \includegraphics[width=\linewidth]{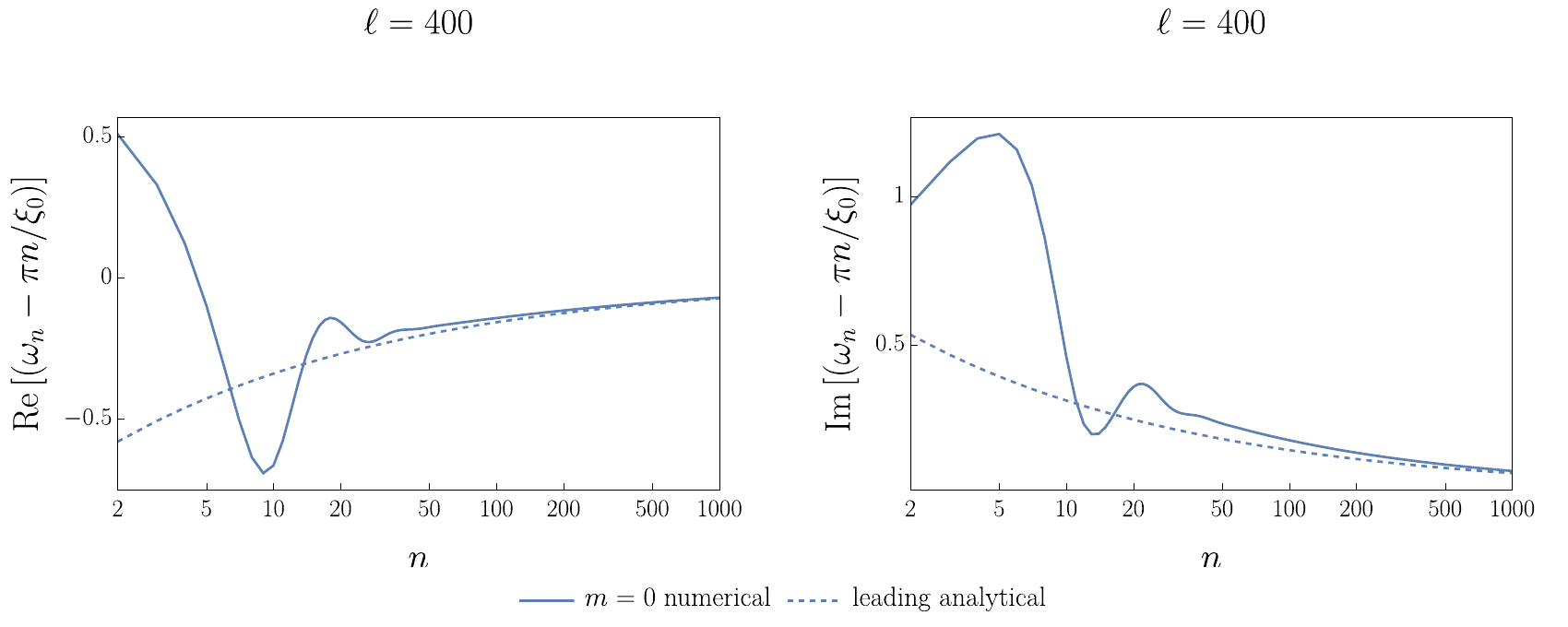}
    \caption{Comparison of the asymptotic QNMs of the charged qBTZ  black hole to the predicted analytical form of the leading correction for specially chosen values of $\ell$ highlighting the oscillatory approach to the asymptotic regime. All displayed modes have $m = 0$. The parameters are $\kappa = -1$, $x_1 = 1$, $\ell_3 = 1$, and $q = 7/100$.}
    \label{fig:compare_numerical_analytical_charged_oscillations}
\end{figure}

Figure~\ref{fig:compare_numerical_analytical_charged_oscillations} compares the numerical frequencies with the analytical approximation after subtracting $\pi n/\xi_0$. For the displayed values of $\ell$, the approach to the asymptotic prediction is oscillatory, with the oscillations decreasing in amplitude as the overtone number increases. This transient behavior complicates the extraction of the series coefficients. The oscillations may reflect a subleading contribution associated with the inner horizon monodromy, whose influence on QNM quantization is known in the asymptotically flat case~\cite{Motl:2003cd}. We do not attempt to derive the oscillatory contribution here.

The leading angular correction enters at order $n^{-2/3}$. Table~\ref{tab:charged-angular} compares its coefficient with the two- and four-point estimates, using $j_*=2$ in the combinations defined above. Both estimates approach the analytical value better as $\ell$ increases, with the four-point combination giving closer agreement in each displayed case. At $\ell=1000$, the relative differences are approximately $1.22\%$ and $0.80\%$, respectively. The angular coefficient remains more sensitive to the fitting choices than the leading $n^{-1/3}$ coefficient.

\begin{table}[!ht]\centering\footnotesize
\setlength{\tabcolsep}{4pt}\renewcommand{\arraystretch}{1.18}
\begin{tabular}{rlrr}\toprule
$\ell$ & Method & Analytical & Fitted \\ \midrule
$100$ & Two-point & $-0.1167806+0.1331529\,i$ & $-0.1349083+0.1296964\,i$ \\
$$ & Four-point & $$ & $-0.1273663+0.1257133\,i$ \\
\addlinespace[3pt]
$250$ & Two-point & $-0.05264928+0.05853734\,i$ & $-0.055852+0.05783507\,i$ \\
$$ & Four-point & $$ & $-0.05456492+0.05729456\,i$ \\
\addlinespace[3pt]
$500$ & Two-point & $-0.02899645+0.03115518\,i$ & $-0.02996499+0.0310039\,i$ \\
$$ & Four-point & $$ & $-0.02960527+0.03089215\,i$ \\
\addlinespace[3pt]
$1000$ & Two-point & $-0.01602948+0.01637128\,i$ & $-0.01630601+0.01633269\,i$ \\
$$ & Four-point & $$ & $-0.01620406+0.01631279\,i$ \\
\bottomrule\end{tabular}
\caption{Charged angular coefficient extracted from the fitted $\widehat{\mathcal C}_2$ using Eqs.~\eqref{eq:angular-two}--\eqref{eq:angular-combinations}. Both combinations have the same analytical value.  The fits use $K=3$ and $500\leq n\leq1000$, with the spacing, offset and all correction amplitudes free.  }
\label{tab:charged-angular}\end{table}

\begin{figure}
    \centering
    \includegraphics[width=\linewidth]{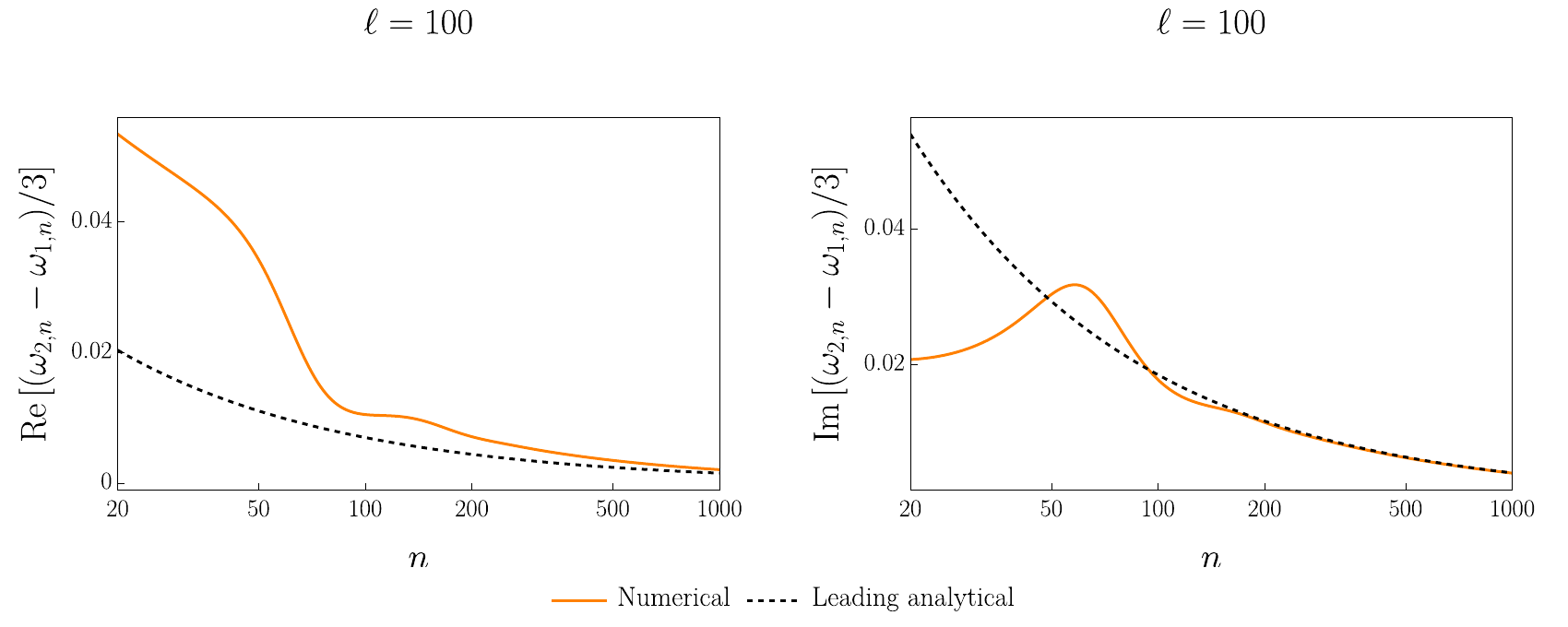}
    \includegraphics[width=\linewidth]{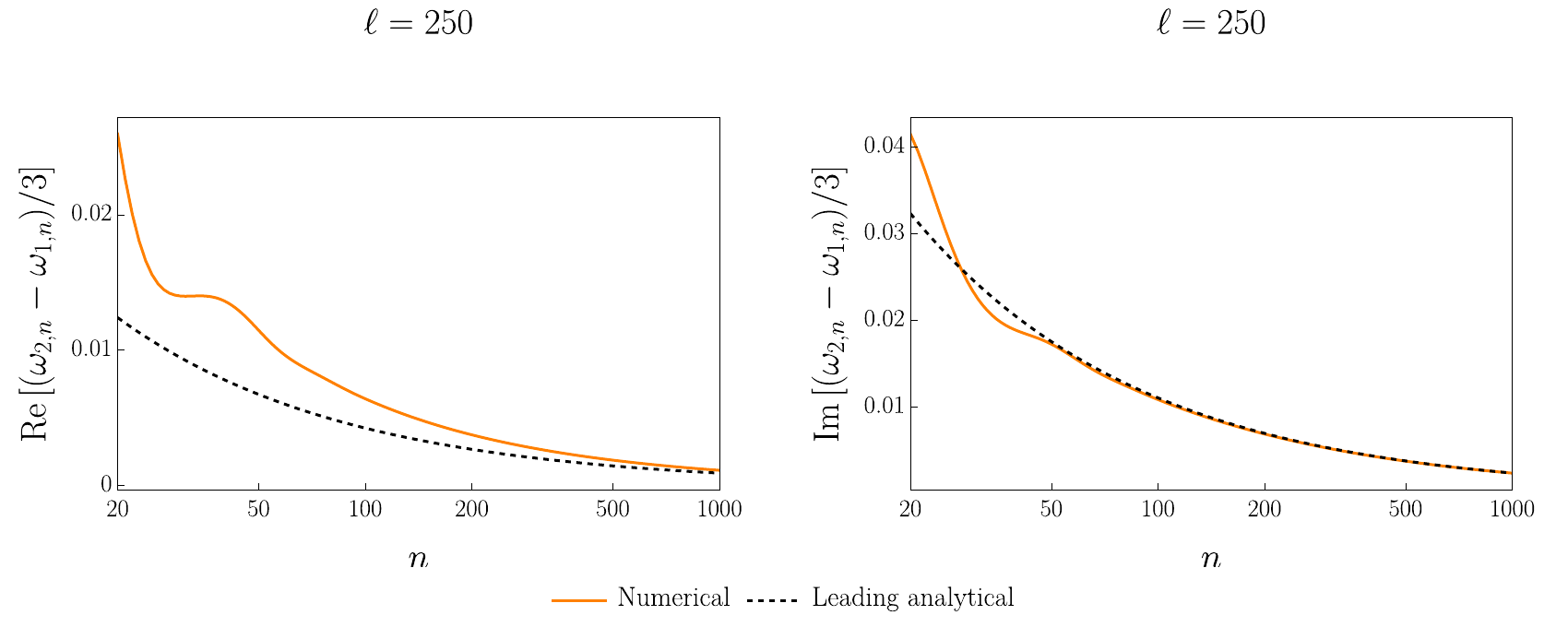}
    \includegraphics[width=\linewidth]{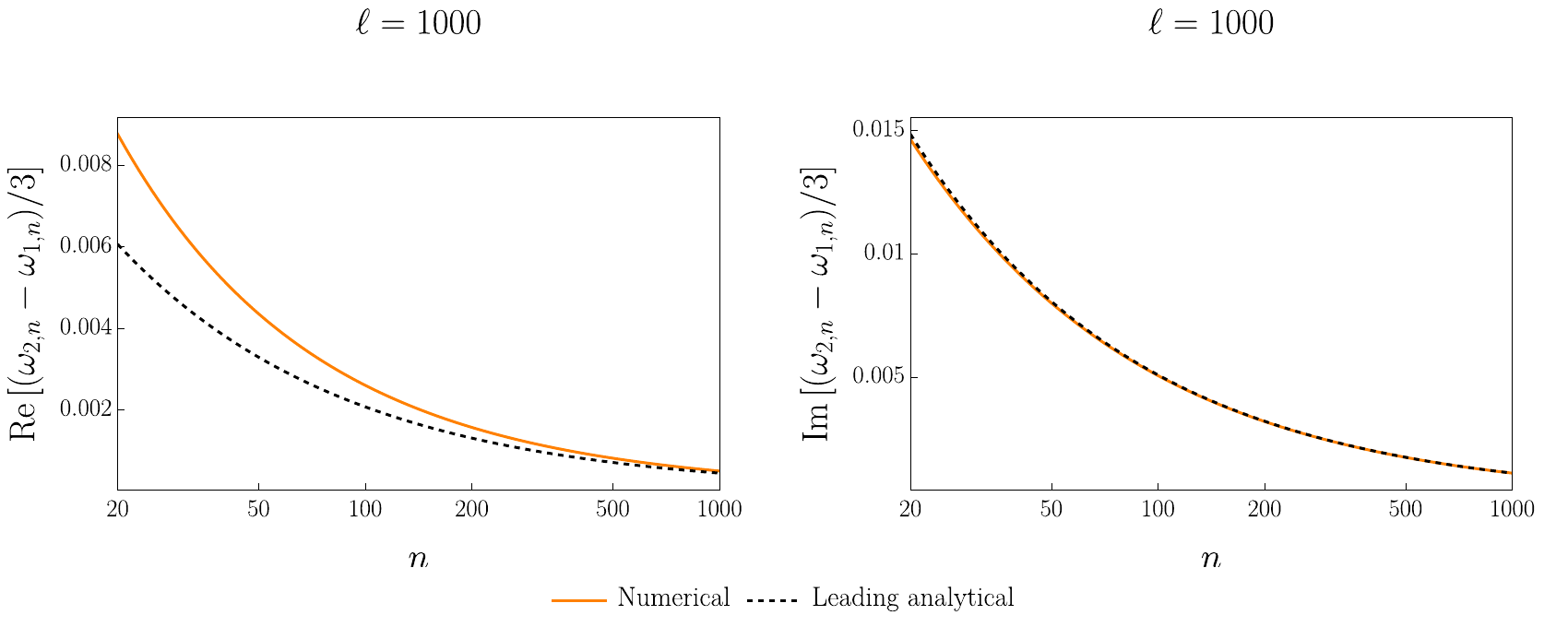}    
    \caption{Real (left) and imaginary (right) parts of the normalized angular difference $(\omega_{2,n}-\omega_{1,n})/3$ for charged qBTZ black holes with $\ell=100,250,1000$ (top to bottom). Solid orange curves show the numerical results, while dotted black curves show the leading analytical prediction. The subtraction cancels all $m$-independent contributions, allowing the leading $m^2$-dependent correction to be probed. Over the range shown, agreement improves with increasing $\ell$ and is generally closer for the imaginary part than for the real part. The parameters are $\kappa = -1$, $x_1 = 1$, $\ell_3 = 1$, and $q = 7/100$.}
    \label{fig:charged-qbtz-subleading-angular}
\end{figure}

Figure~\ref{fig:charged-qbtz-subleading-angular} shows the corresponding comparison directly for the normalized frequency difference $(\omega_{2,n}-\omega_{1,n})/3$. The numerical results approach the leading analytical angular correction as $n$ increases. Over the range shown, agreement improves with increasing $\ell$ and is generally closer for the imaginary part than for the real part.

\FloatBarrier
\section{Interior scaling and the detection of eons}\label{sec:eondetection}

In Section \ref{sec:lookinqBH} we described  how to extract  the Kasner exponents for the near-singularity geometry from the subleading corrections to the asymptotic QNM spectra. Here we apply this technology to the neutral and charged qBTZ black holes. For the neutral black hole, we detect a transition between Kasner eons. For the charged black hole, we identify an oscillatory approach to the timelike singularity.

\subsection{Classical to quantum eons} 

Quantum effects, when manifested as higher-derivative corrections to general relativity, give rise to new transitions  captured by Kasner \emph{eons}. These are periods of time (longer than epochs and eras) dominated by semi-classical or quantum gravitational physics \cite{Bueno:2024fzg}. Deep in the interior, quantum effects are expected to become dominant, which should correspond to a transition from a classical, i.e., ``Einsteinian'' eon to a ``quantum eon'', an eon characterized by the emergent physics.

In the case of the (neutral) qBTZ black hole, for sufficiently small $\ell/\ell_3$, the qBTZ black hole interior approximates  the classical BTZ interior over the radial range $\ell \mu  \ll r \ll  \ell_3$. A separation of scales between the two endpoints is realized provided that $\ell/\ell_3\ll1$ and $\mu$ remains of order unity.  In this regime, the interior takes the Kasner
form (\ref{eq:kasmet3d}) with exponents (\ref{eq:kasexp3d}) $p_t = 0$ and $p_\phi = 1$. These special
exponents describe the approach to the horizon of a Milne patch,
rather than a curvature singularity. Deeper in the interior, quantum corrections dominate and the geometry approaches a Kasner singularity with exponents $p_t = -1/3$ and $p_\phi = 2/3$. This is the qBTZ dominated regime, which will always be reached at sufficiently small $r$. The crossover between these  regimes corresponds to a transition between ``Kasner eons'' in the terminology of~\cite{Bueno:2024fzg}. 

\begin{figure}[t!]
    \centering
    \includegraphics[width = \textwidth]{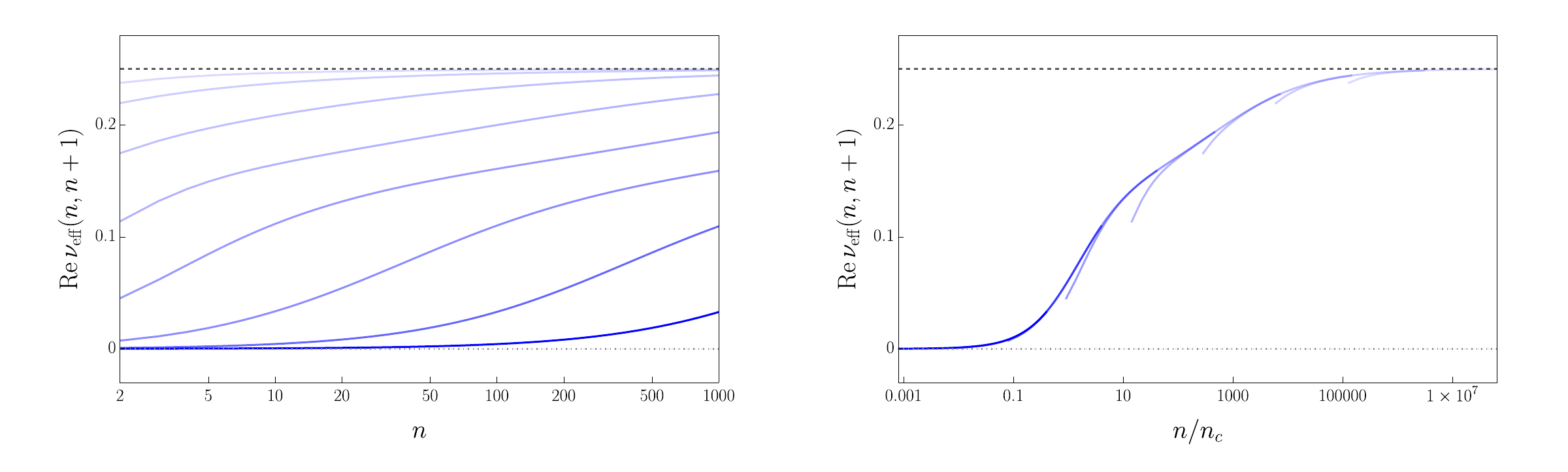}
    \caption{\textit{Left}: The curves correspond to values of $\ell$ from $10^{-4}$
(lowest blue curve) to $10^{3}$ (highest blue curve), increasing by
factors of ten as the opacity decreases. The horizontal lines indicate
the analytically derived scaling exponent $\nu$ for BTZ (lower, dotted)
and for qBTZ in the large-overtone limit (upper, dashed). For small
$\ell$, a pronounced BTZ ``eon'' emerges, with the effective scaling
exponent matching the BTZ value of $\nu = 0$. As the overtone number increases
at fixed $\ell$, the system transitions from a BTZ-dominated to a
qBTZ-dominated eon, corresponding to a transition between
``Kasner eons'' in the quantum black hole. For larger $\ell$, the BTZ
regime is absent, and quantum corrections dominate from the outset. For very small values of $\ell$, observing robust convergence to the qBTZ prediction would take hundreds of thousands of modes, which we have not attempted to compute. \textit{Right}: Rescaled versions of the same curves on the left. The overtone number has been rescaled to $n \to n/n_c$ where $n_c$ is the estimated crossover overtone (see Eq.~\eqref{eq:crossover_overtone} below) where the BTZ and qBTZ regimes compete. 
The parameters in both panels are $\kappa = -1$, $x_1 = 1$, $\ell_3 = 1$, and $q = 0$, and $\nu_{\rm eff}$ was computed using $m = 1$ and $m=2$.   }
\label{fig:eon_detective}
\end{figure}

Should such a separation of scales exist, we would expect the transition between Kasner eons to be reflected in the highly damped QNM spectrum. To probe such a transition, we work with an effective scaling exponent, 
\be 
\nu_{\rm eff}(n, n+1) = - \frac{\log \left( |\Delta_{n+1}|/|\Delta_n| \right)}{2 \log \left((n+1)/n\right)}\;,
\ee
where $\Delta_n$ was defined in Eq.~\eqref{eq:Delta_two_mode}. In a regime where the asymptotic approximation holds, we will have $\nu_{\rm eff} \to \nu$ where $\nu$ is the genuine scaling exponent associated with the singularity. An exact BTZ interior has $\nu = 0$, while for the qBTZ regime we have $\nu = 1/4$.  We show the results in the left panel of Figure~\ref{fig:eon_detective}. The different curves correspond to different values of the parameter $\ell$ at fixed $\ell_3 = 1$ along with fixed solution parameters as described in the caption. We observe that black holes with smaller values of $\ell$ exhibit a clear BTZ-dominated phase before, at sufficiently high overtone, the spectrum begins to exhibit the characteristics of the qBTZ Kasner singularity. On the other hand, for larger values of $\ell$, there is no distinct BTZ phase and the spectrum immediately indicates features suggestive of the approach to the qBTZ Kasner singularity. As a general remark, we observe in our numerics that, when a transition between eons is present, its duration in terms of the number of overtones increases in proportion to the value of $1/\ell$. In some cases, especially those where $\ell$ is very small and hence the solution exhibits a pronounced BTZ-dominated phase, the number of modes required to observe the full transition becomes prohibitive. In the following, we will provide an analytical understanding of this observation.

\paragraph{Analytical estimate of the crossover.} The qBTZ metric function we are considering has the form 
\be 
f(r) = \frac{r^2}{\ell_3^2} - b - \frac{a}{r} \, , \quad \text{with}\quad a=\ell W(M) \quad \text{ and } \quad b=8\mathcal G_3M>0\,.
\ee
The last two terms are important in the interior and control the transition between a BTZ-like interior and a qBTZ-like one. These terms have equal magnitude at the crossover radius $r_c = a/b$. Since the $r^2/\ell_3^2$ term is not important in the interior, let us drop it and write $f(r) \approx -b - a/r$. Then, the complex tortoise coordinate as calculated from the singularity reads
\be 
\xi(r) = \frac{a}{b^2} \left[\log \left(1 + \frac{b r}{a} \right) - \frac{b r}{a} \right] \, .
\ee
Thus, at the crossover radius we have 
\be 
|\xi(r_c) | = \left(1-\log 2\right) \,  \frac{a}{b^2} \, .
\ee
Because the near-singularity equation is naturally written in terms of $z = \omega \xi$, the crossover in this coordinate is $z_c \sim \omega a/b^2$. 

To proceed, let us recall the near-singularity matching problem. In the qBTZ-dominated region, the near-origin radial equation admits Bessel function solutions. To match these to the plane-wave solutions propagated from the horizon and infinity, we use their large argument expansion, which requires $|z|\gg1$. At the same time, the near-origin approximation requires us to remain within the qBTZ-dominated region, where $|z|\ll|z_c|$. The matching calculation therefore requires an overlap region
\be 
1\ll |z|\ll |z_c| \, .
\ee
Such an overlap can exist only when $|\omega|a/b^2\gg1$. When this combination becomes of order unity, both interior terms must be retained in the matching problem. Using the large-overtone relation $\omega_n\simeq\pi n/\xi_0$, we therefore obtain the  overtone estimate for the crossover
\be \label{eq:crossover_overtone}
n_c\sim\frac{ |\xi_0|b^2}{\pi a} = \frac{|\xi_0|}{4 \pi \ell}\, ,
\ee 
where in the last equality we made the replacements for the neutral qBTZ parameters. In the limit of weak backreaction, we further have $|\xi_0| \to \pi \ell_3$ and hence can estimate $n_c \sim \ell_3/(4 \ell)$.  

We can also estimate the width of the transition. To do so, suppose that we are approaching the qBTZ-dominated regime so that we have the angular difference
\be 
\Delta_n = K n^{-1/2} \left[1 + d \left(\frac{n_c}{n}\right)^{1/2} + \cdots \right] \, .
\ee
Then, at large $n$ we have 
\be 
\nu_{\rm eff} = \frac{1}{4}  + \frac{{\rm Re} \, d}{4} \left(\frac{n_c}{n}\right)^{1/2} + \cdots  \, ,
\ee
where we assume here that ${\rm Re}(d) \neq 0$. Therefore, under this correction law approaching the qBTZ value of the scaling exponent within a tolerance $\varepsilon$ requires a number of modes 
\be 
n_{\rm end} \sim n_c \left(\frac{{\rm Re} \, d}{4 \varepsilon}\right)^2 \, .
\ee
For a fixed tolerance we therefore have $n_{\rm end} \sim n_c$. However, we see that as a function of the tolerance $n_{\rm end} \sim 1/\varepsilon^2$. Hence, halving the tolerance would require computing four times as many modes. These features are in agreement with our numerical observations. 

These calculations suggest a different perspective on the data presented in the left panel of Figure~\ref{fig:eon_detective}. Given our data for the effective scaling exponent $\nu_{\rm eff}$, we can perform a rescaling of the data pairs $\{n, \nu_{\rm eff}\} \to \{n/n_c, \nu_{\rm eff}\}$. We show the results of this rescaling in the right panel of Figure~\ref{fig:eon_detective}. Under this rescaling, the curves collapse approximately onto a single curve interpolating between the BTZ and qBTZ eons. The collapse is not perfect for the following reason: The curves corresponding to larger values of $\ell$ are not well described by a BTZ-dominated regime at small overtone number and therefore deviate from the common trend in the small overtone regime. The collapse to a single approximate curve suggests an emergent scaling behaviour. Within the regime where the crossover estimate applies, the dependence on $\ell$ is largely absorbed into the scale $n_c$, yielding an approximately universal curve as a function of $n/n_c$.  As a consequence of this scaling behaviour, one can construct a smooth picture of the transition between Kasner eons by piecing together the appropriately rescaled curves for different values of $\ell$. This is much more efficient than fixing a particular $\ell$ and computing the otherwise necessary hundreds of thousands of modes. It would be interesting to understand this from an analytical perspective.

% should therefore be regarded as an emergent scaling behaviour in the large $n/n_c$-limit

% Nevertheless, by combining the portions of the rescaled curves that lie within their respective regimes of validity, one obtains a smooth picture of the transition between the Kasner eons in the black hole interior. 

\paragraph{More general crossovers.} Our results concerning the overtone at which a crossover between two dominant behaviors occurs are not restricted to the qBTZ example above. We now explain how this works under more general circumstances.\footnote{See also~\cite{Daghigh:2006hu} where a similar problem was addressed for charged black holes.} For this purpose, let us suppose that in the black hole interior there are two competing terms of the form
\be \label{eq:crossover_metric_gen}
f(r) \sim \frac{\alpha_1}{r^{n_1}} + \frac{\alpha_2}{r^{n_2}} \, .
\ee
We assume that $0 \le n_1 < n_2$, but otherwise make no assumptions on the signs of $\alpha_i$. The two terms have equal magnitude at the crossover radius
\be 
r_c = \left|\frac{\alpha_2}{\alpha_1} \right|^{1/(n_2-n_1)} \, .
\ee
For $r \ll r_c$, the second term dominates, regardless of its sign. 

In the region where the second power dominates, the local tortoise coordinate has the form
\be 
\xi(r) \sim \frac{r^{n_2+1}}{(1+n_2) \alpha_2} \, .
\ee
In the near-origin matching problems, we use $z = \omega \xi$. As a result, we have 
\be 
|z| \sim 1 \quad \Rightarrow \quad |r_\omega| \sim \left(\frac{(1+n_2) |\alpha_2|}{|\omega|} \right)^{1/(1+n_2)}\, .
\ee
To perform the matching calculation we need an overlap region where a large $|z|$ series expansion is valid while simultaneously remaining in the region where the second term in~\eqref{eq:crossover_metric_gen} dominates. This requires the hierarchy $|r_\omega| \ll |r_c|$. We therefore can estimate where the crossover occurs by solving for $|r_\omega| \sim |r_c|$. This gives
\be 
n_c \sim \frac{(1+ n_2)|\xi_0|}{\pi} \left( \frac{|\alpha_1|^{1+n_2}}{|\alpha_2|^{1+n_1}} \right)^{1/(n_2-n_1)} \, .
\ee

This estimate for the crossover applies whenever there are two competing terms.  We can now come full circle and reconnect to a motivation made in our introduction. The quantities $\alpha_i$ are dimensionful. If these terms were to both arise from the same underlying physics, we would expect them to share a common length scale $L$. We may therefore define dimensionless parameters $\lambda_i = \alpha_i/L^{n_i}$, so that
\be 
n_c \sim \frac{(1+n_2)|\xi_0| }{L \pi} \left( \frac{|\lambda_1|^{1+n_2}}{|\lambda_2|^{1+n_1}} \right)^{1/(n_2-n_1)} \, .
\ee
Now consider a situation with three competing terms, all associated with the same length scale. There are then two potential crossovers: $n_c^{1\to 2}$, marking the transition between the regimes dominated by $\lambda_1$ and $\lambda_2$, and $n_c^{2\to 3}$, marking the subsequent transition between those dominated by $\lambda_2$ and $\lambda_3$. The result above shows that both crossover overtones are parametrically controlled by the same scale, $|\xi_0|/L$, differing only through dimensionless combinations of the $\lambda_i$.  Unless these parameters are strongly hierarchical, the two crossovers will therefore occur at comparable overtone numbers, making it difficult in general to resolve the intermediate regime. This returns us to the motivation, discussed in the introduction, for studying this question in braneworld quantum black holes. If the correction to the metric were merely the first of an infinite sequence of perturbative corrections, there would in general be no parametrically separated regime in which a calculation retaining only that first correction could be trusted.

\subsection{The `heartbeat' of a charged quantum black hole}

\begin{figure}[ht]
    \centering
    \includegraphics[width=\linewidth]{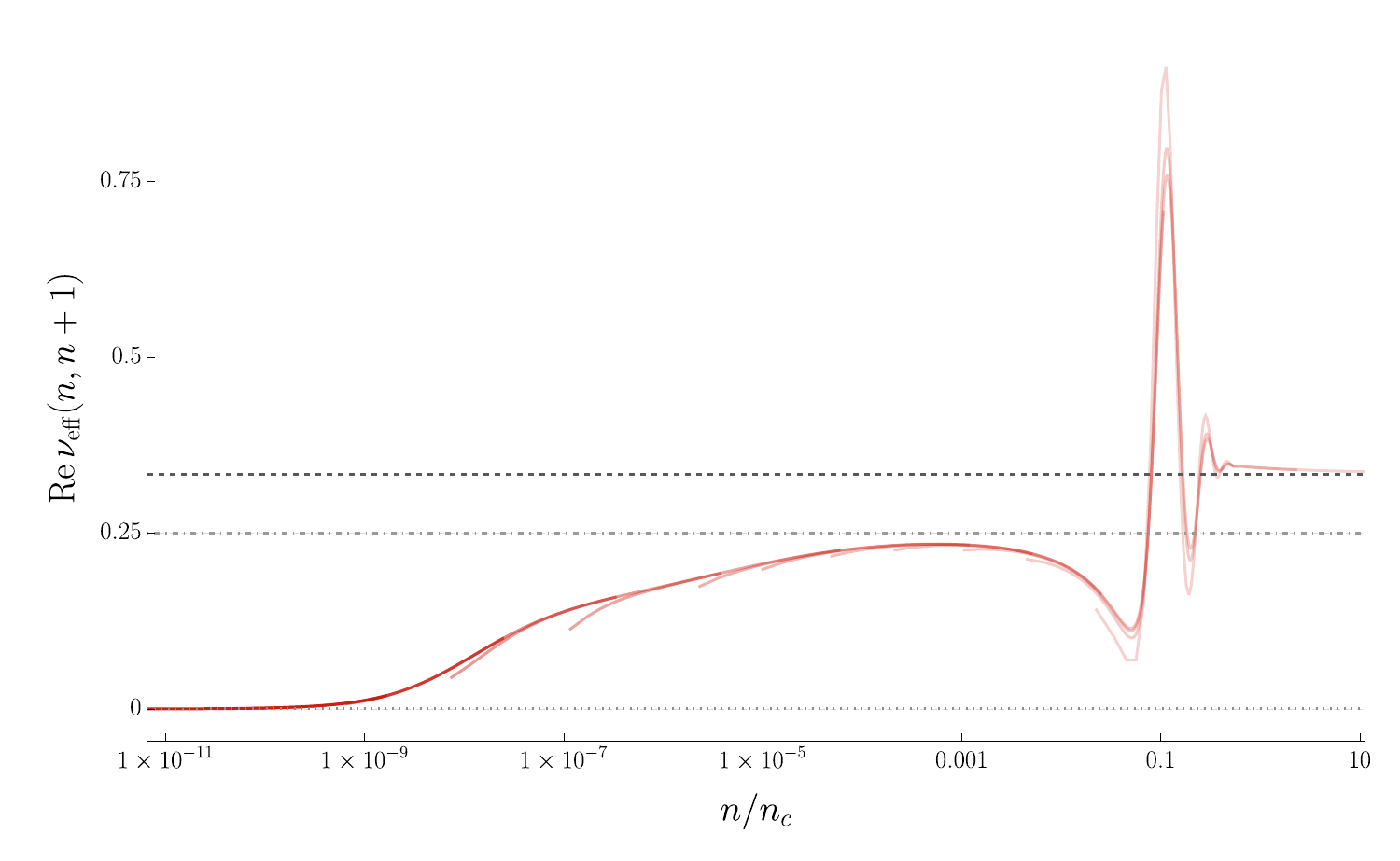}
    \caption{Effective scaling exponent for the charged qBTZ black hole as a function of the rescaled overtone number $n/n_c$, where $n_c$ is the $\ell$-dependent crossover overtone (the right-hand side of Eq.~\eqref{eq:crossover_n_charged}). The curves correspond to $\ell \in \{10^{-4},10^{-3}, 10^{-2}, 10^{-1}, 1, 10, 30, 100, 300, 1000, 3000, 10000 \}$, with opacity decreasing as $\ell$ increases. This rescaling produces an approximate collapse onto a common curve. The remaining parameters are fixed at $\kappa = -1$, $x_1 = 1$, $\ell_3 = 1$, and $q = 0.025$, while $\nu_{\rm eff}$ was computed using $m = 1$ and $m=2$. The horizontal lines mark $\nu=0$, $\nu=1/4$, and $\nu=1/3$, corresponding to the BTZ, neutral qBTZ, and charged qBTZ scaling regimes, respectively.}
    \label{fig:heartbeat}
\end{figure}

Let us now examine the interior of the charged qBTZ black hole. We note that the singularity in the charged solution does not permit a Kasner description (it is timelike). This poses no issue for the extraction of the scaling of the singularity, as our result applies equally well to spacelike or timelike singularities.  We show the effective scaling exponent as a function of overtone number in Figure~\ref{fig:heartbeat}. The plot includes up to 1,000 overtones for twelve different values of $\ell$, ranging from $\ell = 10^{-4}$ to $\ell = 10^{4}$, in units where $\ell_3 = 1$. We have rescaled all curves by the $\ell$-dependent crossover overtone
\be \label{eq:crossover_n_charged}
n_c \sim \frac{3 |\xi_0| W^3}{\pi \ell P^2}  \, .
\ee
The rescaling has the effect of collapsing all of the results to a single approximate curve. 

At small $n/n_c$, the effective scaling exponent remains close
to $\nu=0$, consistent with the BTZ Kasner regime. As the
rescaled overtone number increases, the $1/r$ qBTZ correction
becomes important, and the exponent initially rises monotonically toward the neutral qBTZ value $\nu=1/4$. For the parameters shown, this intermediate regime appears as a broad shoulder below $\nu=1/4$, rather than a fully developed plateau. As charge effects become important, the curve dips downward before exhibiting pronounced oscillations of
decreasing amplitude about the charged qBTZ value $\nu=1/3$. It would be interesting to understand if these `heartbeat'-shaped oscillations are a consequence of the inner horizon. 

The approximate collapse to a single curve observed separately in the neutral and charged qBTZ cases suggests that $n_c$ captures the dominant $\ell$-dependence of the effective scaling exponent. An analytical explanation of this behavior remains to be established. A promising route may be to solve the wave equation in the crossover regimes, dropping all terms in the metric function besides two competing behaviors. In the charged case, an analysis of the resulting reduced equations could help clarify the origin of the oscillatory approach to $\nu=1/3$.

\FloatBarrier

\section{Discussion} \label{sec:disc}

In this article we computed the asymptotic QNMs of a class of exact, static quantum black holes in AdS$_{3}$ (neutral and charged) and Mink$_{3}$. We accomplished this by analytically continuing the radial coordinate into the complex plane and deriving the QNM quantization condition using Stokes line matching or the monodromy method. The leading order QNM frequencies all fit the generic form $\omega=(\text{offset})+n(\text{gap})$, for overtone number $n$. We numerically confirmed our analytical results to high precision. 

From the asymptotic QNM spectra we showed how to reconstruct information about the singularity structure of a broad class of black holes with AdS$_{3}$ asymptotics (though our method easily extends to higher dimensions).  Under an assumption about the global Stokes topology of the complexified radial coordinate, we showed how to extract the scaling exponent of (spacelike or timelike) black hole singularities from the offset. We proceeded to provide a recipe to compute subleading corrections to the asymptotic QNMs, leading to a robust method (independent of the topology of Stokes lines) for extracting the scaling of the metric function in the vicinity of the  singularity. Applying our method to the neutral and charged qBTZ black holes, we computed the asymptotic QNM spectra including subleading corrections. Finally, for the neutral quantum BTZ solution, we explicitly uncovered a transition between its Kasner eons. Thus, our results show how the spectrum reveals emergent physics deeper inside a black hole and how, in principle, an external observer could recover this information.

Our work opens up new avenues to explore, a few of which we now briefly describe. 

\vspace{2mm}

\noindent \textbf{Higher dimensions and including matter.} Here we focused on (2+1)-dimensional black holes, for concreteness. Our general method for computing the subleading corrections to the asymptotic QNM spectra, however, straightforwardly extends to spherically symmetric black holes in higher dimensions. Indeed, subleading order large-overtone QNMs were shown to capture  temporal and spatial Kasner scalings for AdS$_{5}$ black holes, such that the
corresponding metric exponents can be reconstructed \cite{Xiao:2026pir}. Additionally, it has long been known that the local dynamics near a spacelike singularity can be dramatically altered when gravity is coupled to matter, signaled through a modification to the relations between Kasner exponents \cite{Belinskii:1973sud,Belinsky:1981vdw}, as well as novel interior transitions \cite{Caceres:2024edr,Caceres:2026mug}. Our method here could be used to track such transitions.

\vspace{2mm}

\noindent \textbf{Holographic flows into the interior.} For AdS black holes, via AdS/CFT duality, there are a number of holographic ``observables'' with which to indirectly probe the interior geometry. These include analytically continued one-point functions and thermal correlators \cite{Fidkowski:2003nf,Festuccia:2005pi,Grinberg:2020fdj,Parisini:2023nbd,Ceplak:2024bja,Chakravarty:2025ncy,Afkhami-Jeddi:2025wra,Jia:2026ryl,Ceplak:2026ard}, entanglement entropy \cite{Hartman:2013qma,Anegawa:2024kdj,Li:2026bof}, and complexity \cite{Jorstad:2023kmq}. Further, when AdS black holes have scalar hair, the singularity geometry is deformed to a more general Kasner universe \cite{Frenkel:2020ysx}. This initiated a plethora of work focused on dynamics and deformations of Kasner geometry, e.g.,  \cite{Hartnoll:2020rwq,Hartnoll:2020fhc,Sword:2021pfm,Sword:2022oyg,
Caceres:2022smh,Hartnoll:2022snh,Caceres:2022hei,Caceres:2023zhl,Caceres:2023zft,DeClerck:2023fax,Blacker:2023ezy,Carballo:2024hem}, which has been understood in the language of holographic renormalization group flows \cite{Das:2021vjf,Caceres:2022smh,Hartnoll:2022snh}. It would be interesting to connect the asymptotic QNM spectra to these holographic flows by, for example, tracking the asymptotic spectral data as the thermal $a$-function tends to zero as one delves deeper into the interior.

\vspace{2mm}

\noindent \textbf{Asymptotic QNMs and OPE signatures.} In this article we asked what information
about the black hole interior can be recovered from its QNM spectra. For AdS black holes or black branes with a (thermal) CFT dual, an old question has been, given a CFT correlator, what does its operator product expansion (OPE) data reveal about the bulk singularity \cite{Fidkowski:2003nf,Festuccia:2005pi}. In particular, in the large-$N$, planar diagram limit of the dual CFT, thermal CFT correlation functions feature branch-point singularities associated to the (boundary time) difference between endpoints of geodesics that ``bounce'' off the black hole/brane singularity \cite{Parisini:2023nbd,Ceplak:2024bja,Chakravarty:2025ncy}. In \cite{Ceplak:2026ard} the asymptotic behavior of holographic OPE data in scalar two-point functions for CFT states dual to black branes, specifically the power-law exponent of the bouncing singularities, was directly related to the Kasner exponent associated with the time direction, but not the transverse Kasner exponent. It would be interesting to directly relate the asymptotic QNM spectra to such holographic OPE data.

\noindent\section*{Acknowledgments}

We are grateful to Ruth Gregory and Juan Pedraza for insightful feedback. CC and AG are supported by King’s College London through NMES-funded studentships. RAH is supported by a Willmore Fellowship at Durham University. KS is supported by a scholarship from King's-China Scholarship Council. AS is funded by the Royal Society under the grant “Concrete Calculables in Quantum de Sitter” and  further supported by the STFC consolidated grant ST/X000753/1.

\appendix

\section{Numerical methods}
\label{app:numerical_methods}
In this appendix we collect technical details of the numerical methods used in Section \ref{sec:numericmethods}. We employ two complementary approaches: a pseudospectral method and a finite Hill-determinant method inspired by Leaver's approach.

For the pseudospectral method, we begin with a pedagogical introduction to Chebyshev--Gauss--Lobatto (CGL) collocation points, Chebyshev polynomials, and differentiation matrices. We then apply this formalism to the specific problems considered in this work. In particular, we show how the radial differential equation and the associated boundary conditions are discretized to obtain a generalized matrix eigenvalue problem, whose eigenvalues provide numerical approximations to the QNM frequencies.

For the Leaver-inspired method, we first compactify the radial domain and factor out the prescribed asymptotic behavior at the boundaries. We then expand the remaining regular part of the radial function about the event horizon, obtaining a finite-term recurrence relation for the expansion coefficients. Upon truncation, this recurrence can be written as a finite-dimensional matrix equation. The QNM frequencies are determined by locating the zeros of the corresponding finite Hill determinant.

Finally, we compare the QNM spectra obtained using the two methods.

\subsection{Pseudospectral method}\label{app:pseudospectral}

\paragraph{Pedagogical introduction.} Our treatment of the pseudospectral method follows \cite{Boyd:2001,el-baghdady2016}. Suppose we would like to solve $\mathcal{L}u(r)=0$ on some interval, where $\mathcal{L}$ is a linear differential operator and $u(r)$ is an unknown function. We assume we can approximate the unknown function by a finite sum of $N+1$ basis functions,
\begin{equation}
    u(r)\approx u_{N}(r)=\sum_{n=0}^{N}a_{n}\phi_{n}(r)\;.
    \label{eq:ps-basis-expansion}
\end{equation}
We define the \emph{residual} function $\mathcal{R}(r;a_{0},\dots,a_{N})\equiv\mathcal{L}u_{N}(r)$, which measures how far the truncated series is from the true solution. The goal is to choose the series coefficients $a_n$ such that the residual is minimized. In the \emph{pseudospectral} (or \emph{collocation}) method, one simply demands that the residual vanishes at a set of $N+1$ points $r_j$, the so-called collocation nodes,
\begin{equation}
    \mathcal{R}(r_{j};a_{0},\dots,a_{N})=0\;,\qquad j=0,\dots,N,
    \label{eq:ps-collocation-condition}
\end{equation}
which yields $N+1$ algebraic equations for $N+1$ unknowns. In practice, we can work directly with the values $u_j\equiv u(r_j)$ of the function at the nodes rather than with the coefficients $a_n$. In this approach, the action of $\partial_r$ is encoded in a differentiation matrix $\mathbf{D}$, so that the whole operator $\mathcal{L}$ can be represented by an $(N+1)\times (N+1)$ matrix.

There are two free choices to be made: the basis functions $\{\phi_n\}$ and the nodes $\{r_j\}$. For non-periodic problems on a finite interval, it turns out that the Chebyshev polynomials $T_n$ are an optimal, or close to optimal, basis \cite{Boyd:2001}. A natural choice of nodes is the Chebyshev-Gauss-Lobatto (CGL) points, which are the extrema of $T_N$, together with the endpoints. CGL nodes are clustered towards the endpoints of the interval; this suppresses Runge's phenomenon, which afflicts uniformly spaced grids. Another benefit of these choices is that the error of the approximation \eqref{eq:ps-basis-expansion} decays exponentially in $N$ for functions that are analytic in a neighborhood of the interval, rather than as a power of $N$ as in finite-difference or finite-element schemes. Indeed, this is what makes the pseudospectral method so effective for finding QNMs: as we will describe below, the rescaled radial function we solve for is analytic on the domain, and we are able to determine many overtones by using a few hundred nodes.

\vspace{2mm}

\noindent \textit{Chebyshev polynomials.} The Chebyshev polynomials of the first kind, $T_n$, are the eigenfunctions of the Sturm-Liouville problem,
\begin{equation}
    \frac{d}{dx} \left( \sqrt{1-x^2}\frac{dT_n(x)}{dx}\right) + \frac{n^2}{\sqrt{1-x^2}}T_n(x)=0,
\end{equation}
and are given by the recurrence relation,
\begin{equation}
\begin{split}
    T_0(x)&=1, \\
    T_1(x)&=x, \\
    T_{n+1}(x)&=2x T_n(x) - T_{n-1}(x).
\end{split}
\end{equation}
These are mutually orthogonal over the interval $[-1,1]$ with respect to the weight $w(x) = \frac{1}{\sqrt{1-x^2}}$, that is,
\begin{equation}
\begin{split}
    \int^1_{-1} T_n(x) T_m(x) w(x) dx = \frac{d_n \pi}{2}\delta_{nm}, \\
    d_0=2, d_n=1, \quad\forall n\geq 1.
\end{split}
\end{equation}
More commonly seen is the explicit relation of these Chebyshev polynomials to trigonometric functions,
\begin{equation}
    T_n(x) = \cos\left(n \arccos\left(x\right)\right).
\end{equation}

The CGL nodes are the extrema of $T_n(x)$, along with the endpoints. For an order $N$ Chebyshev polynomial, there are $N+1$ collocation points given by
\begin{equation}\label{eq:cgl-nodes}
    z_j^{(N)} = -\cos\left(\frac{j\pi}{N}\right), \quad j=0,1,\ldots,N.
\end{equation}
Note that sometimes the minus sign in front of the cosine is omitted, which simply reverses the order of the CGL points. These points are clustered quadratically towards $x=\pm1$, with spacing $\mathcal{O}(N^{-2})$ near the endpoints. This choice of points is effective at reducing interpolation error, and the clustering reduces Runge's phenomenon, where oscillations near an interval's boundaries increase with higher order polynomial interpolation, similar to Gibbs's phenomenon in Fourier series approximations.

\vspace{2mm}

\noindent \textit{Lagrange interpolating polynomials and differentiation matrices.} Rather than working directly with the spectral coefficients $a_n$ of \eqref{eq:ps-basis-expansion}, it is convenient to work instead with the values of the function at the nodes. This is achieved by using the Lagrange interpolating polynomials associated with the CGL nodes as the basis functions. Given $N+1$ CGL nodes, the Lagrange basis for polynomials of degree $\leq N$ is the set of polynomials $\{\phi_i^{(N)}(z)\}^N_{i=0}$, each of degree $N$,
\begin{equation}\label{eq:lagrange-basis}
    \phi_i^{(N)}(z) = \prod^N_{m=0,m\neq i}\frac{z-z_m^{(N)}}{z_i^{(N)}-z_m^{(N)}}, \quad i=0,\ldots,N,
\end{equation}
with the property
\begin{equation}
    \phi_i^{(N)}\left(z_j^{(N)}\right)=\delta_{ij}.
\end{equation}
The numerator of \eqref{eq:lagrange-basis} has a root at every node except the $i$-th, while the denominator normalizes the polynomial to one at that node. A compact, closed form is given by
\begin{equation}
    \phi_i^{(N)}(z) = \frac{(-1)^{i+N+1} (1-z^2)}{c_i N^2 \left(z-z_i^{(N)}\right)}T'_N(z),
\end{equation}
where 
\begin{equation}
    c_i = 
    \begin{cases}
        2, \quad i=0,N, \\
        1, \quad 1\leq i\leq N-1.
    \end{cases}
\end{equation}

Any function $f$ defined on the interval $[-1,1]$ may be approximated by Lagrange basis polynomials as 
\begin{equation}\label{eq:f-approx}
    f(z) \approx \sum^N_{i=0} f_i \phi_i^{(N)}(z),
\end{equation}
where $f_i \equiv f\left(z_i^{(N)}\right)$. This will be exact if $f(z)$ is a polynomial of degree at most $N$.

In matrix notation, \eqref{eq:f-approx} is
\begin{equation}
    f(z) \approx \mathbf{\Phi}^{(N)} \cdot \mathbf{F},
\end{equation}
where
\begin{align}
    \mathbf{\Phi}^{(N)} &= \left[\phi_0^{(N)}(z),\ldots,\phi_N^{(N)}(z)\right], \\
    \mathbf{F}&=\left[f\left(z_0^{(N)}\right),\ldots,f\left(z_N^{(N)}\right)\right]^T.
\end{align}

The first derivative of \eqref{eq:f-approx} is 
\begin{equation}
    f'(z) \approx \sum^N_{i=0} f_i \phi_i^{\prime (N)}(z)\;,
\end{equation}
where $\phi^{\prime (N)}_i(z)$ is a polynomial of degree $N-1$. Now we can use \eqref{eq:f-approx} to write $\phi'$ in terms of Lagrange basis polynomials,
\begin{equation}
    \phi^{\prime (N)}_i(z) = \sum^N_{k=0}\phi^{\prime \left(N\right)}_i\left(z_k^{\left(N\right)}\right)\phi^{\left(N\right)}_k\left(z\right), \quad i=0,\ldots,N.
\end{equation}
In matrix form,
\begin{equation}\label{eq:lagrange_pol_diff}
    \frac{d}{dz}\mathbf{\Phi}^{(N)}(z) = \mathbf{\Phi}^{(N)}(z) \mathbf{D}_{N+1}\,,
\end{equation}
where $\mathbf{D}_{N+1}$ is the differentiation matrix with dimension $N+1$. We have
\begin{equation}
    \left[\mathbf{D}_{N+1}\right]_{i,k} = \phi'^{ (N)}_k \left(z_i^{(N)}\right),
\end{equation}
where on the right hand side, we have the derivative of the $k^{\mathrm{th}}$ Lagrange basis polynomial evaluated at the $i^{\mathrm{th}}$ CGL node $z_i^{(N)}$.

For the CGL nodes, the entries in the differentiation matrix can be given in closed form, 
\begin{equation}\label{eq:diff-matrix}
    \left[\mathbf{D}_{N+1}\right]_{i,k} = 
    \begin{cases}
        \frac{c_i}{c_k} \frac{(-1)^{i+k}}{z_i^{(N)}-z_k^{(N)}}, \quad &i\neq k, \\
        -\frac{2N^2+1}{6}, \quad &i=k=0, \\
        -\frac{z_i^{(N)}}{2(1-(z_i^{(N)})^2)}, \quad &1\leq i=k \leq N-1, \\
        \frac{2N^2+1}{6}, \quad &i=k=N.
    \end{cases}
\end{equation}
The second order differentiation matrix is given by
\begin{equation}
    \left[\mathbf{D}^2_{N+1}\right]_{i,k} = 
    \begin{cases}
        2[\mathbf{D}_{N+1}]_{i,k}\left([\mathbf{D}_{N+1}]_{i,i}-\frac{1}{z_i^{(N)}-z_k^{(N)}}\right),\quad &i\neq k, \\
        -\sum^N_{k=0,k\neq i}[\mathbf{D}^2_{N+1}]_{i,k},\quad &i=k, 
    \end{cases}
\end{equation}
which coincides with the square of $\mathbf{D}$. Constructing $\mathbf{D}^2$ from this explicit expression is more accurate in finite precision calculations than squaring the matrix $\mathbf{D}$ \cite{baltensperger2003}. Note that $\mathbf{D}$ is fully specified by the grid, and is independent of the equation being solved, and so may be computed once for a particular grid size and reused. 

\vspace{2mm}

\noindent \textit{Extension to arbitrary interval.} The discussion so far has been for the interval $[-1,1]$; however, we can extend this to an arbitrary interval. Any function $h(x)$ defined on an interval $[a,b]$ may be approximated by making the transformation from $z\in [-1,1]$ to $x\in [a,b]$ as
\begin{equation}\label{eq:fn-arbitrary-interval}
    h(x) \approx \sum^N_{i=0} h\left(x_i^{(N)}\right) \phi^{(N)}_{i,[a,b]}(x),
\end{equation}
where 
\begin{equation}
\begin{aligned}
    x_i^{(N)} &= \frac{b-a}{2}(z_i^{(N)}+1) + a,\quad  &i=0,\ldots,N,\\
    \phi^{(N)}_{i,[a,b]}(x)&=\phi_i^{(N)} \left(\frac{2}{b-a} (x-a) -1\right), \quad &i=0,\ldots,N,
\end{aligned}
\end{equation}
are the shifted CGL nodes and Lagrange basis polynomials associated with the interval $[a,b]$. Then the derivative of the transformed Lagrange polynomials is given by a simple modification to \eqref{eq:lagrange_pol_diff},
\begin{equation}\label{eq:scaled-diff}
    \frac{d^n}{dx^n} \mathbf{\Phi}^{(N)}_{[a,b]}(x) = \left(\frac{2}{b-a}\right)^n \mathbf{\Phi}^{(N)}_{[a,b]}(x)\mathbf{D}_{N+1}^n,
\end{equation}
where $\mathbf\Phi^{(N)}_{[a,b]}$ is the row vector of shifted Lagrange basis polynomials.

Finally, putting the pieces together: to solve $\mathcal{L}u(x)=f(x)$ we expand
\begin{equation}
    u(x) \approx \sum^N_{i=0} u\left(x_i^{(N)}\right) \phi_{i,[a,b]}^{(N)}(x),
\end{equation}
and then impose the differential equation at each collocation point,
\begin{equation}
    \left[\mathcal{L} \left(\sum^N_{i=0} u\left(x_i^{(N)}\right) \phi_{i,[a,b]}^{(N)}(x)\right)\right]_{x=x_j^{(N)}} = f\left(x_j^{(N)}\right), \quad j=0,1,\ldots,N.
\end{equation}
Equivalently, if
\begin{equation}
    \mathcal{L}=a_2(x)\partial_x^2+a_1(x)\partial_x+a_0(x),
\end{equation}
then we have
\begin{equation}
    \sum_{i=0}^N\left[a_2(x_j)\left(\tfrac{2}{b-a}\right)^2\left[\mathbf{D}^2\right]_{ji}+a_1(x_j)\left(\tfrac{2}{b-a}\right)\left[\mathbf{D}\right]_{ji}+a_0(x_j)\delta_{ji}\right]u_i=f(x_j).
\end{equation}
This is a system of $N+1$ equations for the $N+1$ unknowns $u_{i}\equiv u(x_{i}^{(N)})$. The derivatives in $\mathcal{L}$ have been replaced by the appropriate differentiation matrix, and the coefficients by diagonal matrices containing the values at the nodes. The differential equation has been transformed into an algebraic system of equations which can be solved numerically.

 For details on other aspects of this method, we refer the reader to more comprehensive sources, such as \cite{Boyd:2001,Trefethen:2000} for an overview, \cite{matthews2004} for error analysis, \cite{mendes2019} for examples, and \cite{baltensperger2003} for numerical tricks which can be used to reduce numerical errors.

Before we can apply the pseudospectral method, the QNM problem must be brought into a suitable form. This involves (i) compactifying the semi-infinite radial domain to a finite interval, (ii) factoring out any singular behavior at the endpoints so that the function approximated by the interpolant is analytic, and (iii) making explicit the dependence of the equation on $\omega$ so that the discretized system of equations can be recast as a generalized matrix eigenvalue problem. We address each of these in turn. Throughout, we keep the blackening factor $f(r)$ general, so that the neutral and charged quantum BTZ black holes, and the neutral, flat quantum black hole, as well as their tensionless limits, can be treated simultaneously. 

\paragraph{Compactification.}
In Section~\ref{sec:scalar-qbh-brane} we brought the radial wave equation into a Schr\"odinger-like form. Here, we proceed by factoring out the ingoing behavior at the black hole horizon. We write
\begin{equation}
    R(r)=e^{-i\omega r_*}\Phi(r),
    \label{eq:ingoing-redefinition-general}
\end{equation}
so that $\Phi(r)$ is regular at a future event horizon. The radial equation becomes
\begin{equation}
    f(r)\Phi''(r)
    +
    \left[
        f'(r)-2i\omega
    \right]\Phi'(r)
    -
    \frac{1}{4r^2}
    \left[
        4m^2-f(r)+2r f'(r)
    \right]\Phi(r)
    =
    0.
    \label{eq:brane-radial-phi-r}
\end{equation}
Next, we introduce the compactified coordinate
\begin{equation}
    y=-\frac{1}{r},
    \label{eq:y-compact-coordinate}
\end{equation}
which maps the black hole exterior $r\in(r_h,\infty)$ onto the finite interval,
\begin{equation}
    y\in[y_h,0], \qquad y_h=-\frac{1}{r_h}<0,
\end{equation}
with the horizon at $y=y_h$ and the AdS$_3$ boundary, or in the flat case, spatial infinity, at $y=0$. This deals with point (i) above. In this compactified coordinate the equation becomes 
% \begin{equation}
% \begin{split}
% 0&=
% y^4 f(y)\Phi''(y)
% +
% y^2
% \left[
%     -2i\omega
%     +2y f(y)
%     +y^2 f'(y)
% \right]\Phi'(y) \\
% &+
% \frac{y^2}{4}
% \left[
%     -4m^2
%     +f(y)
%     +2y f'(y)
% \right]\Phi(y),
% \end{split}
% \label{eq:brane-radial-phi-y}
% \end{equation}
% where primes denote derivatives with respect to $y$. 

% It is convenient to rewrite \eqref{eq:brane-radial-phi-y} in the compact form
\begin{equation}\label{eq:brane-radial-compact}
    a_{2}(y)\Phi''(y) + \left[a_{1}(y)-2i\omega y^{2}\right]\Phi'(y)+a_{0}(y)\Phi(y)=0,
\end{equation}
with the coefficient functions,
\begin{equation}\label{eq:ps-a-coeffs}
\begin{split}
    a_{2}(y)&=y^{4}f(y),\\
    a_{1}(y)&=2y^{3}f(y)+y^{4}f'(y),\\
    a_{0}(y)&=-m^{2}y^{2}+\frac{y^{2}}{4}f(y)+\frac{y^{3}}{2}f'(y).
\end{split}
\end{equation}
% For quantum BTZ black holes, the blackening factor \eqref{eq:charged-qbtz-mass-form-zoo} in the compactified coordinate is
% \begin{equation}\label{eq:blackening-y}
%     f(y)_{\rm qcBTZ}=\frac{1}{\ell_{3}^{2}y^{2}}-8\mathcal{G}_{3}M+\ell W(M,Q)y+\ell^{2}P(M,Q)y^{2},
% \end{equation}
% with $8\mathcal{G}_{3}M$, $W(M,Q)$ and $P(M,Q)$ given in \eqref{eq:charged-qbtz-functions-zoo}. The neutral qBTZ black hole is recovered by setting $P(M,Q)=0$, and the tensionless limit by $\eta\to1$ together with the scaling \eqref{eq:mu-scaling-tensionless-new}.

% For asymptotically flat, neutral quantum black holes, the blackening factor \eqref{eq:charged-qschw-mass-form-zoo} in the compactified coordinate is
% \begin{equation}
%     f_{\rm flat}(y) = 1 - 8 G_3 M + 
%     \ell W(M,Q) y
%     +
%     \ell^2 P(M,Q) y^2 \, ,
%     \label{eq:charged-qschw-blackening-y}
% \end{equation}
% with $8G_{3}M$, $W(M,Q)$ and $P(M,Q)$ given in \eqref{eq:charged-qschw-functions-zoo}.

\paragraph{Boundary conditions.}
We now turn to point (ii). In principle, we could impose the boundary conditions \eqref{eq:asympsolns} directly on the discretized system, by replacing the first and last rows of the matrices with the appropriate constraints. However, it is simpler and more numerically stable to first factor out the behavior at the endpoints analytically, and then solve for the remaining function which is regular at both ends, as per \cite{Jansen:2017oag}. We use the indicial exponents at $y=y_h$ and $y=0$, which follow from the following Frobenius analysis.

Near a non-extremal (non-degenerate) event horizon, we have $f(y_h)=0$ and $f'(y_h)\neq0$. A Frobenius expansion then gives
\begin{equation}
\begin{split}
    \Phi(y)&\sim (y-y_h)^\gamma, \\
    \gamma=0,& \quad \gamma=\frac{2i\omega}{y_h^2 f'(y_h)}.
\end{split}
\label{eq:horizon-frobenius-general}
\end{equation}
The ingoing solution corresponds to $\gamma=0$, since the ingoing phase has already been extracted in \eqref{eq:ingoing-redefinition-general}. For asymptotically AdS geometries with
\begin{equation}
    f(y)\sim \frac{1}{L^2y^2},
    \quad \text{as }y\to0,
    \label{eq:ads-asymptotic-f}
\end{equation}
the two asymptotic behaviors of $\Phi$ are
\begin{equation}
    \Phi(y)\sim y^{-1/2},
    \qquad
    \Phi(y)\sim y^{3/2}.
    \label{eq:boundary-frobenius-general}
\end{equation}
The normalizable branch is therefore $\Phi(y)\sim y^{3/2}$.
In the asymptotically flat case, while the expansion of the field $\Phi(r)$ at the horizon is the same as in the AdS case, at infinity (remembering \eqref{eq:ingoing-redefinition-general}) for an outgoing wave it behaves in the following way:
\begin{equation}
    \Phi(y)\sim\exp\left(\frac{-2 i \omega}{\Delta^2 \kappa\, y}\right) y^{\frac{- 2 i \ell_4 \,\mu \,\omega}{\Delta \kappa^2}}.
\end{equation}
For numerical stability, we then define the rescaled radial function $\psi(y)$ by factorizing out the boundary behavior of $\Phi(y)$ with additional powers of $-1$ in the prefactors to ensure that $\psi(y)$ vanishes linearly at both endpoints, rather than approaching non-zero constants:
\begin{equation}\label{eq:ps-rescaling}
    \Phi_{\rm AdS}(y)=(y-y_h)^{0-1}y^{\frac{3}{2}-1} \psi(y), \qquad \Phi_{\rm flat}(y)=(y-y_h)^{0-1}\exp\left(\frac{-2 i \omega}{\Delta^2 \kappa\, y}\right) y^{\frac{- 2 i \ell_4 \,\mu \,\omega}{\Delta \kappa^2}} \psi(y).
\end{equation}
This allows the boundary conditions to be enforced automatically. 

\paragraph{The generalized eigenvalue problem.}

Finally, we come to point (iii). Inserting \eqref{eq:ps-rescaling} into the wave equation \eqref{eq:brane-radial-compact} for each AdS case and dividing through by the prefactor gives the equation for $\psi$,
\begin{equation}\label{eq:ps-psi-operators}
    \left[\mathcal{L}_{0}+\omega\mathcal{L}_{1}\right]\psi(y)=0,
\end{equation}
with
\begin{equation}
\begin{split}
    \mathcal{L}_0&=a_2 \partial_y^2+\left[2a_2\frac{p'}{p}+a_1\right]\partial_y+\left[a_2\frac{p''}{p}+a_1\frac{p'}{p}+a_0\right],\\
    \mathcal{L}_1&=-2iy^2\left[\partial_y+\frac{p'}{p}\right],
\end{split}
\end{equation}
where we have defined $p(y)$ to be the prefactor in \eqref{eq:ps-rescaling}, with $p_{\rm AdS}(y) = (y-y_h)^{-1}y^{\frac{1}{2}}$. In the asymptotically flat case, the situation is slightly different. The prefactor which implements the outgoing boundary condition at spatial infinity contains the frequency explicitly ($p_{\rm{flat}}(y)=(y-y_h)^{-1}\exp\left(\frac{-2i\omega}{\Delta^2\kappa y}\right)y^{\frac{-2i\ell_4\mu\omega}{\Delta \kappa^2}}$) and consequently, the resulting equation for $\psi(y)$ is no longer linear in $\omega$. In particular, we have a quadratic eigenvalue problem of the type: 
\begin{equation}
   \left[ \mathcal{A}_0+\omega \mathcal{A}_1+\omega^2 \mathcal{A}_2\right] \psi(y)=0.
\end{equation}
We solve this problem by writing it as a generalized eigenvalue problem of twice the size. Introducing the auxiliary vector 
\begin{equation}
    \vec v=\omega \vec \psi
\end{equation}
we can rewrite the quadratic problem as:
\begin{equation}
    \begin{pmatrix}
        0&I\\
        -\mathcal{A}_0 & -\mathcal{A}_1
    \end{pmatrix}
    \begin{pmatrix}
        \vec \psi\\
        \vec u
    \end{pmatrix}
    =\omega\begin{pmatrix}
        I&0\\
        0&\mathcal{A}_2
    \end{pmatrix}
    \begin{pmatrix}
        \vec \psi\\
        \vec u
    \end{pmatrix}
\end{equation}
This linearized problem is equivalent to the original quadratic one, but it can be solved with the same generalized-eigenvalue methods that we use for the AdS cases, at the price of doubling the matrix size of $\mathcal{L}_0$ and $\mathcal{L}_1$ in \eqref{eq:ps-psi-operators}.

At this stage, the $\omega$-dependence is explicit and we can discretize the problem. We map $N+1$ CGL nodes $y_j$ onto the domain $[y_h,0]$ and represent $\psi$ by a vector containing its values at the nodes, $\vec{\psi}=\left(\psi(y_0),\ldots,\psi(y_N)\right)^T$. The derivatives in \eqref{eq:ps-psi-operators} are replaced by Chebyshev differentiation matrices $\mathbf{D}$ and $\mathbf{D}^2$, and the coefficient functions by diagonal matrices. The operators are thereby represented by $(N+1)\times(N+1)$ matrices, and \eqref{eq:ps-psi-operators} becomes the generalized eigenvalue problem,
\begin{equation}
    \mathbf{A}\vec{\Psi}=\omega\,\mathbf{B}\vec{\Psi}\;.
\end{equation}
The QNM frequencies are the eigenvalues of this problem, which we obtain with the inbuilt \texttt{Eigenvalues} function in Wolfram Mathematica. 

An inherent limitation of this method is that the discretized problem has exactly $N+1$ eigenvalues, whereas the actual differential equation has infinitely many quasinormal modes. Additionally, many of the eigenvalues of the discrete problem are spurious: they are associated with the grid rather than with the equation being discretized, and their number increases with $N$. Fortunately in practice, it is simple enough to distinguish between the QNM frequencies and the spurious modes by repeating the calculations with different numbers of CGL nodes. Changing the number of nodes causes the spurious modes to drift systematically, whereas the QNMs converge with increasing $N$.

\subsection{Leaver-inspired method}

Let us now describe our second numerical scheme which was used to obtain the QNMs for very high overtones. The implementation of this method was carried out by ChatGPT with human oversight. Here we explain the basics of the method used. Our exposition here is a result of combining the human-generated instructions with the explanations obtained by  having an independent instance of ChatGPT audit the numerical implementation.

\paragraph{Problem setup.}
In this section we work with a metric of the form
\be 
 {\rm d} s^2=-f(r){\rm d} t^2+\frac{{\rm d} r^2}{f(r)}+r^2{\rm d}\phi^2,
 \qquad
 f(r)=\frac{r^2}{\ell_3^2}-B-\frac{A}{r}+\frac{D}{r^2} \,.
 \label{eq:fast-metric}
\ee
This reproduces our solutions of interest for $A=\ell W$, $B=8\mathcal G_3M$, and $D=\ell^2P$. On this background we wish to extract the QNM frequencies. We begin by factoring out the ingoing horizon behavior. In Eddington--Finkelstein time $v = t + r_*$ we write $\Psi = e^{- i \omega v + i m \phi} u(r)$, in terms of which the scalar equation becomes
\be 
f u'' + \left(f' + \frac{f}{r} -2 i \omega \right) u' - \left(\frac{i \omega}{r} + \frac{m^2}{r^2} \right) u = 0\, .
\label{eq:fast-ef}
\ee
The ingoing solution then has a regular Frobenius expansion at $r_+$, while normalizability at the AdS boundary requires $u = O(r^{-2})$. Next we introduce $x = r_+/r$, factor $u = x^2 y(x)$, and define
\be 
 \alpha=\frac{A\ell_3^2}{r_+^3},\qquad
 \delta=\frac{D\ell_3^2}{r_+^4},\qquad
 \beta=1-\alpha+\delta,\qquad
 w=\frac{\omega\ell_3^2}{r_+},\qquad
 \chi=\frac{m^2\ell_3^2}{r_+^2}\,,
 \label{eq:fast-dimensionless}
\ee 
along with
\be 
p(x)=1-\beta x^2-\alpha x^3+\delta x^4 \,  ,
\ee 
to arrive at a new form of the radial scalar equation, 
\be 
xp\,y''(x)+(3p+xp'+2 i wx)y'(x)
 +(2p'-\chi x+3 i w)y(x)=0 \, .
 \label{eq:fast-regularized}
\ee
Here primes denote derivatives with respect to the displayed argument. In this form, the boundary is at $x = 0$ and the solution must be finite there. The horizon is at $x = 1$, and there we demand analyticity. To proceed further, it will be useful to introduce yet another coordinate $z = 1- x$ and expand the solution as a power series in $z$,
\be 
y(z) = \sum_{k=0}^\infty h_k z^k \, .
\ee
In this coordinate system, the equation takes the form $a(z) y''(z) + b(z) y'(z) + c(z) y(z) = 0$ with
\be 
a(z) = x p(x) \, , \quad b(z) = -3 p(x) - x p'(x) -2 i w x \, , \quad c(z) = 2 p'(x) - \chi x +3 i w \, ,
\ee
where it is understood that $x = 1 - z$.  Let us then define $a_j, b_j,$ and $c_j$ to be the coefficients of the different powers of $z$ in these polynomials. More explicitly,
\be 
a(z) = \sum_{j=1}^5 a_j z^j \, , \quad b(z) = \sum_{j=0}^4 b_j z^j\,, \quad c(z) = \sum_{j=0}^3 c_j z^j \, .
\ee
For a given metric and test frequency, the $\{a_j, b_j, c_j\}$ are known. We do not present the explicit values of these coefficients (there are a lot and they are a bit messy). By contrast, the $h_k$ in $y(z)$ are not known \textit{a priori} and must be determined. Substituting the series for $y(z)$ and setting the coefficient of each power of $z^k$ to zero gives
\be \label{eq:recurr_reln}
\sum_{j=0}^4 C_{k,j} h_{k+1-j} = 0 \, , \quad C_{k,j} = (k+1 - j) (k-j) a_{j+1} + (k+1-j) b_j + c_{j-1} \, ,
\ee
with the convention that $h_j = 0$ for $j < 0$ and any polynomial coefficients outside their ranges set to zero. For example, substituting $k = 0$ in the above yields $b_0 h_1 + c_0 h_0 = 0$, so choosing the (arbitrary) normalization $h_0 = 1$ then fixes $h_1 = -c_0/b_0$ when $b_0 \neq 0$. The recurrence then determines the coefficients at successively higher orders.

We are expanding $y(z)$ around the event horizon at $z = 0$. We will want to impose boundary conditions on the resulting series at the boundary at $z = 1$. For this to be done consistently, the series expansion must converge at the boundary. One potential limitation is if the metric function $f(r)$ contains a second zero (e.g.~an inner horizon) at $r = r_-$. In that case, the differential equation will have a singular point at the corresponding value of $z = z_-$. Convergence of our series expansion near the boundary therefore requires that any such horizon (including complex ones) have $|z| \ge  1$. In turn, because $z_- = 1- r_+/r_-$ this requires $r_-/r_+ \le 1/2$ for an inner horizon. Our code checks this condition, together with the locations of any complex roots, before proceeding with the series method. 

\paragraph{Solving the recurrence relation.} The capability of the implementation derives from the techniques used to solve the recursion relations. For this we rely on long-known techniques dating back to the work of Leaver~\cite{Leaver:1985ax,Majumdar:1989tzg,Leaver:1990zz} and in particular follow the implementation described in~\cite{Benda:2025tni}. Analogous methods were used in~\cite{Cardoso:2003vt}, where it was claimed that up to 5,000 overtones were calculable. We now describe our method. 

Truncate the recurrence relation \eqref{eq:recurr_reln} at order $N$ by setting $h_{N+1} = 0$. This condition is the finite-$N$ approximation of the requirement that $y(z)$ remain finite at the boundary. Define the $(N+1)\times(N+1)$ matrix $H_N(w)$ to have entries $(H_N)_{k, k+1-j} = C_{k, j}$. In terms of this matrix, the system of equations takes the form
\be 
H_N(w) \begin{pmatrix}
h_0 \\
h_1 \\
\vdots \\
h_N
\end{pmatrix}
=0 \, .
\ee 
Hence, candidate QNM frequencies will be solutions to ${\rm det} H_N(w) = 0$ that stabilize as $N$ increases. This is a finite Hill-determinant formulation of the QNM boundary value problem.

For a trial frequency $w$, all entries of $H_N(w)$ are known. We must evaluate its determinant and adjust $w$ until the determinant vanishes. In pursuit of high overtones, the matrix $H_N(w)$ will become large. However, because of Eq.~\eqref{eq:recurr_reln} each row of $H_N(w)$ contains at most five nonzero entries. The numerical implementation utilizes this to build the determinant from smaller ones, without ever storing the full matrix. 

Let $\mathcal{D}_k$ be the determinant of the leading principal submatrix of $H_N(w)$ containing rows and columns $0,\dots, k$. Expanding the determinant along its last row gives the general rule (see also Eq.~(16) of Ref.~\cite{Benda:2025tni})
\be 
 \mathcal D_k=
 \sum_{s=1}^{4}(-1)^{s-1}C_{k,s}
 \left(\,\prod_{t=k-s+1}^{k-1}C_{t,0}\right)\mathcal D_{k-s},
 \qquad \mathcal D_{-1}=1,
\ee
Here $k$ is the final row index of the $(k+1)\times(k+1)$ block. The sum over $s$ tells us
that each new determinant uses only the four preceding values,
$\mathcal D_{k-1},\ldots,\mathcal D_{k-4}$. The value
$\mathcal D_{-1}=1$ initializes the calculation; terms with $k-s<-1$ are
omitted, and a product with no factors is one. Continuing the calculation eventually gives $\mathcal{D}_N = {\rm det} H_N$. The numerical implementation uses several additional speed-ups to evaluate the determinants. We compute determinant recurrences from both ends of the matrix and combine them at an intermediate row. The coefficients are quadratic in the row index, so we update them using finite differences to reduce the arithmetic required at each step. At intermediate stages, we rescale the determinants to avoid numerical overflow. These determinant calculations are implemented in compiled C++ code with arbitrary precision.  

\paragraph{Mode finding and validation.} Our implementation tracks only those modes with ${\rm Re}(\omega) > 0$. The calculation starts, for a given value of $m$, with the first three modes obtained independently via a pseudospectral calculation like the one described in the previous section. Once three successive overtones are available, we then use a simple approximation to obtain a trial seed frequency. At fixed $m$, let $d_{m, n-1} = \omega_{m, n-1} - \omega_{m, n-2}$ be the spacing between the previous two modes and analogously for $d_{m, n-2} = \omega_{m, n-2} - \omega_{m, n-3}$. We take as our seed frequency
\be 
\omega_{m,n}^{\rm seed} = \omega_{m, n-1} + d_{m, n-1} + \left(d_{m, n-1} - d_{m, n-2} \right) =  3 \omega_{m, n-1} - 3 \omega_{m, n-2} + \omega_{m, n-3} \, .
\ee
This assumes that the modes are approximately evenly spaced, but incorporates small variations between modes through the $d_{m, n-1} - d_{m, n-2} $ term. 

Starting from this seed frequency, we solve for vanishing determinant using a complex secant iteration. Each iteration uses the value of the determinant at two nearby trial frequencies to estimate a slope and propose a better frequency. This is iterated, with the determinant reevaluated at each trial frequency. To ensure the root finder converges to the correct next overtone, we monitor the successive spacing between modes and stop the calculation if a mode deviates significantly from the average differences between modes. To confirm that a candidate mode is converging we repeat the calculation for different values of the truncation order $N$ and with different values of the working precision.  In our implementation, we compare runs differing by $\Delta N = {\rm max}(200, n/5)$ where $n$ is the overtone number. Our initial working precision $P$ is taken to be $P = 150 + n$, with the reevaluation performed with $P = 350 + n$. If the consistency checks fail, we repeat the calculation with further increased truncation order and precision.

\subsection{Comparison of numerical methods}

\begin{figure}
    \centering
    \includegraphics[width=\linewidth]{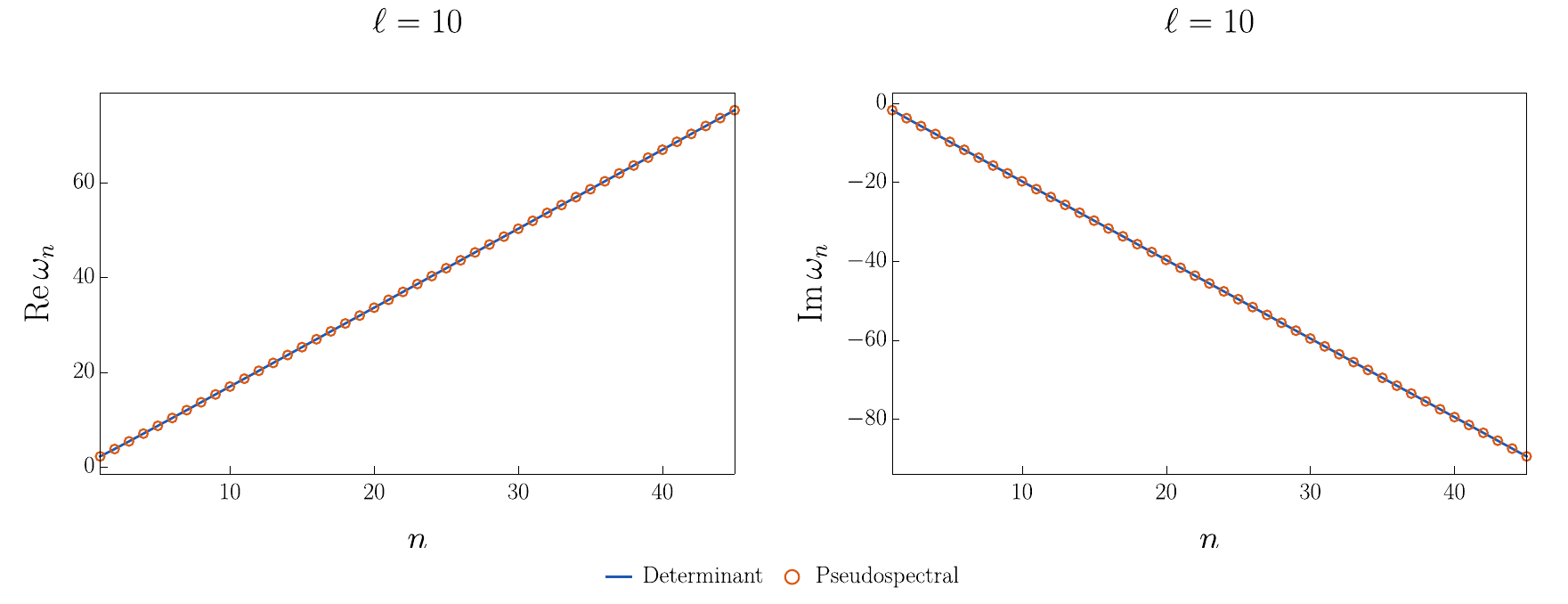}
    \includegraphics[width=\linewidth]{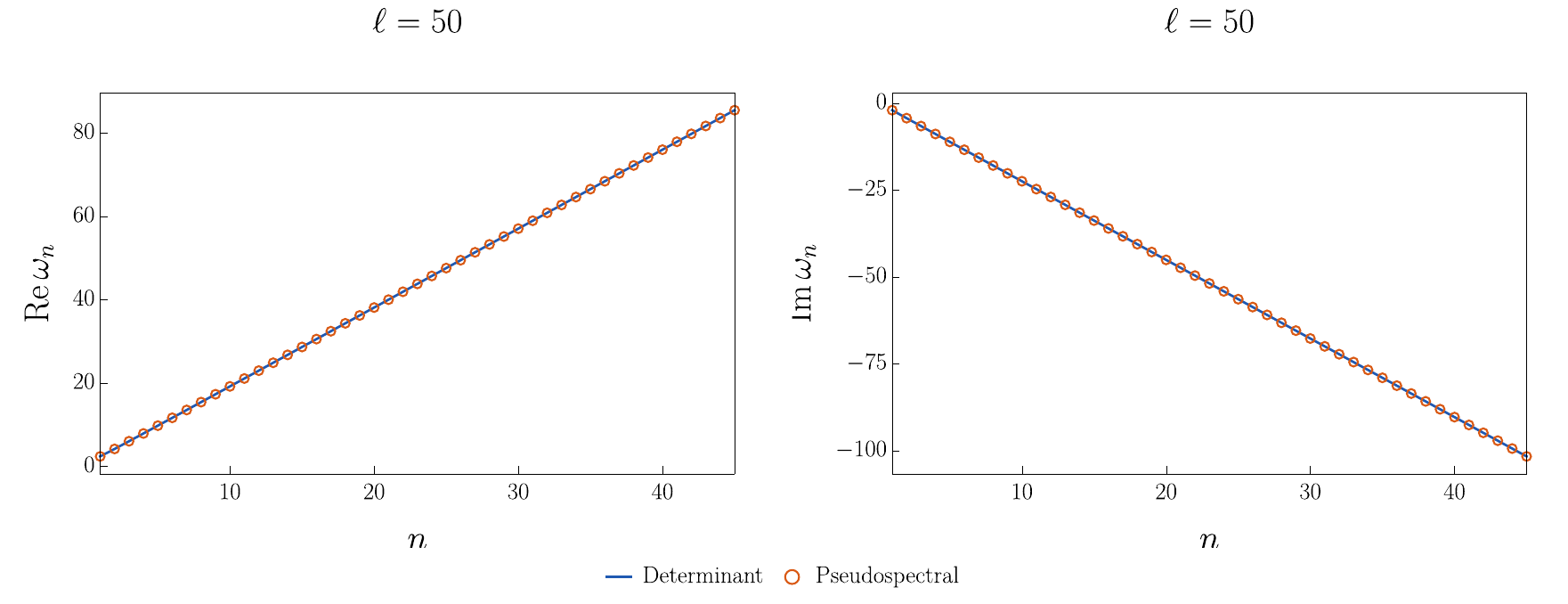}
    \includegraphics[width=\linewidth]{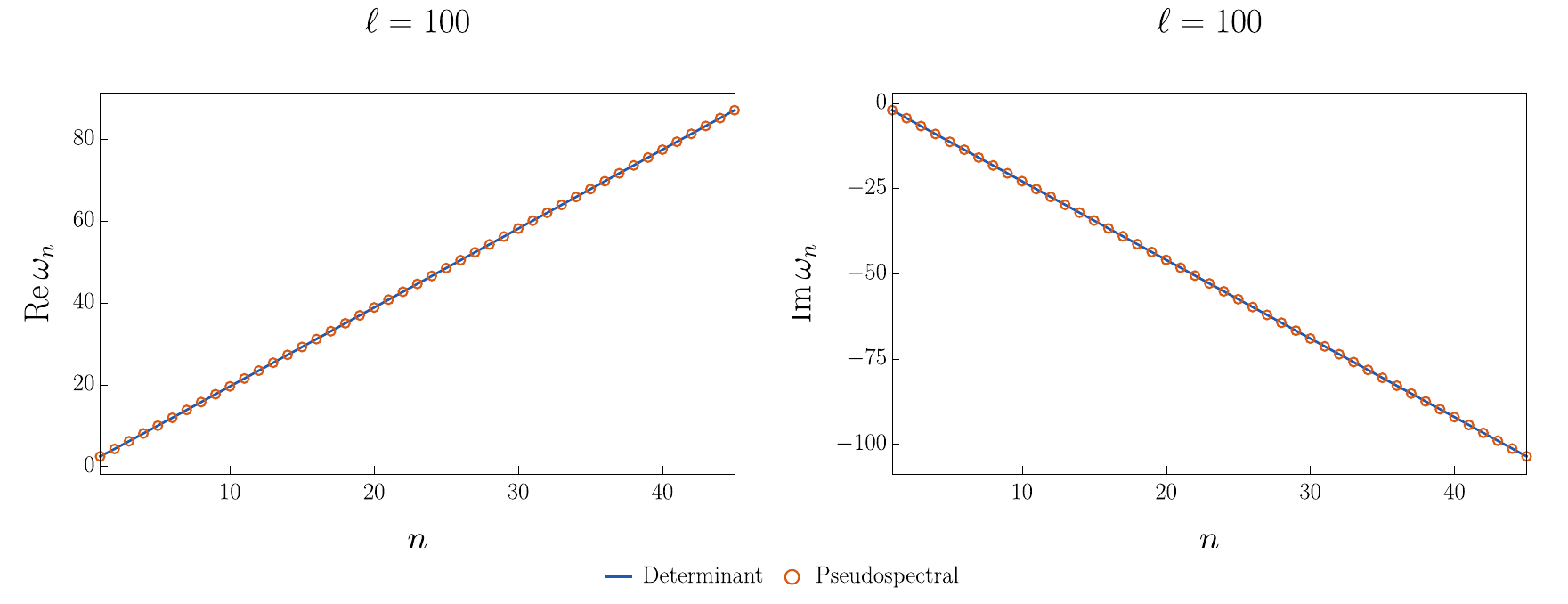}    
    \caption{Comparison of the $m=1$ QNM spectra computed using the pseudospectral and Leaver-inspired determinant methods, with $\kappa=1$, $\widetilde{M}=1$, $\ell_4=1$, and $q=0$. The spectra agree to the six decimal places reported for the pseudospectral data, with differences consistent with rounding at this precision.}
    \label{fig:neutral-numerical-compare}
\end{figure}

\begin{figure}
    \centering
    \includegraphics[width=\linewidth]{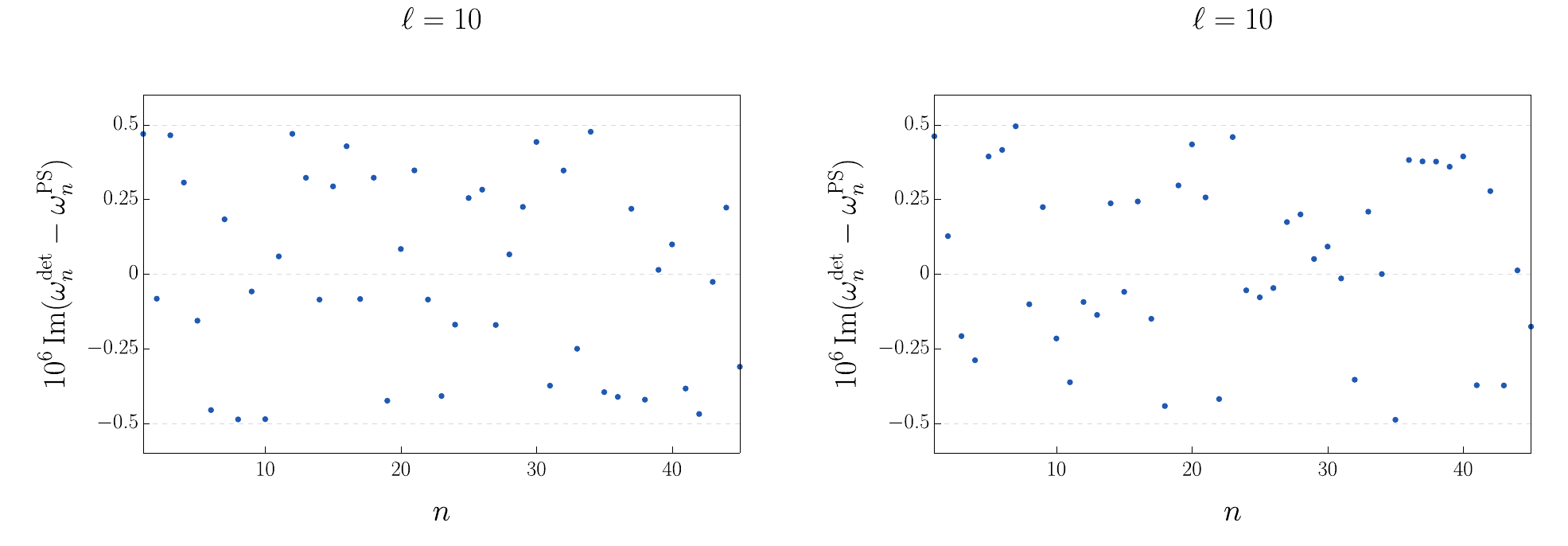}
    \includegraphics[width=\linewidth]{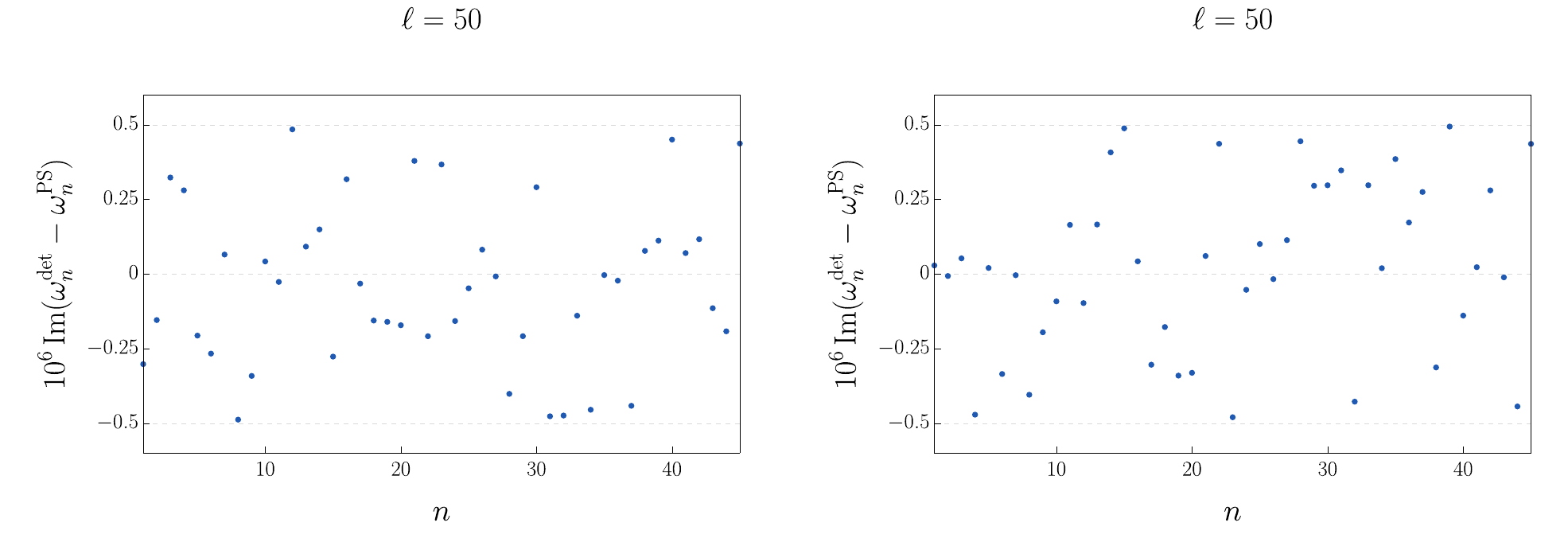}
    \includegraphics[width=\linewidth]{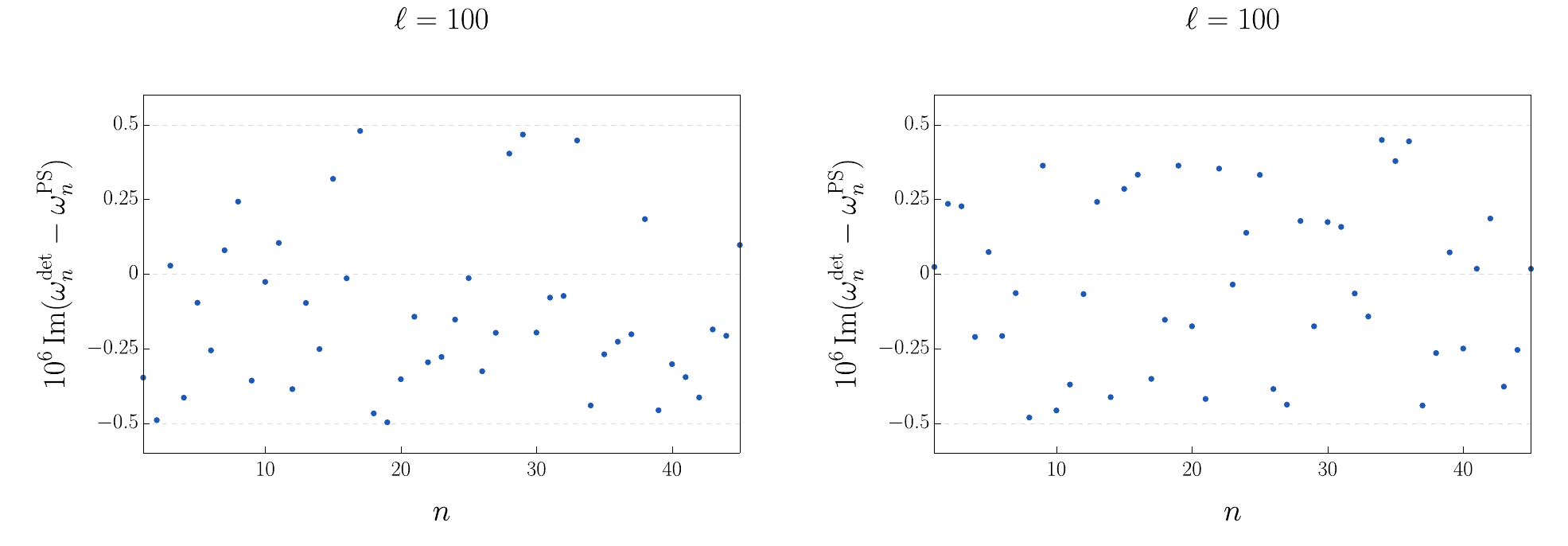}    
    \caption{Comparison of the $m=1$ QNM spectra computed using the pseudospectral and Leaver-inspired determinant methods, with $\kappa=1$, $\widetilde{M}=1$, $\ell_4=1$, and $q=0$. Comparing the subtracted spectra reveals agreement to six decimal places.}
    \label{fig:neutral-numerical-error-compare}
\end{figure}

We close this appendix by providing a brief comparison between the QNMs obtained using our fully human-generated solver and those obtained using our AI-assisted implementation. We present representative comparisons for the neutral qBTZ black hole in Figure~\ref{fig:neutral-numerical-compare}, with explicit plots of the differences shown in Figure~\ref{fig:neutral-numerical-error-compare}. The two methods are found to agree to six decimal places, which is the precision of our pseudospectral results in these plots.

\section{Analytical results for QNMs of the quantum Schwarzschild black hole}
\label{app:zeroReFlat}
In this appendix we collect supplementary analytical results for the QNM spectrum of the quantum Schwarzschild black hole. The metric function we are considering is given in Eq.~\eqref{eq:charged-qschw-mass-form-zoo} with $Q = 0$. For notational simplicity, here we will write the metric function as
\be 
f(r) =  \frac{a \left( r-r_0\right)}{r} \,  \qquad 0 < a < 1 \,, \quad r_0 > 0 \, ,
\ee
where for brevity we have $a=1-8G_3M$ and $r_0=\ell W(M)/a$. We can write the tortoise coordinate as
\be 
r_* = \frac{r - r_0}{a} + \frac{r_0}{a} \log \left(\frac{r-r_0}{r_0}\right) \, ,
\ee
where we chose the additive constant so that the logarithm has a dimensionless argument. In this appendix we are restricting the tortoise coordinate to $r > r_0$ and hence need not worry about its analytic properties.

\subsection{Exact $m=0$ QNMs }

Here we show that in the case of $m = 0$ the QNMs of the quantum Schwarzschild black hole can be obtained analytically.  The solution of the radial equation can be written as
\be \label{eq:R_flat_exact}
R(r) = \frac{1}{\sqrt{f(r)}} \left[C_1 M_{\kappa, \kappa}(z) + C_2 W_{\kappa, \kappa}(z)\right]  \qquad \text{for $m = 0$}\,,
\ee
where $M_{\kappa, \kappa}(z)$ and $W_{\kappa, \kappa}(z)$ are the Whittaker functions and we defined
\be 
\kappa = \frac{ i \omega r_0}{a} \,, \qquad z = - \frac{2 i \omega}{a} (r-r_0)  \, .
\ee
As it will be useful below, we note that
\be 
e^{\pm i \omega r_*} \propto e^{\mp z/2} z^{\pm \kappa} \, .
\ee

\paragraph{Large-$r$ behavior.} For large $|z|$ we have
\be 
W_{\kappa, \kappa}(z) \sim e^{-z/2}z^\kappa  \propto e^{+i \omega r_*} \, .
\ee
This part of the solution represents outgoing waves at infinity. On the other hand, we have
\be 
M_{\kappa,\kappa}(z) \sim \frac{\Gamma(1+ 2 \kappa)}{\Gamma\left(\frac{1}{2} \right)} e^{z/2}z^{-\kappa} + e^{\pm i \frac{\pi}{2}} \frac{\Gamma(1+2\kappa)}{\Gamma\left(\frac{1}{2} + 2 \kappa \right)} e^{-z/2}z^\kappa \, .
\ee 
The first term represents incoming waves, while the second represents outgoing waves.\footnote{The $e^{\pm i \pi/2}$ term depends on ${\rm Arg}(z)$, but it plays no role in our analysis here.} Because the Gamma function has no zeroes, it is not possible to set the ingoing contribution to zero through a judicious choice of $\kappa$. Hence, the QNM boundary conditions require us to set $C_1 = 0$ in~\eqref{eq:R_flat_exact}.

\paragraph{Near-horizon behavior.} The horizon corresponds to $|z| \to 0$. In this limit, after setting $C_1 = 0$ in~\eqref{eq:R_flat_exact}, we have
\be 
R(r) \sim \frac{\Gamma(-2 \kappa)}{\Gamma \left(\frac{1}{2} - 2 \kappa \right)} z^\kappa + \frac{\Gamma(2 \kappa)}{\Gamma \left( \frac{1}{2} \right) } z^{-\kappa} \, .
\ee
Because the horizon is at $|z| \to 0$, we have $e^{\pm i \omega r_*} \propto e^{\mp z/2} z^{\pm \kappa} \to z^{\pm \kappa} $. Hence the first term in the expansion above constitutes outgoing waves and must be set to zero. The Gamma function has poles at nonpositive integer arguments, and hence this term can be consistently set to zero by demanding
\be \label{eq:flat_QNM_quant_condition}
\frac{1}{2} - 2 \kappa = 1 -n  \, , \qquad n =  1, 2, \dots \, ,
\ee
where we choose our conventions for $n$ to match the monodromy calculation in the main text. Crucially, $\Gamma(2\kappa) = \Gamma(n-\tfrac{1}{2})$ is finite for these values of $n$ and we therefore have a consistent solution to the QNM problem. Translating the result in Eq.~\eqref{eq:flat_QNM_quant_condition} back to $\omega$, we obtain
\be 
\omega_n = - \frac{i a}{4 r_0}\left(2n - 1\right)\,, \qquad n = 1, 2, \dots \, .
\ee

\paragraph{Exact QNM frequencies.}

Expressed in terms of the notation of Eq.~\eqref{eq:charged-qschw-mass-form-zoo} with $Q = 0$, the QNM frequencies become
\be 
\omega_n = - i \, 2 \pi  T_{\rm H}\left(n - \frac{1}{2}\right)\,, \qquad n = 1, 2, \dots \,, 
\ee
where the Hawking temperature is
\be \label{eq:qSchw_hawking_temp}
T_{\rm H} = \frac{\left(1-8 G_3 M \right)^2}{4 \pi \ell W(M)} = \frac{\kappa_h}{2 \pi } \, .
\ee
These are the exact $m= 0$ QNM frequencies for the quantum Schwarzschild black hole.

\subsection{Large $|m|$ modes: WKB approach}

In the previous subsection we saw that the $m = 0$ modes of the quantum Schwarzschild black hole can be obtained exactly. For $m \neq 0$, the radial equation becomes of confluent Heun type and analytical progress is limited. We can get an analytical handle on the QNMs by using approximation techniques. Here we wish to understand the behavior of the QNMs for $|m| \neq 0$, and we will use a WKB approximation in the limit of large $|m|$. Note that here we are \textit{not} working in the limit of large overtone number, but instead in the eikonal limit. 

The potential that enters the radial equation has a maximum. Following very well-known techniques~\cite{Schutz:1985km} we can perform a WKB approximation around the maximum of the potential. Let $r_p > r_0$ be the location of the maximum of the potential $V$. Then, the leading-order WKB approximation gives 
\be 
\omega^{(\pm)}_{m,n} = \pm \sqrt{V(r_p)} - i \left(n - \frac{1}{2} \right) f(r_p) \sqrt{-\frac{V''(r_p)}{2 V(r_p)} } \, ,
\ee
where $V''(r) = {\rm d}^2 V/{\rm d}r^2$ and $n = 1, 2, \dots$.  While the above formula makes no assumption on $|m|$, the approximation is controlled only in the limit $|m| \to \infty$~\cite{Iyer:1986np}. Note that the assumptions made in the derivation of this formula mean that it does not apply for asymptotically AdS black holes~\cite{Cardoso:2008bp}, but it can be applied to our asymptotically flat quantum black holes, with or without charge.

For the quantum Schwarzschild black hole we have
\be 
V = \frac{a^2 (r-r_0)(3r_0 - r)}{4 r^4} + \frac{m^2 f(r)}{r^2} \, .
\ee
For large $|m|$, the potential has a maximum at 
\be 
r_p = \frac{3 r_0}{2} - \frac{a r_0}{8 m^2} + O(m^{-4}) \, ,
\ee
coinciding with the photon `circle' of the solution as $|m| \to \infty$. We then have
\be 
V(r_p) = \frac{4 a m^2}{27 r_0^2} + O(1) \, , \qquad V''(r_p) = - \frac{32 a m^2}{81 r_0^4} + O(1) \, .
\ee
Hence, the large $|m|$ spectrum is given by
\be 
\omega^{(\pm)}_{m,n} = \frac{8 \pi T_{\rm H}}{3 \sqrt{3}} \left[ \pm \frac{|m|}{\sqrt{1 - 8 G_3 M}} - i \left(n - \frac{1}{2} \right) \right] + O(|m|^{-1}) \, ,
\ee
where the Hawking temperature was given above in Eq.~\eqref{eq:qSchw_hawking_temp}. 
From the analysis in this appendix, we therefore see that while the $m = 0$ frequencies are purely imaginary, modes with sufficiently large $|m|$ will generically have both real and imaginary parts.

\section{Convergence of Eq.~\eqref{eq:Delta_gen_contour}}
\label{sec:convergence_suffering}

In this appendix we analyze the convergence properties of the integral defined in Eq.~\eqref{eq:Delta_gen_contour}, which we repeat here for convenience:
\be 
\Delta_{\sigma \tau}(q) = \int_\Gamma {\rm d}t \, t^{q-1} J_{\sigma \nu}(t) J_{\tau \nu}(t) \, . 
\ee
The contour runs from $+\infty$ to $\epsilon$ along the positive real axis, follows the semicircle $t=\epsilon e^{i\theta}$ with $0\leq\theta\leq\pi$, and then runs from $-\epsilon$ to $-\infty$ along the negative real axis. At the end, we take $\epsilon \to 0$ if the limit exists. In the case where the limit exists, the result of the integrals on $t \in [0, \infty)$ is given by the formula \eqref{eq:Weber-Schafheitlin}, which requires that $2 \nu < q < 1$.

\paragraph{Breaking up the contour.} Let us keep $\epsilon > 0$. For the sake of notational precision, let us write 
\be 
\Delta_{\sigma \tau}(\epsilon; q) = \Delta^{\Gamma_1}_{\sigma \tau}(\epsilon; q) + \Delta^{\Gamma_2}_{\sigma \tau}(\epsilon; q)+ \Delta_{\sigma \tau}^{\Gamma_3}(\epsilon; q) \, .
\ee
Here we use $\Gamma_1$ to denote the $+\infty \to \epsilon$ portion of the contour, $\Gamma_2$ to denote the semicircle contribution, and $\Gamma_3$ to denote the $-\epsilon \to -\infty$ portion. Define the auxiliary integral
\be 
I_{\sigma \tau}(\epsilon; q) = \int_\epsilon^\infty {\rm d}t \, t^{q-1} J_{\sigma \nu}(t) J_{\tau \nu}(t) \, .
\ee
At large positive argument, the Bessel functions decay as $t^{-1/2}$, so these integrals converge at infinity for $q < 1$. The first semi-infinite portion of the contour then contributes $\Delta_{\sigma \tau}^{\Gamma_1}(\epsilon; q) = - I_{\sigma \tau}(\epsilon; q)$. On the second semi-infinite portion we use $J_{\sigma \nu}(e^{i \pi} t) = e^{i \pi \sigma \nu} J_{\sigma \nu}(t)$ along with the phase of the integration measure to obtain 
\be 
\Delta_{\sigma \tau}^{\Gamma_3}(\epsilon; q) = e^{i \pi k_{\sigma \tau}} I_{\sigma \tau}(\epsilon; q) \,, \qquad k_{\sigma\tau} \equiv q + (\sigma +\tau) \nu \, .
\ee
We therefore have 
\be 
\Delta_{\sigma \tau}(\epsilon; q) = \left(e^{i \pi k_{\sigma \tau}}-1 \right) I_{\sigma \tau}(\epsilon; q) + \Delta^{\Gamma_2}_{\sigma \tau}(\epsilon; q) \, .
\ee
Whether or not the semicircle contribution can be discarded depends on the behavior of the integrand near the origin. 

Using the small argument expansion of the Bessel functions gives
\be 
t^{q-1}  J_{\sigma \nu}(t) J_{\tau \nu}(t) = A_{\sigma \tau} t^{k_{\sigma \tau} - 1} \left[1 + O(t^2) \right] \,, \quad A_{\sigma \tau} = \frac{2^{-(\sigma +\tau) \nu}}{\Gamma(1+\sigma \nu) \Gamma(1+\tau\nu) } \, .
\ee
Substituting $t=\epsilon e^{i \theta}$ we therefore find when $k_{\sigma \tau} \neq 0$ 
\begin{align}
\Delta^{\Gamma_2}_{\sigma \tau}(\epsilon; q) &= i A_{\sigma \tau} \epsilon^{k_{\sigma \tau}} \int_0^\pi {\rm d}\theta \, e^{i k_{\sigma \tau} \theta} + O(\epsilon^{k_{\sigma \tau}+2})
\\ 
&= A_{\sigma \tau} \epsilon^{k_{\sigma \tau}} \frac{e^{i \pi k_{\sigma \tau}} - 1}{k_{\sigma \tau} } + O(\epsilon^{k_{\sigma \tau}+2})  \, . \label{eq:semicircle_dies}
\end{align} 
In the case $k_{\sigma \tau} = 0$ the calculation instead gives
\be 
\Delta^{\Gamma_2}_{\sigma \tau}(\epsilon; q) = i \pi A_{\sigma \tau} + O(\epsilon^2) \, .
\ee
Therefore, in the case $k_{\sigma \tau} = 0$ the semicircle need not give a vanishing contribution. 

\paragraph{The case $2 \nu < q < 1$. } In this regime all three exponents $k_{++} = q + 2 \nu$, $k_{+-} = q$, and $k_{--} = q - 2\nu$ are positive. The integrals along the real axis therefore converge at the origin, and Eq.~\eqref{eq:semicircle_dies} shows that the semicircle contribution vanishes in the $\epsilon \to 0$ limit in all cases. We then obtain 
\be 
\Delta_{\sigma \tau}(q) = \left(e^{i \pi k_{\sigma \tau}}-1 \right) I_{\sigma \tau}(q) \,, 
\ee
with $I_{\sigma \tau}(q)$ given as in Eq.~\eqref{eq:Weber-Schafheitlin}.

\paragraph{The case $q = 2 \nu$. } This is the case relevant to the angular momentum correction. The $++$ and $+-$ integrals are convergent at the origin and can be treated exactly as just above. The only problematic term is the $--$ one, for which the small argument expansion of the integrand becomes
\be 
t^{2\nu-1}  J_{- \nu}(t) J_{-\nu}(t) = \frac{A_{--}}{t} + O(t) \,, \quad A_{--} = \frac{2^{2\nu}}{\Gamma^2(1-\nu)} \, .
\ee
Consequently, $I_{--}(\epsilon;q) = A_{--} \log (1/\epsilon) + O(1)$ is now logarithmically divergent as $\epsilon \to 0$. However, for $q = 2\nu$ the relative phase between $\Delta_{--}^{\Gamma_1}(\epsilon; q)$ and $\Delta_{--}^{\Gamma_3}(\epsilon; q)$ is unity and the integrals cancel pointwise. More precisely, we have 
\be 
\left(e^{i \pi k_{--}}-1 \right) I_{--}(\epsilon; q) \to  \left(e^{i \pi 0}-1 \right) I_{--}(\epsilon; q) = 0 \, ,
\ee
and the entire integral comes from the semicircle,
\be 
\Delta_{--}(2 \nu) = \lim_{\epsilon \to 0} \Delta^{\Gamma_2}_{--}(\epsilon; 2\nu) = i \pi  A_{--} = \frac{i \pi 2^{2\nu}}{\Gamma^2(1-\nu)} \, .
\ee

\paragraph{The case $0 < q < 2 \nu$.} In some cases evaluating the metric corrections requires evaluating the integrals in this domain of $q$. The only problematic term is again the $--$ one. Let us set $k = q - 2 \nu$, so that $-1 < k < 0$. For this case, the leading term in the small $t$ expansion of the integrand produces a power law divergence. However, there is only a single divergent term, and the first subleading contribution is integrable. We can therefore define the finite part of the integral as
\be 
I_{--}^{\rm fp}(q) \equiv \lim_{\epsilon \to 0} \left[I_{--}(\epsilon;q) + \frac{A_{--}}{k} \epsilon^k \right] \, .
\ee
Combining the full finite-$\epsilon$ contour before taking the limit we have
\begin{align}
\Delta_{--}(\epsilon; q) &= \left(e^{i \pi k} - 1 \right) \left[I_{--}^{\rm fp}(q) - \frac{A_{--}}{k} \epsilon^k \right] + \frac{A_{--}}{k} \epsilon^k \left(e^{i \pi k} -1 \right) + O\left( \epsilon^{k+2} \right) \,
\\
    &= \left(e^{i \pi k} - 1 \right) I_{--}^{\rm fp}(q) + O\left( \epsilon^{k+2} \right)\, .
\end{align} 
Thus the divergent semicircle contribution cancels the divergence of $I_{--}(\epsilon; q)$ along the combined semi-infinite portions $\Gamma_{1,3}$ and what remains is finite. More specifically, the finite part coincides with the analytic continuation of the Weber--Schafheitlin expression:
\be
I_{--}^{\rm fp}(q)
=
\frac{
2^{q-1}\Gamma(1-q)\Gamma(q/2-\nu)
}{
\Gamma^2(1-q/2)\Gamma(1-\nu-q/2)
}
\,,
\qquad 0<q<2\nu<1 \, .
\ee
Consequently,
\be
\Delta_{--}(q)
=
\left[e^{i\pi(q-2\nu)}-1\right]I_{--}^{\rm fp}(q)
\, .
\ee

\bibliography{qBHrefs}

\end{document}